\documentclass[longauth]{aa}  

\usepackage{graphicx}
\usepackage{txfonts}
\usepackage[dvipsnames]{xcolor}
\usepackage{url}
\usepackage{hyperref}
\usepackage[yyyymmdd]{datetime}
\usepackage{rotating}
\usepackage{tablefootnote}

\usepackage{amsmath,amssymb}
\hypersetup{
    colorlinks=true,
    linkcolor=Blue,
    urlcolor=Blue,
    citecolor=Blue
}

\def\deg{\ensuremath{^\circ}}

\def\arcsec{\ensuremath{''}}

\providecommand{\bcg}{\ensuremath{\mathrm{BC}_G}\xspace}
\providecommand{\bcgsol}{\ensuremath{\mathrm{BC}_{G\odot}}\xspace}

\providecommand{\dist}{\ensuremath{r}\xspace} 

\providecommand{\teff}{\ensuremath{T_{\mathrm{{eff}}}}\xspace}

\providecommand{\lum}{\ensuremath{{L}}\xspace}

\providecommand{\logg}{\ensuremath{\log\,g}\xspace}
\providecommand{\mass}{\ensuremath{{M}}\xspace}
\providecommand{\radius}{\ensuremath{{R}}\xspace}

\providecommand{\mh}{[M/H]\xspace}
\providecommand{\feh}{[Fe/H]\xspace}
\providecommand{\vsini}{\ensuremath{v\sin i}\xspace}

\providecommand{\azero}{\ensuremath{A_0}\xspace}
\providecommand{\ag}{\ensuremath{A_G}\xspace}
\providecommand{\abp}{\ensuremath{A_\mathrm{BP}}\xspace}
\providecommand{\arp}{\ensuremath{A_\mathrm{RP}}\xspace}
\providecommand{\av}{\ensuremath{A_\mathrm{V}}\xspace}

\newcommand{\bpminrp}{\ensuremath{G_\mathrm{BP}-G_\mathrm{RP}}\xspace}
\providecommand{\ebpminrp}{\ensuremath{E(G_{\rm BP} - G_{\rm RP})}\xspace}
\providecommand{\gmag}{\ensuremath{G}}
\providecommand{\bpmag}{\ensuremath{G_\mathrm{BP}}}
\providecommand{\rpmag}{\ensuremath{G_\mathrm{RP}}}
\providecommand{\mg}{M$_\gmag$}
\providecommand{\gravshift}{\ensuremath{rv_{\rm GR}}}

\providecommand{\parallax}{\ensuremath{\varpi}}

\providecommand{\Msun}{\ensuremath{\,{\mass}_{\odot}}\xspace}

\providecommand{\hpix}{HEALPix}
\providecommand{\cairt}{\ion{Ca}{ii}\,IRT\xspace}

\providecommand{\modulename}[1]{#1\xspace}
\providecommand{\apsis}{\modulename{Apsis}}
\providecommand{\smsgen}{\modulename{SMSgen}}
\providecommand{\dsc}{\modulename{DSC}}
\providecommand{\gspphot}{\modulename{GSP-Phot}}

\providecommand{\gspspec}{\modulename{GSP-Spec}}
\providecommand{\msc}{\modulename{MSC}}
\providecommand{\flame}{\modulename{FLAME}}
\providecommand{\espels}{\modulename{ESP-ELS}}
\providecommand{\esphs}{\modulename{ESP-HS}}
\providecommand{\espcs}{\modulename{ESP-CS}}
\providecommand{\espucd}{\modulename{ESP-UCD}}
\providecommand{\ugc}{\modulename{UGC}}
\providecommand{\oa}{\modulename{OA}}

\providecommand{\qsoc}{\modulename{QSOC}}
\providecommand{\tge}{\modulename{TGE}}
\providecommand{\mgalgo}{Matisse-Gauguin}
\providecommand{\basti}{BaSTI}

\providecommand\gaia{\textit{Gaia}\xspace}
\providecommand{\gdr}[1]{Gaia~DR{#1}}
\providecommand{\gedr}[1]{Gaia~eDR{#1}}

\newcommand{\bporrp}{BP/RP\xspace}
\newcommand{\orcit}[1]{\protect\href{https://orcid.org/#1}{\protect\includegraphics[width=8pt]{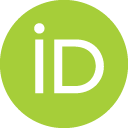}}}

\providecommand{\red}{\textcolor{red}}

\providecommand{\totalfields}{{538}\xspace}
\providecommand{\numberuniqueparameters}{{43}\xspace}

\providecommand{\linktodoc}{https://gea.esac.esa.int/archive/documentation/GDR3}

\providecommand{\linktoapparam}[2]{\href{\linktodoc/Gaia_archive/chap_datamodel/sec_dm_astrophysical_parameter_tables/ssec_dm_#1.html\##1-#2}{\fieldName{#2}\xspace}}
\providecommand{\linktogalparam}[2]{\href{\linktodoc/Gaia_archive/chap_datamodel/sec_dm_extra--galactic_tables/ssec_dm_#1.html\##1-#2}{\fieldName{#2}\xspace}}
\providecommand{\linktogsparam}[2]{\href{\linktodoc/Gaia_archive/chap_datamodel/sec_dm_main_source_tables/ssec_dm_#1.html\##1-#2}{\fieldName{#2}\xspace}}

\providecommand{\linktotable}[1]{\href{\linktodoc/Gaia_archive/chap_datamodel/sec_dm_astrophysical_parameter_tables/ssec_dm_#1.html}{\fieldName{#1}\xspace}}
\providecommand{\aptable}{\linktotable{astrophysical_parameters}}
\providecommand{\apsupptable}{\linktotable{astrophysical_parameters_supp}}

\providecommand{\linktogaltable}[1]{\href{\linktodoc/Gaia_archive/chap_datamodel/sec_dm_extra--galactic_tables/ssec_dm_#1.html}{\fieldName{#1}\xspace}}

\providecommand{\linktogstable}[1]{\href{\linktodoc/Gaia_archive/chap_datamodel/sec_dm_main_source_catalogue/ssec_dm_#1.html}{\fieldName{#1}\xspace}}

\providecommand{\linksubsec}[2]{\href{\linktodoc/Data_analysis/chap_cu8par/#1.html}{#2\xspace}}
\providecommand{\linksec}[2]{\href{\linktodoc/Data_analysis/chap_cu8par/#1}{#2\xspace}}

\providecommand{\linkfig}[1]{\href{\linktodoc/Data_analysis/chap_cu8par/#1.html}{see table\xspace}}

\makeatletter
\DeclareRobustCommand*{\fieldName}[1]{%
  \begingroup\@fieldName\scantokens{\texttt{\small {#1}}\noexpand}\endgroup}
\begingroup\lccode`\~=`\_\relax
   \lowercase{\endgroup\def\@fieldName{\catcode`\_=\active \let~\_}}
\makeatother

\begin{document} 

\title{Gaia Data Release 3: Astrophysical parameters inference system (Apsis) I - methods and content overview}
\titlerunning{Overview of astrophysical parameters in \gdr{3}}

\author{
       O.L.~                       Creevey\orcit{0000-0003-1853-6631}\inst{\ref{inst:0001}}
\and         R.~                         Sordo\orcit{0000-0003-4979-0659}\inst{\ref{inst:0002}}
\and         F.~                       Pailler\orcit{0000-0002-4834-481X}\inst{\ref{inst:0003}}
\and         Y.~                    Fr\'{e}mat\orcit{0000-0002-4645-6017}\inst{\ref{inst:0004}}
\and         U.~                        Heiter\orcit{0000-0001-6825-1066}\inst{\ref{inst:0005}}
\and         F.~                  Th\'{e}venin\inst{\ref{inst:0001}}
\and         R.~                        Andrae\orcit{0000-0001-8006-6365}\inst{\ref{inst:0007}}
\and         M.~                     Fouesneau\orcit{0000-0001-9256-5516}\inst{\ref{inst:0007}}
\and         A.~                         Lobel\orcit{0000-0001-5030-019X}\inst{\ref{inst:0004}}
\and     C.A.L.~                  Bailer-Jones\inst{\ref{inst:0007}}
\and         D.~                      Garabato\orcit{0000-0002-7133-6623}\inst{\ref{inst:0011}}
\and         I.~                Bellas-Velidis\inst{\ref{inst:0012}}
\and         E.~                    Brugaletta\orcit{0000-0003-2598-6737}\inst{\ref{inst:0013}}
\and         A.~                         Lorca\inst{\ref{inst:0014}}
\and         C.~                     Ordenovic\inst{\ref{inst:0001}}
\and       P.A.~                       Palicio\orcit{0000-0002-7432-8709}\inst{\ref{inst:0001}}
\and       L.M.~                         Sarro\orcit{0000-0002-5622-5191}\inst{\ref{inst:0017}}
\and         L.~                    Delchambre\orcit{0000-0003-2559-408X}\inst{\ref{inst:0018}}
\and         R.~                       Drimmel\orcit{0000-0002-1777-5502}\inst{\ref{inst:0019}}
\and         J.~                       Rybizki\orcit{0000-0002-0993-6089}\inst{\ref{inst:0007}}
\and         G.~                Torralba Elipe\orcit{0000-0001-8738-194X}\inst{\ref{inst:0011}}
\and       A.J.~                          Korn\orcit{0000-0002-3881-6756}\inst{\ref{inst:0005}}
\and         A.~                  Recio-Blanco\orcit{0000-0002-6550-7377}\inst{\ref{inst:0001}}
\and       M.S.~                    Schultheis\orcit{0000-0002-6590-1657}\inst{\ref{inst:0001}}
\and         F.~                     De Angeli\orcit{0000-0003-1879-0488}\inst{\ref{inst:0025}}
\and         P.~                   Montegriffo\orcit{0000-0001-5013-5948}\inst{\ref{inst:0026}}
\and        
A.~                Abreu Aramburu\inst{\ref{inst:0027}}
\and       
S. Accart\inst{\ref{inst:0003}} \and
M.A.~                   \'{A}lvarez\orcit{0000-0002-6786-2620}\inst{\ref{inst:0011}}
\and J.~Bakker\inst{\ref{inst:esa}}
\and         N.~                     Brouillet\orcit{0000-0002-3274-7024}\inst{\ref{inst:0030}}
\and         A.~                       Burlacu\inst{\ref{inst:0031}}
\and         R.~                      Carballo\orcit{0000-0001-7412-2498}\inst{\ref{inst:0032}}
\and         L.~                   Casamiquela\orcit{0000-0001-5238-8674}\inst{\ref{inst:0030},\ref{inst:0034}}
\and         A.~                     Chiavassa\orcit{0000-0003-3891-7554}\inst{\ref{inst:0001}}
\and         G.~                      Contursi\orcit{0000-0001-5370-1511}\inst{\ref{inst:0001}}
\and       W.J.~                        Cooper\orcit{0000-0003-3501-8967}\inst{\ref{inst:0038},\ref{inst:0019}}
\and         C.~                       Dafonte\orcit{0000-0003-4693-7555}\inst{\ref{inst:0011}}
\and         A.~                    Dapergolas\inst{\ref{inst:0012}}
\and         P.~                    de Laverny\orcit{0000-0002-2817-4104}\inst{\ref{inst:0001}}
\and       T.E.~                 Dharmawardena\orcit{0000-0002-9583-5216}\inst{\ref{inst:0007}}
\and         B.~                    Edvardsson\inst{\ref{inst:0045}}
Y. Le Fustec \inst{\ref{inst:0031}} 
\and         P.~              Garc\'{i}a-Lario\orcit{0000-0003-4039-8212}\inst{\ref{inst:0046}}
\and         M.~             Garc\'{i}a-Torres\orcit{0000-0002-6867-7080}\inst{\ref{inst:0047}}
\and         A.~                         Gomez\orcit{0000-0002-3796-3690}\inst{\ref{inst:0011}}
\and         I.~   Gonz\'{a}lez-Santamar\'{i}a\orcit{0000-0002-8537-9384}\inst{\ref{inst:0011}}
\and         D.~                Hatzidimitriou\orcit{0000-0002-5415-0464}\inst{\ref{inst:0051},\ref{inst:0012}}
\and         A.~          Jean-Antoine Piccolo\orcit{0000-0001-8622-212X}\inst{\ref{inst:0003}}
\and         M.~                      Kontizas\orcit{0000-0001-7177-0158}\inst{\ref{inst:0051}}
\and         G.~                    Kordopatis\orcit{0000-0002-9035-3920}\inst{\ref{inst:0001}}
\and       A.C.~                     Lanzafame\orcit{0000-0002-2697-3607}\inst{\ref{inst:0013},\ref{inst:0057}}
\and         Y.~                      Lebreton\orcit{0000-0002-4834-2144}\inst{\ref{inst:0058},\ref{inst:0059}}
\and       E.L.~                        Licata\orcit{0000-0002-5203-0135}\inst{\ref{inst:0019}}
\and     H.E.P.~                  Lindstr{\o}m\inst{\ref{inst:0019},\ref{inst:0062},\ref{inst:0063}}
\and         E.~                       Livanou\orcit{0000-0003-0628-2347}\inst{\ref{inst:0051}}
\and         A.~               Magdaleno Romeo\inst{\ref{inst:0031}}
\and         M.~                      Manteiga\orcit{0000-0002-7711-5581}\inst{\ref{inst:0066}}
\and         F.~                       Marocco\orcit{0000-0001-7519-1700}\inst{\ref{inst:0067}}
\and       D.J.~                      Marshall\orcit{0000-0003-3956-3524}\inst{\ref{inst:0068}}
\and         N.~                          Mary\inst{\ref{inst:0069}}
\and         C.~                       Nicolas\inst{\ref{inst:0003}}
\and         L.~               Pallas-Quintela\orcit{0000-0001-9296-3100}\inst{\ref{inst:0011}}
\and         C.~                         Panem\inst{\ref{inst:0003}}
\and         B.~                        Pichon\orcit{0000 0000 0062 1449}\inst{\ref{inst:0001}}
\and         E.~                        Poggio\orcit{0000-0003-3793-8505}\inst{\ref{inst:0001},\ref{inst:0019}}
\and         F.~                        Riclet\inst{\ref{inst:0003}}
\and         C.~                         Robin\inst{\ref{inst:0069}}
\and         R.~                 Santove\~{n}a\orcit{0000-0002-9257-2131}\inst{\ref{inst:0011}}
\and         A.~                       Silvelo\orcit{0000-0002-5126-6365}\inst{\ref{inst:0011}}
\and         I.~                        Slezak\inst{\ref{inst:0001}}
\and       R.L.~                         Smart\orcit{0000-0002-4424-4766}\inst{\ref{inst:0019}}
\and         C.~                      Soubiran\orcit{0000-0003-3304-8134}\inst{\ref{inst:0030}}
\and         M.~                  S\"{ u}veges\orcit{0000-0003-3017-5322}\inst{\ref{inst:0084}}
\and         A.~                          Ulla\orcit{0000-0001-6424-5005}\inst{\ref{inst:0085}}
\and E.~Utrilla\inst{\ref{inst:0014}}
\and         A.~                     Vallenari\orcit{0000-0003-0014-519X}\inst{\ref{inst:0002}}
\and         H.~                          Zhao\orcit{0000-0003-2645-6869}\inst{\ref{inst:0001}}
\and         J.~                         Zorec\inst{\ref{inst:0089}} \and 
D. Barrado\inst{\ref{inst:barrado}} \and 
A. Bijaoui\inst{\ref{inst:0001}} \and  
J.-C. Bouret\inst{\ref{inst:bouret}} \and  
R. Blomme\inst{\ref{inst:0004}} \and  
I. Brott\inst{\ref{inst:brott}} \and  
S. Cassisi\inst{\ref{inst:cassisi1},\ref{inst:cassisi2}} \and  
O. Kochukhov\inst{\ref{inst:0005}} \and  
C. Martayan\inst{\ref{inst:martayan},\ref{inst:0004}} \and  
D. Shulyak\inst{\ref{inst:shulyak1},\ref{inst:shulyak2}} \and  
J. Silvester\inst{\ref{inst:0005}}  
}
\institute{
     Universit\'{e} C\^{o}te d'Azur, Observatoire de la C\^{o}te d'Azur, CNRS, Laboratoire Lagrange, Bd de l'Observatoire, CS 34229, 06304 Nice Cedex 4, France\relax                                                                                                                                                                                              \label{inst:0001}
\and INAF - Osservatorio astronomico di Padova, Vicolo Osservatorio 5, 35122 Padova, Italy\relax                                                                                                                                                                                                                                                                   \label{inst:0002}\vfill
\and CNES Centre Spatial de Toulouse, 18 avenue Edouard Belin, 31401 Toulouse Cedex 9, France\relax                                                                                                                                                                                                                                                                \label{inst:0003}\vfill
\and Royal Observatory of Belgium, Ringlaan 3, 1180 Brussels, Belgium\relax                                                                                                                                                                                                                                                                                        \label{inst:0004}\vfill
\and Observational Astrophysics, Division of Astronomy and Space Physics, Department of Physics and Astronomy, Uppsala University, Box 516, 751 20 Uppsala, Sweden\relax                                                                                                                                                                                           \label{inst:0005}\vfill
\and Max Planck Institute for Astronomy, K\"{ o}nigstuhl 17, 69117 Heidelberg, Germany\relax                                                                                                                                                                                                                                                                       \label{inst:0007}\vfill
\and CIGUS CITIC - Department of Computer Science and Information Technologies, University of A Coru\~{n}a, Campus de Elvi\~{n}a s/n, A Coru\~{n}a, 15071, Spain\relax                                                                                                                                                                                             \label{inst:0011}\vfill
\and National Observatory of Athens, I. Metaxa and Vas. Pavlou, Palaia Penteli, 15236 Athens, Greece\relax                                                                                                                                                                                                                                                         \label{inst:0012}\vfill
\and INAF - Osservatorio Astrofisico di Catania, via S. Sofia 78, 95123 Catania, Italy\relax                                                                                                                                                                                                                                                                       \label{inst:0013}\vfill
\and Aurora Technology for European Space Agency (ESA), Camino bajo del Castillo, s/n, Urbanizacion Villafranca del Castillo, Villanueva de la Ca\~{n}ada, 28692 Madrid, Spain\relax                                                                                                                                                                               \label{inst:0014}\vfill
\and Dpto. de Inteligencia Artificial, UNED, c/ Juan del Rosal 16, 28040 Madrid, Spain\relax                                                                                                                                                                                                                                                                       \label{inst:0017}\vfill
\and Institut d'Astrophysique et de G\'{e}ophysique, Universit\'{e} de Li\`{e}ge, 19c, All\'{e}e du 6 Ao\^{u}t, B-4000 Li\`{e}ge, Belgium\relax                                                                                                                                                                                                                    \label{inst:0018}\vfill
\and INAF - Osservatorio Astrofisico di Torino, via Osservatorio 20, 10025 Pino Torinese (TO), Italy\relax                                                                                                                                                                                                                                                         \label{inst:0019}\vfill
\and Institute of Astronomy, University of Cambridge, Madingley Road, Cambridge CB3 0HA, United Kingdom\relax                                                                                                                                                                                                                                                      \label{inst:0025}\vfill
\and INAF - Osservatorio di Astrofisica e Scienza dello Spazio di Bologna, via Piero Gobetti 93/3, 40129 Bologna, Italy\relax                                                                                                                                                                                                                                      \label{inst:0026}\vfill
\and ATG Europe for European Space Agency (ESA), Camino bajo del Castillo, s/n, Urbanizacion Villafranca del Castillo, Villanueva de la Ca\~{n}ada, 28692 Madrid, Spain\relax                                                                                                                                                                                      \label{inst:0027}\vfill
\and European Space Agency (ESA), European Space Astronomy Centre (ESAC), Camino bajo del Castillo, s/n, Urbanizacion Villafranca del Castillo, Villanueva de la Ca\~{n}ada, 28692 Madrid, Spain\relax                                                                                        \label{inst:esa}\vfill
\and Laboratoire d'astrophysique de Bordeaux, Univ. Bordeaux, CNRS, B18N, all{\'e}e Geoffroy Saint-Hilaire, 33615 Pessac, France\relax                                                                                                                                                                                                                             \label{inst:0030}\vfill
\and Telespazio for CNES Centre Spatial de Toulouse, 18 avenue Edouard Belin, 31401 Toulouse Cedex 9, France\relax                                                                                                                                                                                                                                                 \label{inst:0031}\vfill
\and Dpto. de Matem\'{a}tica Aplicada y Ciencias de la Computaci\'{o}n, Univ. de Cantabria, ETS Ingenieros de Caminos, Canales y Puertos, Avda. de los Castros s/n, 39005 Santander, Spain\relax                                                                                                                                                                   \label{inst:0032}\vfill
\and GEPI, Observatoire de Paris, Universit\'{e} PSL, CNRS, 5 Place Jules Janssen, 92190 Meudon, France\relax                                                                                                                                                                                                                                                      \label{inst:0034}\vfill
\and Centre for Astrophysics Research, University of Hertfordshire, College Lane, AL10 9AB, Hatfield, United Kingdom\relax                                                                                                                                                                                                                                         \label{inst:0038}\vfill
\and APAVE SUDEUROPE SAS for CNES Centre Spatial de Toulouse, 18 avenue Edouard Belin, 31401 Toulouse Cedex 9, France\relax                                                                                                                                                                                                                                        \label{inst:0043}\vfill
\and Theoretical Astrophysics, Division of Astronomy and Space Physics, Department of Physics and Astronomy, Uppsala University, Box 516, 751 20 Uppsala, Sweden\relax                                                                                                                                                                                             \label{inst:0045}\vfill
\and European Space Agency (ESA), European Space Astronomy Centre (ESAC), Camino bajo del Castillo, s/n, Urbanizacion Villafranca del Castillo, Villanueva de la Ca\~{n}ada, 28692 Madrid, Spain\relax                                                                                                                                                             \label{inst:0046}\vfill
\and Data Science and Big Data Lab, Pablo de Olavide University, 41013, Seville, Spain\relax                                                                                                                                                                                                                                                                       \label{inst:0047}\vfill
\and Department of Astrophysics, Astronomy and Mechanics, National and Kapodistrian University of Athens, Panepistimiopolis, Zografos, 15783 Athens, Greece\relax                                                                                                                                                                                                  \label{inst:0051}\vfill
\and Dipartimento di Fisica e Astronomia ""Ettore Majorana"", Universit\`{a} di Catania, Via S. Sofia 64, 95123 Catania, Italy\relax                                                                                                                                                                                                                               \label{inst:0057}\vfill
\and LESIA, Observatoire de Paris, Universit\'{e} PSL, CNRS, Sorbonne Universit\'{e}, Universit\'{e} de Paris, 5 Place Jules Janssen, 92190 Meudon, France\relax                                                                                                                                                                                                   \label{inst:0058}\vfill
\and Universit\'{e} Rennes, CNRS, IPR (Institut de Physique de Rennes) - UMR 6251, 35000 Rennes, France\relax                                                                                                                                                                                                                                                      \label{inst:0059}\vfill
\and Niels Bohr Institute, University of Copenhagen, Juliane Maries Vej 30, 2100 Copenhagen {\O}, Denmark\relax                                                                                                                                                                                                                                                    \label{inst:0062}\vfill
\and DXC Technology, Retortvej 8, 2500 Valby, Denmark\relax                                                                                                                                                                                                                                                                                                        \label{inst:0063}\vfill
\and CIGUS CITIC, Department of Nautical Sciences and Marine Engineering, University of A Coru\~{n}a, Paseo de Ronda 51, 15071, A Coru\~{n}a, Spain\relax                                                                                                                                                                                                          \label{inst:0066}\vfill
\and IPAC, Mail Code 100-22, California Institute of Technology, 1200 E. California Blvd., Pasadena, CA 91125, USA\relax                                                                                                                                                                                                                                           \label{inst:0067}\vfill
\and IRAP, Universit\'{e} de Toulouse, CNRS, UPS, CNES, 9 Av. colonel Roche, BP 44346, 31028 Toulouse Cedex 4, France\relax                                                                                                                                                                                                                                        \label{inst:0068}\vfill
\and Thales Services for CNES Centre Spatial de Toulouse, 18 avenue Edouard Belin, 31401 Toulouse Cedex 9, France\relax                                                                                                                                                                                                                                            \label{inst:0069}\vfill
\and Institute of Global Health, University of Geneva\relax                                                                                                                                                                                                                                                                                                        \label{inst:0084}\vfill
\and Applied Physics Department, Universidade de Vigo, 36310 Vigo, Spain\relax                                                                                                                                                                                                                                                                                     \label{inst:0085}\vfill
\and Sorbonne Universit\'{e}, CNRS, UMR7095, Institut d'Astrophysique de Paris, 98bis bd. Arago, 75014 Paris, France\relax         \label{inst:0089}\vfill
\and Departamento de Astrof\'{i}sica, Centro de Astrobiolog\'{i}a (CSIC-INTA), ESA-ESAC. Camino Bajo del Castillo s/n. 28692 Villanueva de la Ca\~{n}ada, Madrid, Spain\relax  \label{inst:barrado}\vfill
\and Aix Marseille Univ, CNRS, CNES, LAM, Marseille, France\relax\label{inst:bouret}\vfill 
\and Hamburger Sternwarte, Gojenbergsweg 112, 21029 Hamburg, Germany\label{inst:brott}\vfill
\and INAF-Osservatorio Astronomico d’Abruzzo, via M. Maggini, sn. 64100, Teramo, Italy\relax\label{inst:cassisi1}\vfill
\and INFN – Sezione di Pisa, Largo Pontecorvo 3, 56127 Pisa, Italy\relax\label{inst:cassisi2}\vfill
\and European Organisation for Astronomical Research in the Southern Hemisphere, Alonso de C\'{o}rdova 3107, Vitacura, 19001 Casilla, Santiago de Chile, Chile\relax \label{inst:martayan}\vfill
\and Institut für Astrophysik, Georg-August-Universit\"{a}t, D-37077 G\"{o}ttingen, Germany\relax\label{inst:shulyak1}\vfill
\and Instituto de Astrof\'{i}sica de Andaluc\'{i}a (Consejo Superior de Investigaciones Cient\'{i}ficas), E-18008 Granada, Spain\relax\label{inst:shulyak2}\vfill}


\date{Version: \today~\currenttime}
\abstract{
Gaia Data Release 3 contains a wealth of new data products for the community. 
Astrophysical parameters are a major component of this release.
They were produced by the Astrophysical parameters inference system (\apsis) within the Gaia Data Processing and Analysis Consortium.
The aim of this paper is to describe the overall content of the astrophysical parameters in \gdr{3}  and how they were produced.
In \apsis\ we use the mean BP/RP and mean RVS spectra along with astrometry and photometry, and we derive the following parameters: 
source classification and probabilities for 1.6 billion objects,  interstellar medium characterisation and distances
for up to 470 million sources, including a 2D total Galactic extinction map,  6 million redshifts of quasar candidates and 1.4 million redshifts of galaxy candidates,  along with an analysis of 50 million outlier sources through an unsupervised classification.
The astrophysical parameters also include many stellar spectroscopic and evolutionary parameters for up to 470 million sources.  These comprise \teff, \logg, and \mh (470 million using BP/RP, 6 million using RVS), radius (470 million), mass (140 million), age (120 million), 
chemical abundances (up to 5 million), diffuse interstellar band analysis (0.5 million), activity indices (2 million), H$\alpha$ equivalent widths (200 million), and further classification of spectral types (220 million) and emission-line stars (50 thousand).  
This paper is part of a series of three papers: this Paper I focusses on describing the global content of the parameters in \gdr{3}.  The accompanying papers II and III focus on the validation and use of the stellar and non-stellar products, respectively. This catalogue is the most extensive homogeneous database of astrophysical parameters to date, and it is based uniquely on Gaia data.  It will only be superseded in Gaia Data Release 4.  
This catalogue will therefore remain a key reference over the next four years, providing astrophysical parameters independent of other ground- and space-based data.
}


\maketitle
\section{Introduction}\label{sec:introduction}

Physical characterisation of astrophysical objects is a key input for understanding the 
structure and evolution of astrophysical systems.  
By physical characterisation we mean intrinsic properties for a stellar object such as its effective temperature \teff, age, and chemical element composition, as well as other inferred properties such as redshifts of distant sources and object classification. All of these parameters, collectively, we refer to as astrophysical parameters (APs). In the context of \gaia \citep{DR1-DPACP-8,DR2-DPACP-36,DR3-DPACP-185}, APs are complementary to multi-dimensional position and velocity information for achieving a better understanding of the dynamical evolution of the Milky Way.  Characterisation of a significant sample of our Galaxy's stars  also allows studies of individual stellar populations, stellar systems including planets, and a better understanding of the structure and properties of stars themselves.  
\gaia also observes objects both within our own solar system as well as beyond the Milky Way, and characterising these objects in a homogeneous way promises to open new windows \citep{DR3-DPACP-101, DR3-DPACP-150, DR3-DPACP-153}.

The \gaia Data Processing and Analysis Consortium (\gaia-DPAC) is tasked with the analysis of \gaia data to provide a catalogue of astrometric, photometric, and spectroscopic data to the public.  The role of the Coordination Unit 8 (CU8),  "Astrophysical Parameters", is to provide a catalogue of derived APs to the community based on the {\it mean} astrometric, photometric and spectroscopic data.  The Astrophysical Parameters Inference System (Apsis) is the pipeline that was designed and is executed at the Data Processing Center CNES (DPCC), Toulouse, France, which produces APs for all sources in the \gaia catalogue.   These APs are not only destined for \gaia releases, but they are also used internally in DPAC systems, e.g. for determining the radial velocity (RV) template in the RV data reduction and analysis \citep{DR2-DPACP-47}.

The CU8 Apsis pipeline was first described in \citet{Apsis2013} before the launch of \gaia.  Apsis comprises 13 modules that use different input data and/or models to produce APs for sub-stellar objects, stars, galaxies, and quasars.
In \gaia Data Release 2, only two of the thirteen modules processed data to produce five stellar parameters (\teff, extinction \ag, colour-excess \ebpminrp, radius \radius, luminosity \lum) based on parallaxes and integrated photometry \citep{DR2-DPACP-43}.   
Now, in \gaia Data Release 3, all of the 13 Apsis modules processed data and have contributed to the catalogue to provide \numberuniqueparameters primary APs along with auxiliary parameters that appear in a total of  \totalfields archive fields.   

APs produced by CU8 appear in ten tables of the \gaia archive, with a subset of these also appearing in {\tt gaia\_source}.  These data comprise both individual parameters (in four tables) and multi-dimensional data (in six tables).   The individual parameters are properties such as atmospheric properties, evolutionary parameters, chemical element abundances, and extinction parameters for stars,  
along with class probabilities and redshifts of distant sources.
The multi-dimensional data comprise a self-organised map of outliers, with prototype spectra, a 2D total galactic extinction map at four healpix levels as well as an optimal-level map, and Markov Chain Monte Carlo samples for two of the Apsis modules.

This paper is the first in a series of three papers. The goal of the present paper is to describe the production and content of the CU8 data products available in the \gdr{3} archive.  More details on the validation and use of the stellar and non-stellar products can be found in the accompanying papers II \citep{dr3-dpacp-160} and III \citep{DR3-DPACP-158} respectively. More complete descriptions of specific products and methods can be found in the following papers: \citet{2018MNRAS.473.1785D, DR3-DPACP-156,DR3-DPACP-175} and \cite{DR3-DPACP-186}, and the \href{https://gea.esac.esa.int/archive/documentation/GDR3}{official online documentation}\footnote{\url{https://gea.esac.esa.int/archive/documentation/GDR3}}.
The AP content of \gaia DR3 represents one of the most extensive homogeneous databases of APs to date for exploitation in many domains of astrophysics, 
see e.g. \cite{DR3-DPACP-101,DR3-DPACP-123,DR3-DPACP-75,DR3-DPACP-104,DR3-DPACP-144}.

This paper is structured as follows: Sections~\ref{sec:data}, \ref{sec:methods}, and \ref{sec:models} describe the input data, provide an overview of the methods used in \apsis, and describe the stellar models, respectively.  
Section \ref{sec:results} describes the 
general content and scope of the 10 \gdr{3} archive tables with CU8 parameters and contains useful reference tables for guidance, while  
Section~\ref{sec:content} describes all of the APs from CU8 grouped by astrophysical category.
An overview of the validation process is described in  Section~\ref{sec:validation}, while readers are referred to the  
accompanying papers II  \citep{dr3-dpacp-160} and III \citep{DR3-DPACP-158} 
for detailed validation of the \apsis\ results.
In Section~\ref{sec:discussioncaveats} we describe the main caveats and known issues, and we conclude in  Section~\ref{sec:conclusions}.   The appendix contains additional information on the empirical methods that were employed, use of the multi-dimensional tables, 
selection function information, and some tools that have been made available to the community to aid in the exploitation of these products.

\section{Input data }\label{sec:data}

\begin{figure*}
    \centering
    \includegraphics[width=0.8\textwidth]{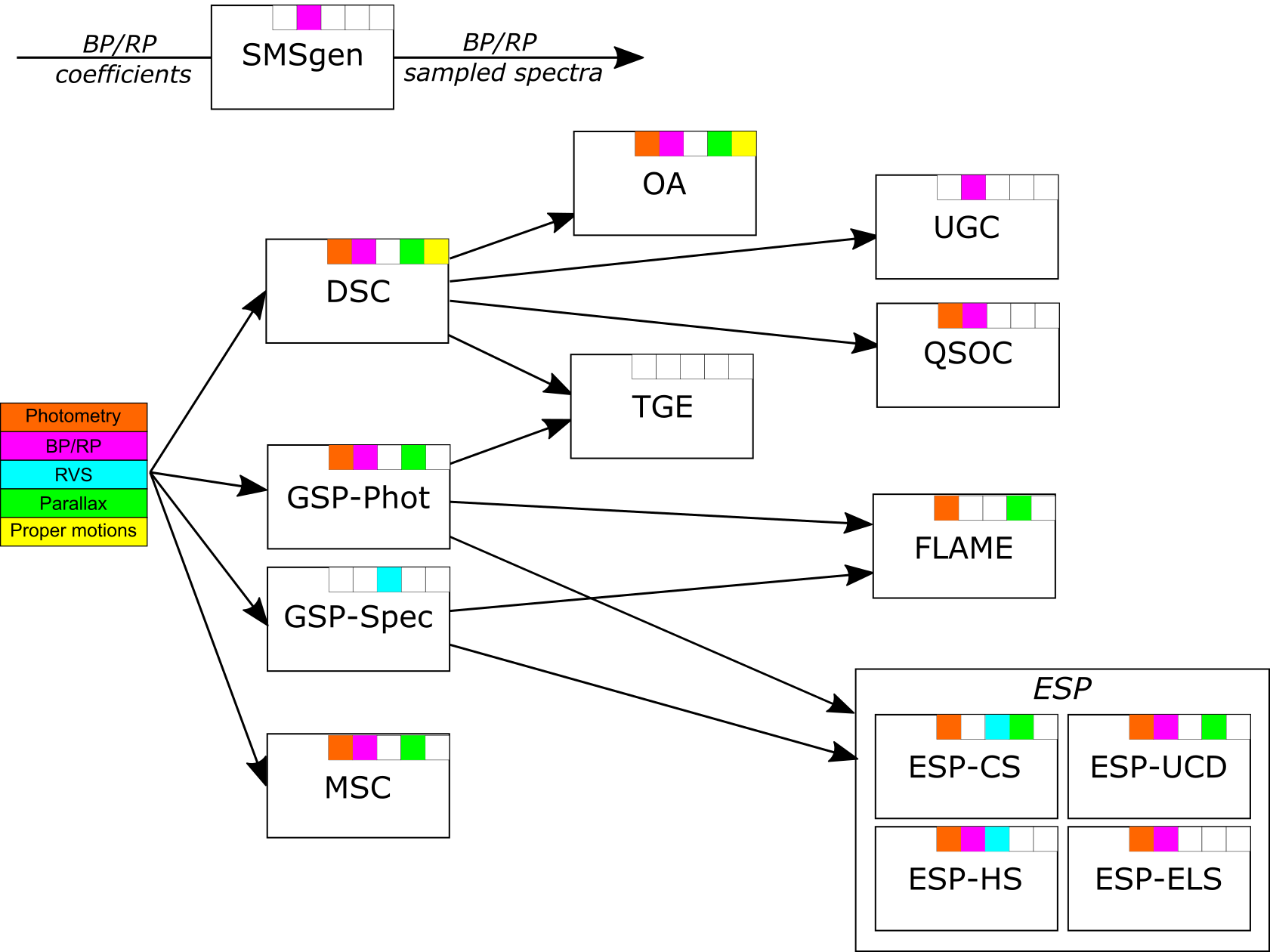}
    \caption{Apsis workflow showing the input data (colour-code) used by the 13 modules producing APs in \gdr{3} along with the dependencies among these modules (arrows). The input BP/RP spectra in \apsis\ are in the form of sampled spectra, produced by \smsgen, see Fig.~\ref{fig:CU5-passbands-and-new-sampling-scheme}.}
    \label{fig:apsisworkflow}
\end{figure*}

The results from \apsis\ in \gdr{3} are based solely on Gaia input data, and these are 
described in this section.
Figure~\ref{fig:apsisworkflow} illustrates the
input data that are used by the different modules in \apsis.   

\subsection{Input astrometry and photometry} 
We used the proper motions and parallaxes from Gaia in the processing of some of the \apsis\ modules. 
As some stellar-based modules are sensitive to the parallax zero-point, 
we implemented the systematic correction to the parallaxes 
as proposed by 
\cite{EDR3-DPACP-132}, who reports biases that vary with magnitude, colour, and ecliptic latitude.

\begin{figure}
    \centering
    \includegraphics[width=0.48\textwidth]{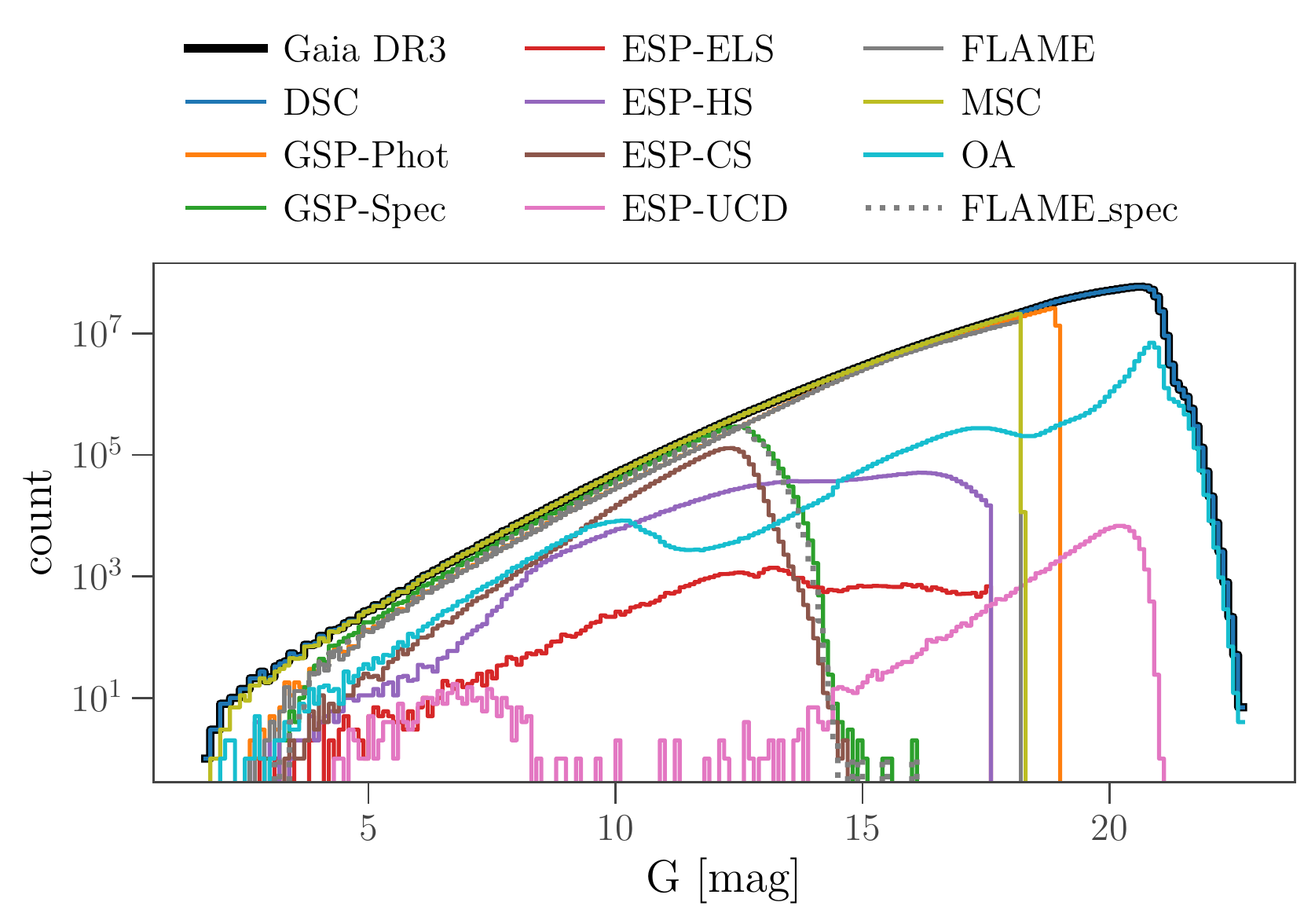}
    \includegraphics[width=0.48\textwidth]{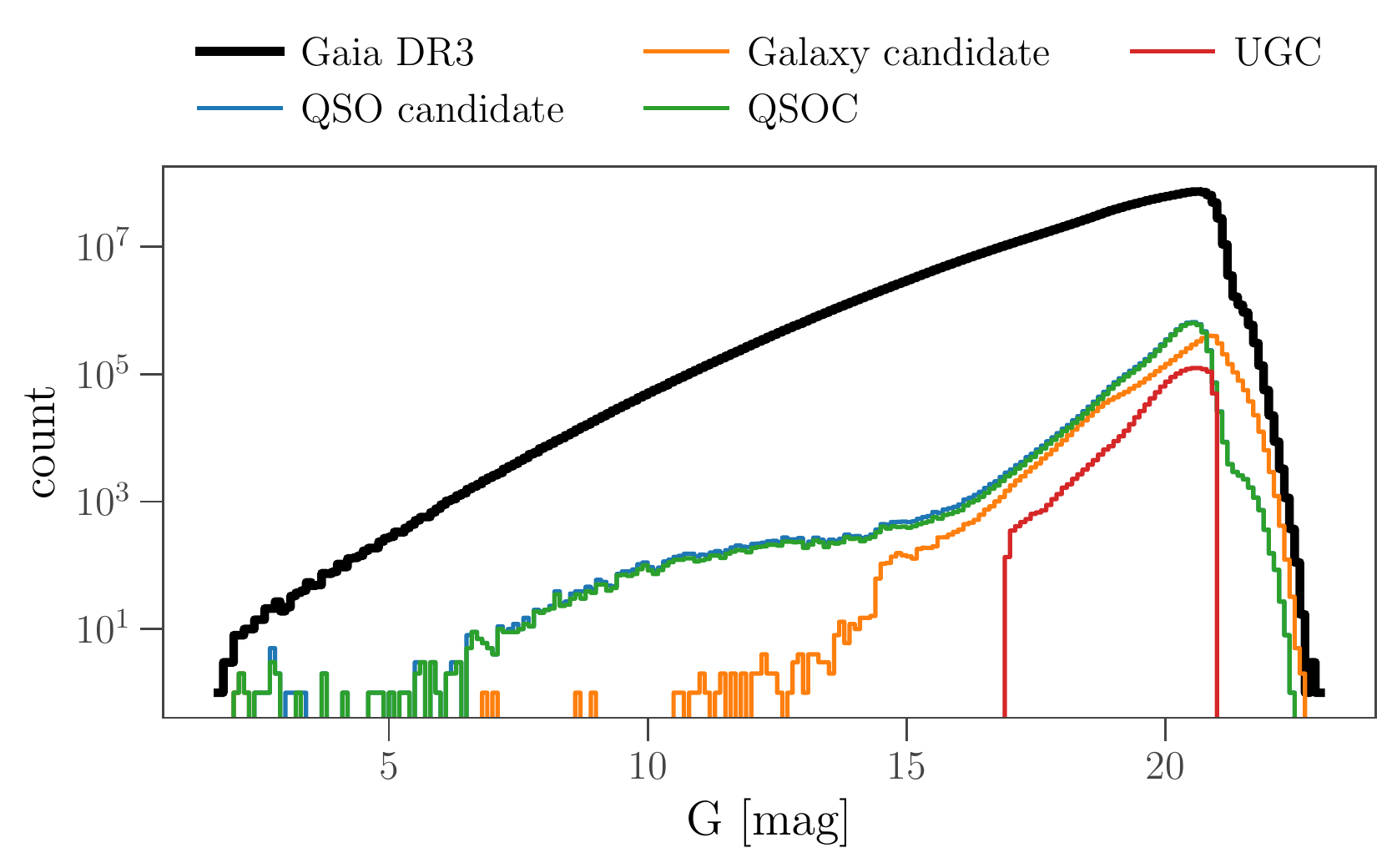}
    \caption{\gmag-magnitude distribution of the sources processed by CU8.  {\sl Top:} The distribution of sources that appear in the \aptable\ and \apsupptable\ tables grouped by module, illustrated by the different colours. 
    The \apsupptable\ contains results from \gspphot, \gspspec\ and \flame\ only, and only those from \flame\ are indicated in this top panel by the dashed lines, because the distributions in both tables are identical for \gspphot\ and \gspspec.
    {\sl Lower:} The sources with a CU8 result in the  \linktotable{qso_candidate} or \linktotable{galaxy_candidate} tables (blue/orange) and the sources in those tables with a redshift from \qsoc\ (green) or \ugc\ (red).}  
    \label{fig:gmag-distribution}
\end{figure}

Some of the \apsis\ modules use integrated photometry in the \gmag, \bpmag\ and/or \rpmag\ bands, using the zero-points provided directly by the Coordination Unit 5 (CU5) in \gedr{3} \citep{EDR3-DPACP-117}.  
They also recommended to correct some of the \gedr{3} photometry, and this was implemented in our processing.
This same correction to the eDR3 photometry has been fixed in \gdr{3} \citep{edr3-correction}.  Figure~\ref{fig:gmag-distribution} shows the distribution in \gmag\  magnitude of all of the individual products from CU8 from the four 1D tables.

\begin{figure*}[h]
\centering
\includegraphics[width=0.95\linewidth]{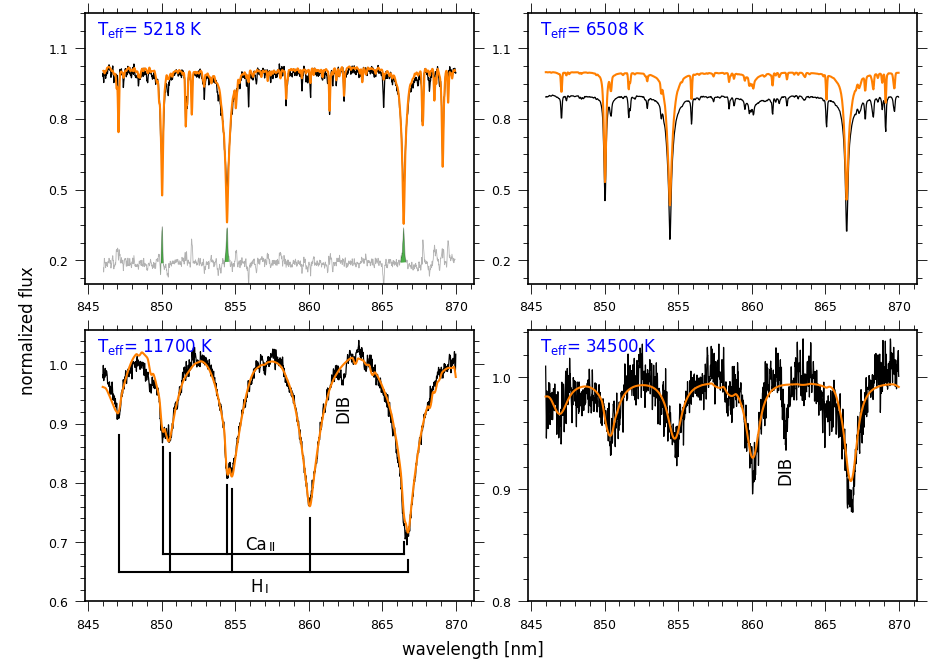}
\caption{Examples of the observed RVS spectra (black curve) analysed by various modules of the \apsis\ pipeline. The effective temperatures estimated by \gspspec\ (upper panels) and by \esphs\ (lower panels) are given in blue, while the best fitting synthetic spectrum is shown in orange. Upper left panel: adopting the APs by \gspspec\ (orange spectrum), \espcs\ derives an activity index from the residuals (gray lines: residuals vertically shifted by $+$0.2) summed up around the calcium triplet line cores (shaded green area). Upper right panel: synthetic and observed (shifted by --0.1 for readability) spectrum corresponding to the \gspspec\ APs.  The spectrum is then used to derive chemical abundances. Lower panels: determination of APs of stars hotter than 7\,500~K, by analysing the RVS and BP/RP data and assuming a solar chemical composition using \esphs.  We overplot the $\lambda$862~nm Diffuse Interstellar Band (DIB) which is also measured by \gspspec.
}
\label{fig:cu6_input_spectra}
\end{figure*}

\subsection{RVS spectra} 

Some of the products from CU8 are based on the RVS spectra
that are processed by Coordination Unit 6 (CU6, \cite{DR3-DPACP-154}).   
The CU6 pipeline provides wavelength-calibrated epoch spectra using standard spectroscopic techniques.  The mean spectrum can therefore be obtained by a simple stacking of the spectra.  
CU8 processed these mean spectra as provided by CU6.
A fraction of these spectra result from a deblending process of overlapping sources.
All spectra were corrected for the stars' radial velocities. They were cosmic-ray-clipped, normalized at the local (pseudo-)continuum (\teff\ $\ge$~3\,500~K), and re-sampled from 846 to 870 nm with a spacing of 0.01 nm. 
The median resolving power $R = \lambda/\Delta\lambda$~=~11\,500 \citep[][]{DR2-DPACP-46}. 
Fig.~\ref{fig:cu6_input_spectra} shows examples of input spectra (black) of different \teff\ and identifies the main spectroscopic features.  Some fits to models are also shown in orange.    These figures are further described in Secs.~\ref{sec:methods} and \ref{sec:content}. 
A more detailed description of the RVS data and its treatment is provided by \citet{DR3-DPACP-154}.

While over 37 million combined spectra were available to CU8, the median signal-to-noise-ratio (SNR) is $\sim6.5$. The \apsis\ modules processing RVS data then applied their own SNR thresholds 
and quality checks before processing.  Therefore, in practice, there were only about 10 million spectra processed, while after applying the module specific post-processing filters $\sim$6.3 million of these led to published astrophysical parameters.
The \gmag\ magnitude range covered by the remaining data varies from 2 to 15.2 mag.

\begin{figure*}
    \centering
    \includegraphics[width=0.98\textwidth]{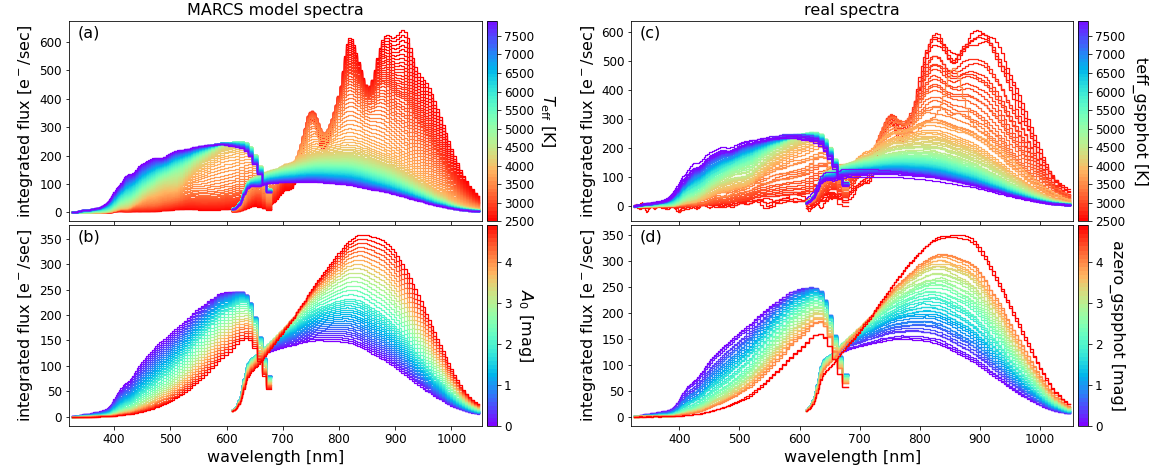}
\caption{Example BP/RP model spectra (left) and real spectra (right). All BP/RP spectra have been rescaled to an apparent magnitude of $G=15$ in order to make their flux levels comparable. Panels (a) and (c) show the variation with \teff, panels (b) and (d) the variation with \azero. Panels (a) and (b) show synthetic BP/RP spectra based on MARCS models (see Sect.~\ref{ssec:synthetic-spectra}). Panels (c) and (d) show BP/RP spectra obtained by Gaia where the APs were produced by the \gspphot\ module in the \apsis\ pipeline. BP spectra approximately cover the wavelength range from 325~nm to 680~nm and RP spectra from 610~nm to 1050~nm, see Fig.~\ref{fig:CU5-passbands-and-new-sampling-scheme}.}
\label{fig:xp-spectra-variation-with-teff-and-a0}
\end{figure*}

\subsection{BP and RP spectra}
Most of the \apsis\ modules produce APs based on the mean blue and red prism spectra (BP and RP), which are available for all of the sources in \gdr{3}.  
Examples of these prism spectra are shown
in Fig.~\ref{fig:xp-spectra-variation-with-teff-and-a0}  for stars with different \teff\ and extinction (\azero). These mean low resolution spectra \citep[$20 \leq R \leq 60$ for BP,  $30 \leq R \leq 50$ for RP, see Fig.~18 of][]{EDR3-DPACP-120} allow one to extract the atmospheric parameters (\teff, \logg, \ag, \mh), but also contain enough resolution to explore specific features such as the H$\alpha$ line for  emission-line stars and extragalactic objects. The RP spectra of very cool stars also show molecular absorption bands from TiO and VO \citep[e.g.][]{2007A&A...473..245R}.
The BP and RP spectra are processed by CU5 and then
adapted within the \apsis\ pipeline in the form of sampled spectra, as explained below.

\subsubsection{Production of data by CU5} 

The production of internally calibrated mean BP/RP spectra by CU5 is described in detail in \citet{2021A&A...652A..86C} and \citet{EDR3-DPACP-118}. 
We emphasise that these mean BP/RP spectra are averaged over time, i.e.\ any intrinsic variability of sources is lost.
One should be aware of this point when using APs from \apsis\ for stars with important variability.
Due to varying geometry over the field of view, occasional sub-optimal centering of the window on the observed target and variations of the instrument response and optics across the focal plane and in time, the epoch spectra of all transits of a given source cannot be simply stacked\footnote{Other normal ageing effects impacting all observations are the contamination level, the radiation damage, the changes in small-scale effects such as hot/cold columns, the changes in CCD and filter response.}.
Instead, they need to be carefully calibrated, resulting in each epoch spectrum having its own pixel sampling. The combined mean spectrum is a continuous mathematical function that can be evaluated at any pixel position. For this, CU5 adopts a linear basis representation in terms of Gauss-Hermite polynomials. The resulting expansion coefficients and their covariance matrix are the fundamental CU5 data products and are the input to the \apsis\ pipeline.  

\subsubsection{Sampled Mean Spectrum Generator (\smsgen)}
\label{ssec:cu8par_apsis_smsgen}

\begin{figure*}
    \centering
    \includegraphics[width=\textwidth]{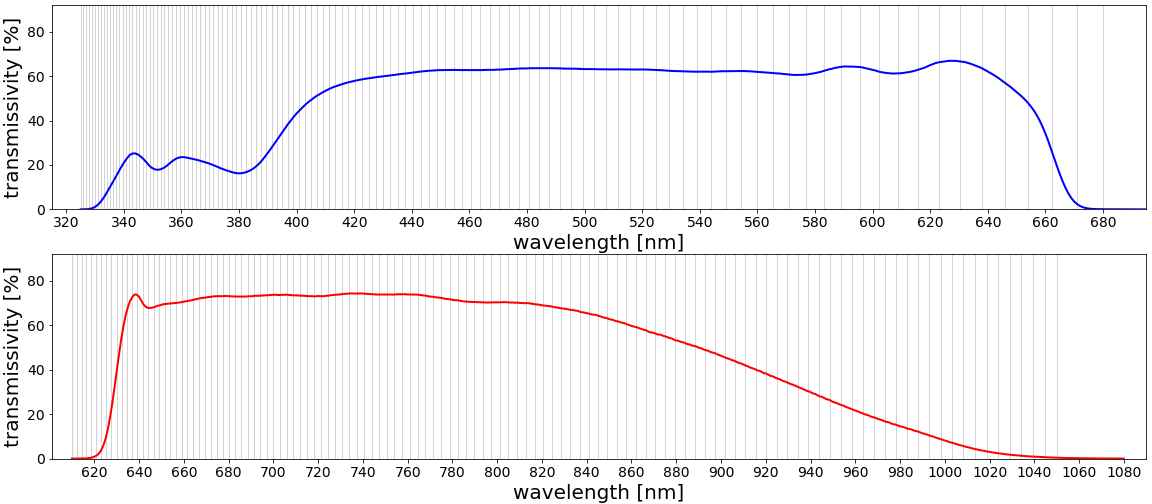}
    \caption{CU8 sampling scheme for BP/RP spectra. Vertical grey lines show the wavelengths of 121 pixel edges defining 120 pixels for BP (top panel) and RP (bottom panel). The blue and red lines show the BP and RP transmission curves from Gaia Early DR3, respectively. }
    \label{fig:CU5-passbands-and-new-sampling-scheme}
\end{figure*}

The modules in the \apsis~pipeline use the internally calibrated mean BP/RP spectra in the format of sampled spectra (integrated flux vs.~pixel). Computing these sampled spectra from the CU5 coefficients is the task of the Sampled Mean Spectrum Generator (hereafter \smsgen).
To this end, \smsgen\ takes the CU5 definition of the basis functions and integrates the spectral flux densities for a fixed wavelength grid. This wavelength grid defines 120 pixels for each BP and RP spectrum that cover the range of non-zero transmission in each spectrum as shown in Fig.~\ref{fig:CU5-passbands-and-new-sampling-scheme}\footnote{The sampling scheme is available 
at \url{https://www.cosmos.esa.int/web/gaia/dr3-aps-wavelength-sampling}.}. 
Here, we use the most recent EDR3 passbands\footnote{\url{https://www.cosmos.esa.int/web/gaia/edr3-passbands}}. 
The wavelength sampling is approximately uniform in pixel space but non-uniform in wavelengths. \smsgen\ then numerically integrates the flux densities in order to obtain integrated fluxes in each pixel. Note that BP/RP spectra can exhibit non-zero flux in pixels which have no transmission due to the LSF smearing effect of the BP/RP prisms, although this is negligible in practice. \apsis~modules using BP/RP spectra anyway discard several pixels at the edges that typically have very low flux and very low SNR.

\begin{figure}
    \centering
    \includegraphics[width=0.49\textwidth]{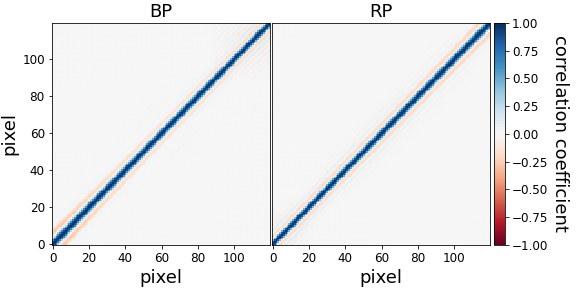}
    \caption{Random example (\texttt{source\_id}$=$5336426878835464960) of correlation matrices of pixel flux uncertainties for BP (left panel) and RP (right panel) for the CU8 sampling scheme shown in Fig.~\ref{fig:CU5-passbands-and-new-sampling-scheme}.}
    \label{fig:CU8-paper-example-XP-pixel-correlation-matrix}
\end{figure}

The sampling process of the CU5 basis functions as well as the flux integration are strictly linear operations. Consequently, \smsgen\ can easily propagate the CU5 uncertainty estimates on the coefficients into uncertainties of the sampled BP/RP spectrum. However, Apsis modules currently ignore any correlations between pixels, so \smsgen\ only provides standard deviations for the flux uncertainties of each pixel. This approximation ensures lower computational cost, which is a limiting factor during CU8 operations. 
Unfortunately, as illustrated in Fig.~\ref{fig:CU8-paper-example-XP-pixel-correlation-matrix}, notable long-range correlations between pixels do exist in BP and RP spectra
\citep[see][]{DR3-DPACP-127}. Ignoring these correlations therefore causes several \apsis~modules to systematically underestimate the uncertainties in their parameters,
although most modules have inflated their uncertainties to account for this effect. 

\section{Parameter estimation methods}\label{sec:methods}
\label{ssec:apsismodules}

\begin{figure}
    \centering
    \includegraphics[width=0.5\textwidth]{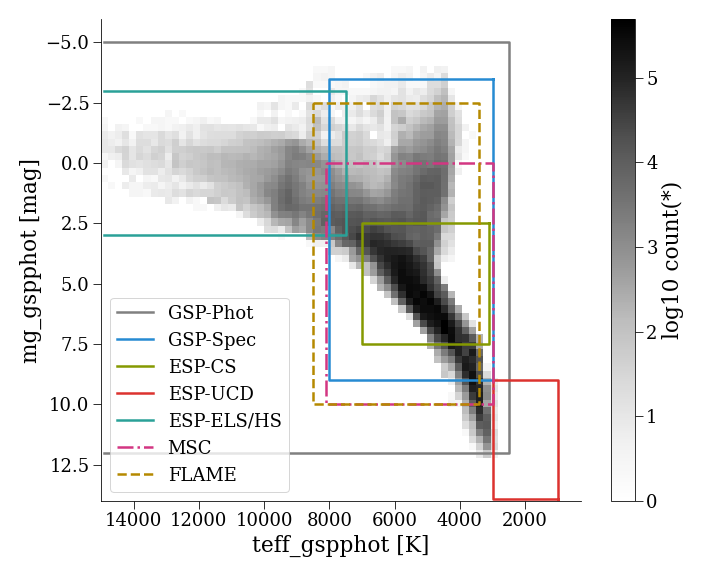}
    \caption{HR diagram showing the parameter spaces covered by the different stellar modules in \apsis. 
    The solid and dashed-dotted lines represent the modules that derive spectroscopic parameters.
    \gspphot\ and \gspspec\ are the general stellar parametrizers that use BP/RP and RVS data, respectively.  The {\it esp}
    modules work in specific stellar regimes: the ultra-cool dwarfs (\espucd), cool stars (\espcs), hot stars (\esphs) and emission-line stars (\espels).  \esphs, \espels\ and \gspphot\ provide results on stars up to 50\,000 K (not shown).   
    The green dashed-dotted line shows the regime of \msc\ which analyses the BP/RP spectrum as a combination of two components of an unresolved binary.  The dashed line shows the parameter space of \flame\ that derives evolutionary parameters only.
    The grey data are the stellar parameters from \gspphot.}
    \label{fig:hrdiag_apsis}
\end{figure}

The \apsis\ chain produces
all of the data from CU8 for \gdr{3}. 
\apsis\ is composed of 14 modules, 13 of which produce data for the release.
All of the modules are described individually and in more technical detail in  \linksec{}{Section 3 of Chapter 11 {\it Astrophysical Parameters} in the  online documentation for \gdr{3}}.
The first module providing the BP/RP spectra in the CU8 format (\smsgen) is summarised above in Sect.~\ref{ssec:cu8par_apsis_smsgen}.
Here, we provide a brief overview of the other modules in order for the reader to
have a basic understanding of the underlying methods, along with the dependencies among modules and dependencies on models and training data.   
Both Figure~\ref{fig:apsisworkflow} and Table~\ref{tab:apsisalgorithms} 
provide an overview of these details, which together, describe the different categories of parameters, the object type, the CU8 and non-CU8 input data, the dependencies,
the models and training data that are used, the approximate number of sources in \gdr{3} and their \gmag\ magnitude range for which a result can be found, see also Fig.~\ref{fig:gmag-distribution} for the distribution of \gmag\ magnitude.
In addition, Fig.~\ref{fig:hrdiag_apsis} shows a HR diagram illustrating the parameter spaces in which the different stellar-based modules are applied.  The background HR diagram is a representative random sample of 10 million \teff\ and \mg\ from \apsis.

\subsection{Discrete Source Classifier}\label{ssec:dsc}
The  {Discrete Source Classifier, \dsc} (\linksubsec{sec_cu8par_apsis/ssec_cu8par_apsis_dsc}{Section~11.3.2 of the online documentation}, \citealt{DR3-DPACP-158}, \citealt{LL:CBJ-094}), classifies sources probabilistically into five classes -- quasar, galaxy, star, white dwarf, physical binary star -- although it is primarily intended to identify extragalactic sources. \dsc\ comprises three classifiers: (1) Specmod, an ExtraTrees method  using the BP/RP spectrum; (2) Allosmod \citep{2019MNRAS.490.5615B}, a Gaussian Mixture Model using several photometric and astrometric features; (3)
Combmod, which combines the probabilities from the other two classifiers. The classes are defined empirically. \dsc\ incorporates a global class prior that reflects the intrinsic rareness of extragalactic objects.
All classifiers produce posterior class probabilities. 

\subsection{Outlier Analysis}\label{ssec:oa}
The {Outlier Analysis module, \oa} (\linksubsec{sec_cu8par_apsis/ssec_cu8par_apsis_oa}{Section~11.3.12 of the online documentation}), aims to complement the overall classification performed by the \dsc module by processing those sources with the lowest combined classification probabilities from \dsc. In order to analyse outliers, \oa performs an unsupervised classification (clustering) by means of Self-Organizing Maps \citep[SOM,][]{kohonen_self-organizing_2001}, grouping similar objects according to their BP/RP spectra. 
Each group of similar objects is referred to as a neuron. 
In addition, \oa characterises each neuron by reporting statistics of various parameters within them, such as magnitudes, galactic latitudes, parallaxes, and number of transits. 

\subsection{Unresolved Galaxy Classifier}\label{ssec:ugc}
The {Unresolved Galaxy Classifier, \ugc} (\linksubsec{sec_cu8par_apsis/ssec_cu8par_apsis_ugc}{Section~11.3.13 of the online documentation}, \citealt{DR3-DPACP-158}, \citealt{DR3-DPACP-101}), aims to estimate the redshift of unresolved galaxies observed by \gaia. The module processes every source that has a combined probability greater than or equal to 0.25 of being a galaxy according to \dsc, i.e. 
\hyperlink{gaia_source-	
classprob_dsc_combmod_quasar}{\fieldName{classprob_dsc_combmod_galaxy}} $\geq 0.25$, and which has a magnitude within the range $13\leq{}\gmag\leq21$ (after postprocessing there are no results with $G<15$). 
\ugc\ predicts the redshift of the source by applying a supervised machine learning model based on Support Vector Machines (SVM, \citealt{CortesVapnik95}) to its sampled BP/RP spectrum. The module is trained 
on a set of Gaia spectra of galaxies with redshifts provided by an external catalogue (see Sect.~\ref{ssec:training-ugc}) and 
predicts redshifts in the range $0.0\leq{}z\leq0.6$. 

\subsection{QSO Classifier}\label{ssec:qsoc}
The {QSO Classifier, \qsoc} (\linksubsec{sec_cu8par_apsis/ssec_cu8par_apsis_qsoc}{Section~11.3.14 of the online documentation}, \citealt{DR3-DPACP-158}, \citealt{DR3-DPACP-101}), aims to determine the redshift of the sources that are classified as quasars by the \dsc module, though it uses a loose cut of \hyperlink{gaia_source-	
classprob_dsc_combmod_quasar}{\fieldName{classprob_dsc_combmod_quasar}} $\geq 0.01$ in order to be as complete as possible. The method is based on a chi-square approach whereby the cross-correlation function between a rest-frame quasar template and an observed BP/RP spectrum is evaluated at a range of trial redshifts. The module predicts redshifts in the range $0.0826 < z < 6.1295$ and also provides an uncertainty and quality measurements from which flags are derived.

\subsection{General Stellar Parametrizer from Photometry}\label{ssec:gspphot}
The {General Stellar Parametrizer from Photometry, \gspphot} (\linksubsec{sec_cu8par_apsis/ssec_cu8par_apsis_gspphot}{Section~11.3.3 of the online documentation}, \citealt{DR3-DPACP-156,2012MNRAS.426.2463L,2011MNRAS.411..435B}), estimates effective temperature \teff, logarithm of surface gravity \logg, metallicity \mh, absolute magnitude \mg, radius \radius, distance \dist, line-of-sight extinctions \azero, \ag, \abp and \arp, as well as the reddening \ebpminrp\ by forward-modelling the  
BP/RP spectra, apparent $G$ magnitude and parallax using a a Markov Chain Monte Carlo (MCMC) method.
To this end, \gspphot\ employs PARSEC 1.2S Colibri S37 models \citep[][and references therein]{2014MNRAS.445.4287T,2015MNRAS.452.1068C,2020MNRAS.498.3283P} in a forward-model interpolation in order to obtain self-consistent temperatures, surface gravities, metallicities, radii and absolute magnitudes. For full details, we refer readers to \citet{DR3-DPACP-156}.
\gspphot\ results come from four stellar synthetic spectra ``libraries'' using different grids of atmospheric models (MARCS, PHOENIX, A stars, OB stars, see Table~\ref{tab:apsisalgorithms}) that cover different temperature ranges. A ``best'' library is recommended according to which library achieves the highest mean log-posterior value averaged over the MCMC samples.

\subsection{General Stellar Parametrizer from Spectroscopy}\label{ssec:gspspec}
The {General Stellar Parametrizer from spectroscopy, \gspspec} (\linksubsec{sec_cu8par_apsis/ssec_cu8par_apsis_gspspec}{Section~11.3.4 of the online documentation}, \citealt{DR3-DPACP-186}) estimates, from combined RVS spectra of single stars, stellar atmospheric parameters (\teff, \logg, \mh, [$\alpha$/Fe]), individual chemical abundances ([N/Fe], [Mg/Fe], [Si/Fe], [S/Fe], [Ca/Fe], [Ti/Fe], [Cr/Fe], [Fe/M], [FeII/M], [Ni/Fe], [Zr/Fe], [Ce/Fe], [Nd/Fe]), Diffuse Interstellar Band (DIB) parameters, and a CN under/over-abundance proxy with auxiliary parameters. No additional information (astrometric, photometric or BP/RP data) is considered, allowing a purely spectroscopic treatment. \gspspec\ uses specific synthetic spectra grids computed from MARCS models, see Sect.~\ref{ssec:synthetic-spectra}, and two different algorithms, \mgalgo\ and ANN, which are described in  \cite{2016A&A...585A..93R}, see also \citet{2006MNRAS.370..141R} for the Matisse algorithm.  Both algorithms are applied for atmospheric parameter estimates. Individual abundances and DIB parameters are estimated only from the \mgalgo\ algorithm using the approaches described in \cite{2016A&A...585A..93R} and \cite{2021A&A...645A..14Z},  respectively.

\subsection{Extended Stellar Parametrizer for Emission-Line Stars}\label{ssec:espels}
\begin{figure}
    \centering
    \includegraphics[width=0.5\textwidth]{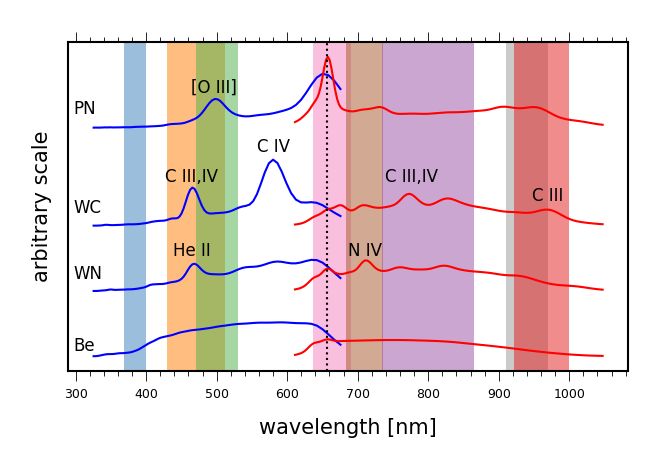}
    \caption{BP (blue) and RP (red) spectra typical of planetary nebula (PN), Wolf-Rayet (WC, WN) stars, and Be stars. The wavelength position of the strongest features is noted, while the H$\alpha$ line is represented by the vertical broken line. The wavelength domains considered for ELS classification are shown with colour shades, see Sects.~\ref{sec:results-classification-espels} and \ref{sect:training.espels}.}
    \label{fig:els_features}
\end{figure}

The {Extended Stellar Parametrizer for Emission-Line Stars, \espels} (\linksubsec{sec_cu8par_apsis/ssec_cu8par_apsis_espels}{Section~11.3.7 of the online documentation}), identifies the BP/RP spectra of emission-line stars brighter than magnitude $G=17.65$.  It then proposes a class label chosen among the following: Be, Herbig Ae/Be, Wolf Rayet (WC or WN), T Tauri, active M dwarf (dMe) stars, and planetary nebulae (PN). Figure\,\ref{fig:els_features} shows typical BP/RP spectra of some of these classes. The module uses three random forest classifiers (RFC, Sect.\,\ref{sect:training.espels}), and a measure of the pseudo-equivalent width (pEW) of the H$\alpha$ line. A first classifier (ELSRFC1) trained on synthetic BP/RP spectra is used to get a first coarse temperature estimate and assigns to each target one of the following spectral type tags: O, B, A, F, G, K, M, CSTAR (candidate carbon star, see also \citealt{DR3-DPACP-123}). Only non-``CSTAR'' targets that received a spectral type tag are further processed by the module. The second RFC (ELSRFC2) identifies the spectra of PN and of Wolf Rayet WC and WN stars. All the targets that are not identified as PN, WC, or WN are further processed. If significant H$\alpha$ emission is suspected based on the pEW value, a third RFC (ELSRFC3) is applied to the data in order to identify Be, Herbig Ae/Be, T Tauri, and dMe stars. In this process, the astrophysical parameters derived by \gspphot\ are used to help disentangle the candidate members of the four classes. 

\subsection{Extended Stellar Parametrizer for Hot Stars}\label{ssec:esphs}
The {Extended Stellar Parametrizer for Hot Stars, \esphs} ( \linksubsec{sec_cu8par_apsis/ssec_cu8par_apsis_esphs}{Section~11.3.8 of the online documentation}), derives \teff, \logg, \azero, \ag, \ebpminrp, and \vsini\ (broadening) of stars with \teff\ between 7\,500~K and 50\,000 K, based on either BP/RP+RVS spectra, or BP/RP alone, by assuming solar composition for stars with $G\le 17.65$. 
The target selection is based on receiving an A, B or O spectral type tag derived by \espels (see Sect.\,\ref{ssec:espels}). 
The BP/RP spectra (over the range 340 to 800~nm) are compared to synthetic spectra processed by \smsgen\ and rebinned into 40 wavelength bins, and fit in a multi-step $\chi^2$-minimisation. 
The flux uncertainties were multiplied by a factor of five to account for the amplitude of the systematic differences found between the observations and the simulations based on synthetic spectra.

Note that gravitational darkening due to rapid rotation in hot stars is expected to affect the parameter determination based on BP/RP and/or RVS spectra (e.g. \citealt{fremat2005}). It is however beyond the scope of the automatic pipeline to take these effects into account.

\subsection{Extended Stellar Parametrizer for Cool Stars}\label{ssec:espcs}

The {Extended Stellar Parametrizer for Cool Stars, \espcs} (\linksubsec{sec_cu8par_apsis/ssec_cu8par_apsis_espcs}{Section~11.3.9 of the online documentation}), computes a chromospheric activity index from the analysis of the \cairt\ (calcium infrared triplet) in the RVS spectra.
The activity index is derived by comparing the observed RVS spectrum with a purely photospheric model (assuming radiative equilibrium) with \teff, \logg, and \mh\ from either \gspspec\ or \gspphot, and from \linktogsparam{gaia_source}{vbroad} when available from CU6 (provided in \linktogstable{gaia_source}), see Fig.~\ref{fig:cu6_input_spectra} top left panel.
An excess equivalent width factor in the core of the  \cairt lines, computed on the observed-to-template ratio spectrum in a $\pm \Delta \lambda = 0.15~\mathrm{nm}$ interval around the core of each of the triplet lines, is taken as an index of the stellar chromospheric activity or, in more extreme cases, of the mass accretion rate in pre-main sequence stars.

\subsection{Extended Stellar Parametrizer for Ultra Cool Dwarfs}\label{ssec:espucd}
The {Extended Stellar Parametrizer for Ultra Cool Dwarfs, \espucd} (\linksubsec{sec_cu8par_apsis/ssec_cu8par_apsis_espucd}{Section~11.3.10 of the online documentation}), 
provides \teff\ of 
\gaia~sources cooler than 2500~K. This is an arbitrary definition that includes stellar objects and brown dwarfs. In practice, \teff predictions up to 2700~K have been included in the catalogue in order to accommodate uncertainties. 
As UCDs are detected at very short distances, typically less than 200 pc, extinction should be very small and therefore we ignored this parameter for these objects in \gdr{3}.
The \espucd module consists of a Gaussian Process regression module that takes RP spectra as input and assigns \teff estimates. 
The RP spectra used as input to the \espucd module were reconstructed from the continuous representation using a truncation procedure described in \cite{2021A&A...652A..86C}. 
We use $a$=3 where $a$ is the threshold coefficient in Equation 27 of \cite{2021A&A...652A..86C}.  

\subsection{Final Luminosity Age Mass Estimator}\label{ssec:flame}
\hypertarget{sec31flame}{The {Final Luminosity Age Mass Estimator}}, {\flame} (\linksubsec{sec_cu8par_apsis/ssec_cu8par_apsis_flame}{Section~11.3.6 of the online documentation}, \cite{LL:OLC-035}),
aims to produce {the stellar mass and} evolutionary parameters for each 
Gaia source that has been analysed by \gspphot\ and/or \gspspec, that is, \flame\  produces two results for some sources.
The \flame\ parameters comprise the
radius \radius, 
luminosity \lum, and 
gravitational redshift $rv_{\rm GR}$,
along with the 
mass \mass, 
age $\tau$, and 
evolutionary stage $\epsilon$.
\flame\ uses as input data \teff, \logg, and [M/H] from \gspphot\ ``best library'' and, when available, these same parameters from \gspspec \mgalgo, along with a distance estimate, $G$-band photometry (Sect.~\ref{sec:data}), and extinction from \gspphot.  A bolometric correction is evaluated on a grid of models, see Sect.~\ref{ssec:bolometric_correction}.  
To infer  \mass, $\tau$, and $\epsilon$, the \basti\footnote{\url{http://basti-iac.oa-abruzzo.inaf.it}} \citep{2018ApJ...856..125H} solar-metallicity stellar evolution models are employed, which consider a mass range from 0.5 -- 10~\Msun and evolution stage from Zero-Age-Main-Sequence (ZAMS) until the tip of the red giant branch \cite{EDR3-DPACP-120}.

\subsection{Multiple Star Classifier}\label{ssec:msc}
The {Multiple Star Classifier, \msc}, (\linksubsec{sec_cu8par_apsis/ssec_cu8par_apsis_msc}{Section~11.3.5 of the online documentation}) 
infers stellar parameters by assuming the BP/RP is a composite spectrum of an unresolved coeval binary system  and that the two components have a flux ratio in the BP/RP spectrum between 1 and 5. The primary is defined as the brighter source in the BP+RP spectrum total flux. \msc uses an empirical BP/RP model (Sect.~\ref{ssec:empiricaltraining-msc}) within an MCMC method to sample the posterior over its parameter space: \teff\ and \logg\ of its primary and secondary components, as well as a common metallicity, extinction, and distance. 
\msc\ produces results for all sources with BP/RP spectra, a parallax, and $G\le 18.25$. 

\subsection{Total Galactic Extinction}\label{ssec:tge}
The {Total Galactic Extinction} module, \tge, (\linksubsec{sec_cu8par_apsis/ssec_cu8par_apsis_tge}{Section~11.3.11}, \citealt{DR3-DPACP-158}), 
uses a subset of giants with extinction estimates provided by \gspphot as extinction tracers, to construct all-sky maps at various resolutions of the total foreground extinction from the Milky Way. The maps specify the median extinction $A_0$ of the tracers per HEALPix, where $A_0$ is the extinction parameter of the adopted extinction curve of \citet{1999PASP..111...63F}, see Sect.~\ref{ssec:extinction} for details. 
Sky coverage is 97.2\% at HEALPix level 6 (0.84 square degrees per HEALPix), with missing extinction estimates for some HEALPixes at galactic latitude $|b| < 5 \deg$. Sky coverage is less at higher resolution, due to the limited number of tracers per HEALPix.

\begin{table*}[]
    \caption{Input data, models, training data, data products, and dependencies of the \apsis\ algorithms.}
    \label{tab:apsisalgorithms}
    \centering
       \scalebox{0.90}{
       
    \begin{tabular}{lllllllrrrrc}
    \hline\hline
    
module & category & object & \multicolumn{2}{c}{input data} & \multicolumn{2}{c}{inference model} & number & $G$\\ 
 && type &non-Apsis & Apsis  &models & empirical& (millions)\\
\hline
\dsc & probabilities, & SEB &  XP, $\varpi$,&   &       & ExtraTrees/ &  1\,591 &$<21$\\
& classification & & pm, $G$  & & & Gauss.Mix&\\
\oa & classification & O & XP & (\dsc) & &SOM  & 56 & $<21$\\
\ugc & galaxy redshift & E & XP & (\dsc) & &SVM & 2 & $<21$\\
\qsoc & quasar redshift & E & XP & (\dsc) & &  & 6 & $<21$\\
\gspphot & spectroscopic, & SI & XP, $\varpi,G$ & & MARCS/PHOENIX & & 470&$<19$\\
& interstellar, evolution & & \bpmag,\rpmag& & OB/A/PARSEC \\
\gspspec & spectroscopic, & SI & RVS & & MARCS & & 6&$<16$\\
 & abundances, DIB& & & & & & \\
\msc&spectroscopic, & BI & XP & &  & ExtraTrees& 350&$<18.3$\\
&interstellar &  & &  (\gspphot) &\\
\espels & probabilities, & S & XP & \gspphot & & Random & 0.01& $<17.65$ \\
& classification, & & & & & Forest\\
& H$\alpha$ pEW & & & & &&235& \\
\esphs & spectroscopic & S & XP, RVS & \espels & A/OB & & 2 & $<17.6$\\
\espcs & spectroscopic & S & RVS & (\gspspec) & MARCS & & 2 & $<15$\\
\espucd & spectroscopic & S & RP,$\varpi,G$ & & & Gauss.Proc. & 0.1 & $ < 21 $ \\
\flame &  evolution &S & $\varpi, G$ & \gspphot & MARCS/\basti & & 280&$<18.3$\\
 &  && & \gspspec & MARCS/\basti & & 5&$<13$\\
\tge & map & I & & \gspphot\ & & & -- & $<19$\\
\hline
    \end{tabular}
    }
    
    \tablefoot{
     The following notation is used: XP = BP/RP spectra (through sampled mean spectra), RVS = RVS spectra;  
    $\varpi$ = parallax, pm = proper motions, $G$ = Gaia photometry which implies \gmag, \bpmag, and/or \rpmag.  
    Object Type is S, B, E, I, O for star, binary, extragalactic, interstellar, and outlier, respectively.  Under Apsis Input Data, values in "()" mean that APs are used to initialise the analysis or as selection criteria only.
    }
\end{table*}

\section{Models and training data}\label{sec:models}

The \apsis\ modules require models and training data to infer APs.   In this section we describe these auxiliary data.

\subsection{Synthetic spectra} 
\label{ssec:synthetic-spectra}

\begin{table*}[]
\caption[CU8 synthetic stellar libraries]{CU8 synthetic stellar libraries list of BP/RP spectra. Minimum grid spacing is listed in parentheses.
  \label{tab:cu8par_inputdata_xp_SimuLibTab}
  }
\centering
\scalebox{0.90}{
\begin{tabular}{ l c c c c c c l}
\hline
\hline
LIBRARY NAME & $N$ models &\teff\ [K]        & ~~ &\logg [dex]& ~~ &\feh [dex] & provider/contact \\      

\hline
A (LL models) &2332&   ~~6000 $-$ 16000 (250)        &    &2.5 $-$ 4.5 (0.25)       &    &~~$-$1.5 $-$ +0.5 (0.5) & O. Kochukhov/D.\ Shulyak \\
MARCS         &27951&     2800 $-$ 8000 (25)         &    &$-$0.5 $-$ 5.0 (0.5) &    &~~$-$5 $-$ +1 (0.25) & B.\ Edvardsson \\
PHOENIX       &4651&   ~~3000 $-$ 10000 (100)       &    &$-$0.5 $-$ 5.5 (0.5) &    &$-$2.5 $-$ +0.5 (0.5) & I.\ Brott/P.\  Hauschildt \\
OB            &2162&    15000 $-$ 55000 (1000)       &    &1.75 $-$ 4.75 (0.25)     &    &~~0.0 $-$ 0.6 (0.1) & Y.\  Fr\'{e}mat \\
BTSettl   &2084&    ~~400 $-$ 3000         &    &$-$0.5 $-$ 5.5~~  &    &~~$-$4.0 $-$ +0.5 & F.\ Allard\tablefootmark{$\dagger$}/L.\ Sarro \\
(HotSpot       &31957&   3000 $-$ 7000           &    & 3 $-$ 5          && -0.5 $-$ +1.0 & A.\ Lanzafame) \\
(WD            &186&   ~~6000 $-$ 90000        &    &7 $-$ 9           &    &  +0.0  & D.\ Koester)    \\
\hline                      
\end{tabular}
}
\tablefoot{
\tablefoottext{$\dagger$}{France Allard (1963-2020)}
}
\end{table*}

For the estimation of stellar APs, extensive synthetic spectral libraries based on atmospheric models have been computed for the \gmag, \bpmag\ and \rpmag\ filter ranges and the  BP/RP and RVS wavelength ranges. These libraries were used to simulate \gaia-observed spectra through the Gaia instrument models, with noise and extinction added (see Section~\ref{ssec:MIOG-simulations} and \citealt{EDR3-DPACP-120}).

\begin{figure}
    \centering
    \includegraphics[width=0.48\textwidth]{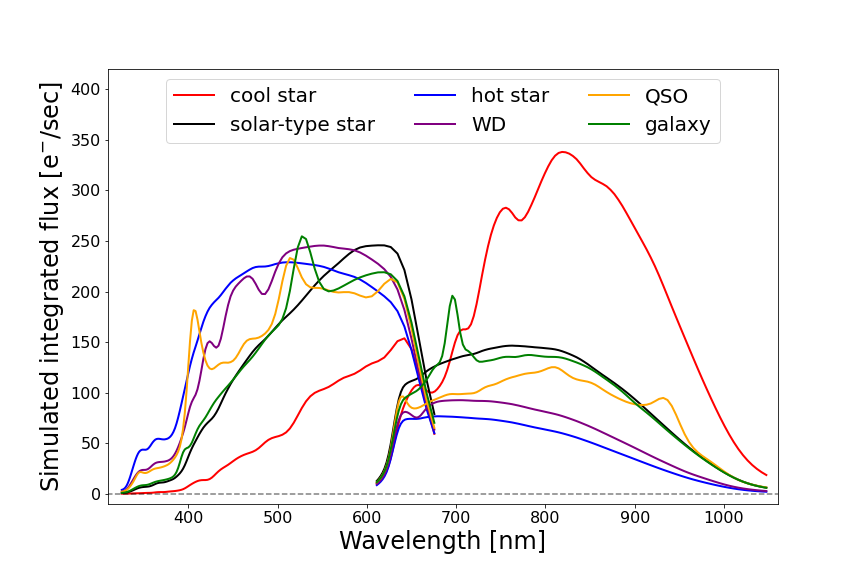}
\caption{Examples of BP/RP simulations of different type of sources. All sources have been simulated at $G$=15. Red and black lines show spectra from MARCS library, having \teff= 3500 and 6000~K. In blue an OB spectrum with \teff=30000~K, in purple a WDA spectrum having \teff=15000~K and \logg=8.0.  In orange and green, an SDSS QSO and an SDSS Galaxy, having redshift $z$=2.3 and 0.06 respectively (randomly selected).}
\label{fig:xp-spectra-different-source-type}
\end{figure}

\subsubsection{Synthetic fluxes for BP/RP} 

Stellar fluxes have been simulated using standard 1D stellar-atmosphere codes, covering all spectral types of normal stars. Several grids have been produced by different code families, each different in physics and assumptions, with large overlaps in the parameter space.
The providers of these libraries were free to compute models following their own expertise and preferences while paying attention to the challenges of the respective stellar types (e.g.\ dust formation, molecular absorption, treatment of convection, chemical peculiarities, departures from local thermodynamic equilibrium (LTE), stellar winds). 
For example, models for OB-type stars take into account non-LTE effects both in the computation of the model and of the spectrum. For the MARCS models
\citep{2008AA...486..951G}, the chemical abundances compared to the Sun have been varied over several orders of magnitude by enhancing or reducing all metals (atomic mass $A > 4$) with $\alpha$-elements
roughly following the Galactic trend changing linearly from [$\alpha$/Fe]\,=\,0.0 at [Fe/H]\,=\,0.0 (solar) to [$\alpha$/Fe]\,=\,0.4 below [Fe/H]\,=\,$-1.0$. 
Some differences in the assumed solar reference composition exist between individual libraries, reflecting choices of modellers at the time of computation. 
Cool stars (\teff\,$<$\,4500~K) with prominent molecular bands react sensitively to different assumptions concerning the chemical mixture. The assumed composition should thus be considered when comparing results derived using different libraries at this low \teff.

Spacing between grid points also varies, both between and within libraries, and can be as low as 25\,K in $\Delta$\teff for the MARCS models (see Table~\ref{tab:cu8par_inputdata_xp_SimuLibTab}). 
\gspphot\ relies on linear interpolation between grid points (for computational cost reasons).
Since the spectral flux does not change linearly with changes in the parameters, see e.g. \cite{Zwitter2004}, finer grids will result in better performance than coarser ones.

An overview of the parameter space, number of models and stellar model providers is given in  Table \ref{tab:cu8par_inputdata_xp_SimuLibTab}, and some examples of synthetic spectra are shown in Fig.~\ref{fig:xp-spectra-different-source-type} for different objects.
While several libraries cover the physical parameter space of horizontal-branch stars, only the \esphs\ module provides APs for these (Sect.~\ref{sec:results-spectroscopic-esphs}). Libraries `HotSpot' and `WD' were finally not used for the production of the data in DR3.

The computation of each of the libraries requires basic information such as input stellar parameters, key individual abundances, mass fractions of H, He and metals etc.  For the MARCS, PHOENIX, A and OB libraries, these parameter files can be retrieved from the  \href{https://www.cosmos.esa.int/web/gaia/dr3-astrophysical-parameter-inference}{\gdr{3} auxiliary data web pages}\footnote{\url{https://www.cosmos.esa.int/web/gaia/dr3-astrophysical-parameter-inference}}. 

\subsubsection{Synthetic spectra for RVS} 

For the parametrisation of the infrared RVS spectra within the \gspspec\ module, large grids of synthetic spectra were computed. These spectra were calculated from MARCS atmospheric models for FGKM-type stars using the TURBOSPECTRUM code \citep{Plez2012} and
specific atomic and molecular line lists \citep{Contursi21}. The covered parameter space of these grids is:
2600 to 8000~K for \teff, $-$0.5 to 5.5 for \logg\ ($g$ in cm/s$^2$) and $-$5.0 to 1.0~dex for the mean metallicity, with varying $\alpha$-element enrichment with respect to iron, as explained above. 
Individual chemical-abundance variations were also considered to derive abundances of N, Mg, Si, S, Ca, Ti, Cr, Fe, Ni, Zr, Ce, Nd. The adopted solar abundances are those of \cite{Grevesse07}.
The computation of these grids of synthetic spectra is discussed in \cite{DR3-DPACP-186}.

For the other modules using RVS data (i.e. \espcs\ and \esphs), the same model atmosphere grids used to prepare the synthetic BP/RP spectra (Table~\ref{tab:cu8par_inputdata_xp_SimuLibTab}) were adopted to compute the flux in the 846 - 870 nm wavelength domain. The library used by \esphs\ was prepared assuming a Solar chemical composition for \teff$>$ 7000~K, while for \espcs\ the MARCS models were considered for \teff\ ranging from 3000 to 7000~K, \logg\ from 3 to 5 dex, and \feh\ from $-$0.5 to $+$0.75.

\subsection{Extinction}\label{ssec:extinction}

Observed spectra are attenuated by the amount of interstellar dust present in the line-of-sight between the observer and the source. In this sense, extinction can be considered an astrophysical parameter of a given source, and can be inferred from the spectra. 
To estimate this parameter from the algorithms we use simulations of the BP/RP spectra that 
cover a wide range of extinction values.

For \apsis\ simulations, we adopted the wavelength dependent extinction law by \cite{1999PASP..111...63F}, see \linksec{}{Section~11.2.3 in the online documentation}. We use the parameter \azero, which is the  monochromatic extinction  at $\lambda_0$ = 541.4~nm\footnote{The wavelength can be derived from Table 3 in Fitzpatrick (1999), which gives $A(\lambda)/E(B-V)$ as a function of wavelength. $\lambda_0$  is the wavelength $\lambda$ for which $A(\lambda)/E(B-V)/3.1$ is equal to 1}. \azero and \av are often confused in literature, the latter being the actual extinction computed in the $V$ band, and as such intrinsically dependent on the spectral shape of the emitting source. This dependence is often, justifiably, neglected in the Johnson $V$ band, but it is particularly evident in the very wide Gaia bands and therefore should not be neglected.
 
Simulations are provided covering a semi-regular grid of 56 values of \azero, from 0 to 10 magnitudes, while the parameter $R$ is kept fixed at 3.1 (see \citealt{1999PASP..111...63F}, their Table~3). For each spectrum and for each \azero\ the extinction in a given band (\ag, \abp, \arp) is computed by comparing the un-reddened and the attenuated flux in the given \gaia\  passband. The values of extinction in these bands, and in addition in the $V$-band, for different APs and \azero\ values are made available to the community in the parameter files (Sect.~\ref{ssec:synthetic-spectra}) on the \href{https://www.cosmos.esa.int/web/gaia/dr3-astrophysical-parameter-inference}{\gdr{3} auxiliary data web pages}\footnote{\url{https://www.cosmos.esa.int/web/gaia/dr3-astrophysical-parameter-inference}}.

\subsection{Bolometric corrections } \label{ssec:bolometric_correction}
In order to derive the bolometric luminosity of stars, specifically in the \flame\ module, we complemented 
the observed photometric $G$ magnitude with a bolometric correction, \bcg.  
The \bcg\ was derived from the MARCS synthetic stellar spectra
as a function of \teff, \logg, \feh\ and [$\alpha$/Fe].
For this data release, we assumed [$\alpha$/Fe] = 0.0 when calculating the correction for all stars because [$\alpha$/Fe] is only estimated for a small fraction of the sources.
A tool is made available to the community to calculate the \bcg as a function of \teff, \logg, \mh, and [$\alpha$/Fe] and can be found  \href{https://www.cosmos.esa.int/web/gaia/dr3-bolometric-correction-tool}{on the Gaia DR3 tools webpages}\footnote{\url{https://www.cosmos.esa.int/web/gaia/dr3-bolometric-correction-tool}}.

We extended the \teff\ range to intermediate-temperature stars by using the A star models.
Their \bcg\ values show a slight offset  
relative to the MARCS grid (due at least in part to different opacities used in the 
two sets of models). We therefore added an offset in magnitude units to achieve  
continuity at 8\,000 K.
The adopted value for the bolometric correction for the Sun is \bcgsol\ = +0.08 mag, where $M_{\rm bol\odot} = 4.74$\footnote{The IAU 2015 Resolution B2 can be found here: \url{https://www.iau.org/static/resolutions/IAU2015_English.pdf}} which yields an absolute magnitude of the Sun M$_{G,\odot}= 4.66$ mag. We estimate an external accuracy on this zeropoint of $\pm 0.015$ mag from comparison with known solar analogues ($M_G = 4.63 - 4.69$ mag), stellar models ($M_G = 0.465$ mag) and colour transformations using \cite{EDR3-DPACP-117} ($V-G = 0.148 \pm 0.003$ mag where $M_V = 4.817$ mag). 

To complement this analysis on the solar reference magnitudes, we estimate the solar colours in \cite{DR3-DPACP-123} using a set of solar analogues although we note that these colours were not used in \apsis\ processing: 
$(\bpmag-\rpmag)_\odot = 0.818  \pm  0.029\,\textrm{mag}, 
(\bpmag-\gmag)_\odot = 0.324  \pm  0.016\,\textrm{mag}$, and 
$(G-\rpmag)_\odot = 0.494  \pm  0.020\,\textrm{mag}$.

\subsection{Stellar evolution models } 
Stellar evolution models are used in two of the \apsis\ modules,
\gspphot\ and \flame.  
For \gspphot\, the published APs 
are astrophysically self-consistent within the PARSEC 1.2S Colibri S37 models \citep[][and references therein]{2014MNRAS.445.4287T,2015MNRAS.452.1068C,2020MNRAS.498.3283P}. Imposing these isochrones ensures that \gspphot\ can simultaneously fit the observed apparent magnitude (using the absolute magnitude) and the amplitude of low-resolution BP/RP spectra \citep[using the radius, see][]{DR3-DPACP-156}. Moreover, the isochrones ensure that only astrophysically reasonable parameter combinations are possible.

For \flame\ the mass, age and 
evolutionary stage 
are based on the use of the BASTI stellar models \citep{2018ApJ...856..125H}.
In \flame these models cover the zero-age main sequence until the tip of the red-giant branch, corresponding to evolutionary indices of between 100 and 1300 (main sequence $<390$; turn-off = 390; sub-giant: 420 to 490, and giant $>490$), and masses between 0.5 and 10~\Msun. 
We furthermore imposed a solar-metallicity prior, see Sect.~\ref{ssec:flame_evol_mrl} for discussion on this assumption.

\subsection{Empirical training} 

One of the drawbacks of training machine learning algorithms on synthetic data is that good results require (a) adequate source models from which to generate the synthetic data, (b) sufficient coverage of the parameter space by the source models, and (c) a good match between the synthetic data (\gaia\ simulations) and the real \gaia\ data of the corresponding objects. For five Apsis modules, specifically 
\dsc, \msc, \ugc, \espucd, and \espels,
one or more of these conditions could not be achieved, so for these we use empirical training. This involves training the algorithm on real \gaia\ data, with classes or astrophysical parameters for the training data obtained from external sources. Typically this involves cross-matching \gaia\ to external catalogues, such as the Sloan Digital Sky Survey (SDSS), and using class labels or APs obtained by others, e.g.\ from higher resolution spectra. Details of the empirical training used by the five Apsis modules are given in appendix \ref{ssec:empiricaltraining}.

\subsection{Simulations with MIOG} 
\label{ssec:MIOG-simulations}
The Mean Instrument Object Generator (hereafter MIOG) simulates low-resolution BP/RP spectra from given model spectral energy distributions (SEDs).  This was developed by CU5 and is only available internal to DPAC systems. MIOG implements the instrument model and the dispersion law, as derived by  CU5 as part of the external calibration process \citep{EDR3-DPACP-120}.  This external calibration
relies on the flux calibration of the Spectro-Photometric Standard Stars by \cite{2012MNRAS.426.1767P, 2015AN....336..515A,2016MNRAS.462.3616M}.

All synthetic libraries described in Sect.~\ref{ssec:synthetic-spectra} were simulated with MIOG.
An example of the simulated spectra for stars of different \teff\ and \azero\ is shown on the left panels of Fig.~\ref{fig:xp-spectra-variation-with-teff-and-a0}. The corresponding real observed spectra are shown on the right panels.  In Fig~\ref{fig:xp-spectra-different-source-type}, simulated stellar spectra at different temperatures (and from different libraries) are shown, together with extra-galactic sources spectra and a white dwarf spectrum.

CU5 have provided  a simplified version of this tool to the community, \href{hhttps://gaia-dpci.github.io/GaiaXPy-website/}{\tt GaiaXPy}\footnote{\url{https://gaia-dpci.github.io/GaiaXPy-website/}}  that simulates the low-resolution spectra from model SEDs, which is fully compatible with the internal DPAC MIOG simulator \citep{EDR3-DPACP-120}.

\section{Catalogue description }\label{sec:results}

\begin{figure*}
    \centering
    \includegraphics[width=0.95\textwidth]{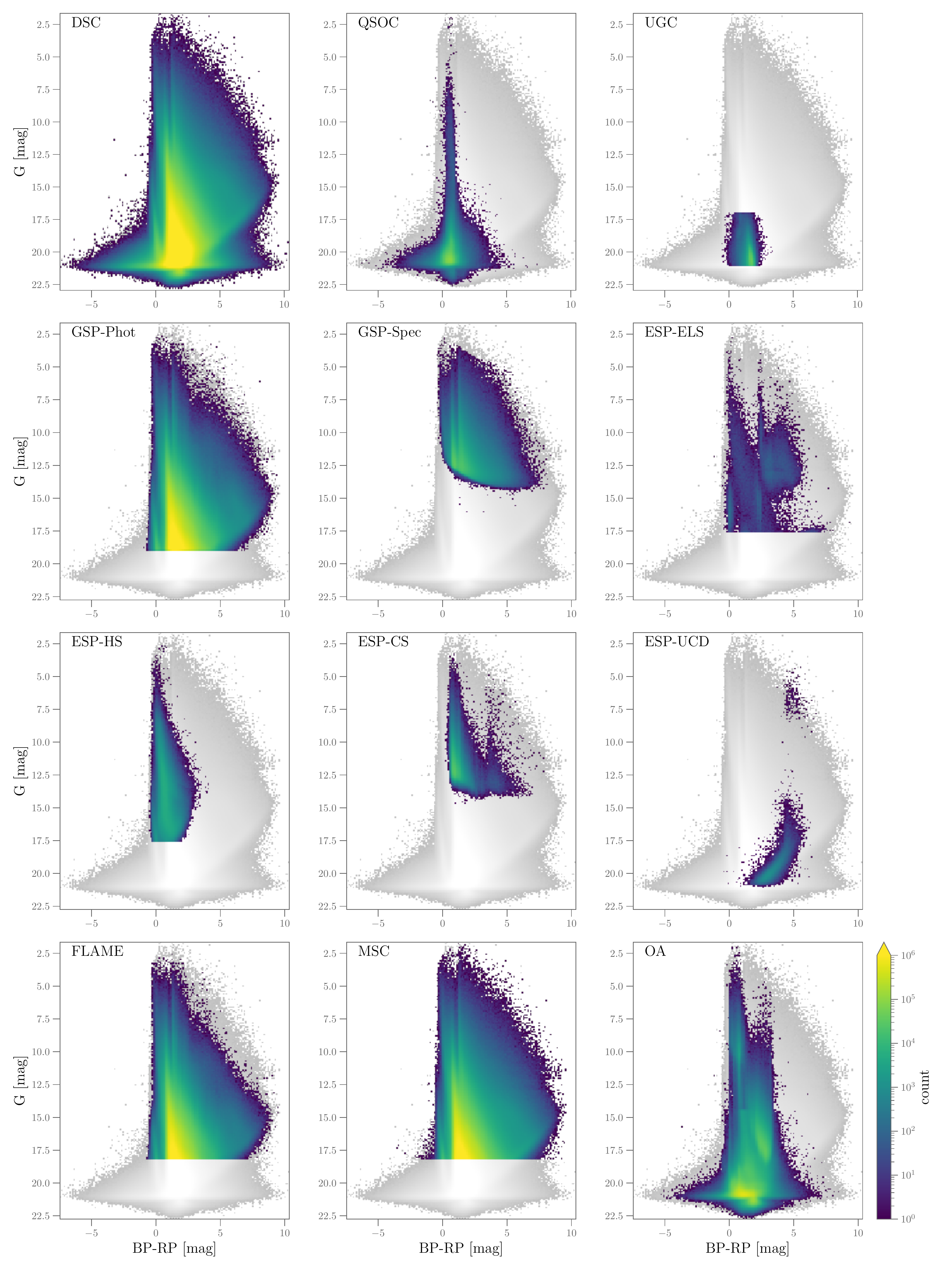}
    \caption{Distribution in colour-magnitude space of the sources with products from CU8 in \gdr{3}, separated by module.
    The colours represent the results per module, and the colour-code is density of sources. The distribution shown in gray in all panels indicates the whole \gdr{3} sample for reference.
    These products are found in the 
    \aptable, \apsupptable,  \linktogaltable{galaxy_candidates} and 
         \linktogaltable{qso_candidates}
    tables.}
    \label{fig:cmd-stellar}
\end{figure*}

The astrophysical parameters produced by CU8 fall under the following categories: 
(a) classification products comprising class probabilities and class labels of objects and emission-line stars, and stellar spectral types,  
(b) interstellar medium characterisation and distances, including 2D total Galactic extinction maps, 
(c) stellar spectroscopic and evolutionary properties, including binary star characterisation,  
(d) redshifts of extragalactic objects, 
(e) outlier analysis products, and 
(f) auxiliary data.  
Most of these products are individual parameters produced on a source-by-source basis.
Multi-dimensional (MD) products are also produced, such as the two 2D total Galactic extinction maps,  
two dedicated outlier tables, and 
Markov Chain Monte Carlo samples from \gspphot\ and \msc\ containing stellar and interstellar medium parameters and distances. 
All of these data products are found in one of ten tables in the Gaia DR3 archive, with a subset of these also copied to the main archive table (\linktogstable{gaia_source}), see Sect.~\ref{sec:tables}. 

\subsection{Operations}\label{sec:content-operations}
The operations run to produce data for DR3 took a total of $\sim$92 days of continuous processing time (1\,021\,219 CPU hours).  This had been preceded by several month-long testing and validation runs, and allowed for sufficient post-operation validation time.  
With a strict delivery date for production and validation of these data of June 2021, which would ensure \gdr{3} in the first half of 2022, 
we had to impose processing limitations in some of the modules that produce stellar parameters.  This was done either on an observed $G$ magnitude basis  or RVS SNR basis. The processing limits that were imposed are the following, for \gspphot: $G \le 19$,  \flame: $G \le 18.25$, \msc: $G \le 18.25$, \espels: $G \le 17.65$, \esphs: $G \le 17.65$, \espcs: $G \le 16.62$, and for \espucd, in addition to all sources with $G \le <17.65$, we also  processed a pre-defined list of around 50 million sources with $G>17.65$.  These limits in magnitude ensured roughly the same number of objects in each magnitude bin ($\sim$130 million) and enabled the schedule to be optimized.
For \gspspec we imposed a minimum SNR~$=20$ in the RVS spectra.
This information is also provided in Table~\ref{tab:apsisalgorithms}. 

For \ugc, \qsoc, and \oa\ no limit was necessary because they process relatively few sources.  \dsc\ also had no limitations imposed because it was designed to run fast in order to process all 2 billion\footnote{The word billion implies 1\,000 million.} sources in \gaia.  The \tge\ module is very quick as it works on a HEALPix basis, but as it processes sources from \gspphot, no sources with \gmag~ $>19$ were included.
In  Fig.~\ref{fig:cmd-stellar} we show the distribution in observed colour - magnitude [(\bpminrp), \gmag] space for the 12 modules producing data on individual sources.  

\subsection{CU8 data tables in \gdr{3}}
\label{sec:tables}

The names and dimensions of the tables with CU8 parameters are summarised in Table~\ref{tab:generaltableinfo}.  
The first four tables
contain APs for which processing was done  on a source-by-source basis, such as \teff or redshifts.   
Most of the stellar parameters, classifications, individual extinction measurements, and auxiliary data are found in the \aptable\ and \apsupptable\ tables, which contain
only CU8 products.  
The former table contains one main result from each of the \apsis\ modules, while the latter table provides supplementary results in the form of specific libraries (\gspphot), methods (\gspspec), or input source types (\flame).
Some of the parameters from \dsc\ and \gspphot\ from the \aptable\ are copied to \linktogstable{gaia_source} for convenience to the user.

The \href{\linktodoc/Gaia_archive/chap_datamodel/sec_dm_main_tables/ssec_dm_galaxy.html}{\fieldName{galaxy_candidates}} and \href{\linktodoc/Gaia_archive/chap_datamodel/sec_dm_main_tables/ssec_dm_qso.html}{\fieldName{qso_candidates}} tables focus on extragalactic objects and consolidate results from different CUs.  In these tables, CU8 was responsible for the galaxy and QSO {redshifts} produced by \ugc\ and \qsoc, respectively, along with the extragalactic class probabilities and labels from \dsc.

To supplement the AP estimates from \gspphot\ and \msc, a sample of the MCMC is also provided as a datalink product. These tables, \linktotable{mcmc_samples_gsp_phot} and \linktotable{mcmc_samples_msc}, contain in addition to the sampled APs, the log posterior and log likelihood, so that the user can re-analyse the samples for their own use case, see \linksec{}{Section~11.3.3 in the online documentation} for details.  In Appendix~\ref{ssec:mcmcchains} we provide information on retrieving the MCMC data.

The primary result from the \oa\ analysis is a self-organising map of $30\times30$ neurons with a statistical description of each neuron, called \linktotable{oa_neuron_information}. 
Additionally a template spectrum for each neuron is provided in the 
\linktotable{oa_neuron_xp_spectra} table. For the sources identified as outliers the \aptable\ table contains the neuron membership information. Examples of how to exploit these data are given in Appendix~\ref{ssec:exploitoa_oa}.

Finally, the results from \tge\ are given in the \linktotable{total_galactic_extinction_map} table in the form of a two-dimensional total Galactic extinction map at 4 \hpix\ levels. The additional \linktotable{total_galactic_extinction_map_opt} table contains a \hpix\ level 9 map but based on the {\it optimal} \hpix\ level.

To help the user navigate to the appropriate table in the archive, we provide in Table~\ref{tab:overviewproducts-tables} an overview of the contents of each
table but organised by the six astrophysical parameter categories mentioned above.  
For example, if one is interested in classifications, Table~\ref{tab:overviewproducts-tables} provides the link to three relevant tables: 
\aptable, \href{\linktodoc/Gaia_archive/chap_datamodel/sec_dm_main_tables/ssec_dm_galaxy.html}{\fieldName{galaxy_candidates}}, and \href{\linktodoc/Gaia_archive/chap_datamodel/sec_dm_main_tables/ssec_dm_qso.html}{\fieldName{qso_candidates}}.
In the last column we give an overview of what type of content is found in each of
those tables for that category.
As another example, if one is interested in monochromatic extinction or extinction in the \gmag\ band, then one should query the \aptable\ and \apsupptable\ tables.

\subsection{Parameters and fields}
\label{sec:parameters_fields}

In the ten archive tables (excluding \fieldName{gaia_source}), there are a total of \totalfields fields produced by CU8 (excluding \fieldName{solution_id} and \fieldName{source_id})\footnote{This count includes two fields that are reproduced in three tables (\fieldName{classprob_dsc_combmod_quasar, classprob_dsc_combmod_galaxy} in \fieldName{astrophysical_parameters, galaxy_candidates, qso_candidates}) and three fields that are reproduced in two tables (\fieldName{classlabel_dsc, classlabel_dsc_joint, classlabel_oa} in \fieldName{galaxy_candidates, qso_candidates}).}. Each field has a field name associated with it, along with a data type, unit, and a simple and detailed description. Some of these field names are related. For example, for the chromospheric activity index there are three related \fieldName{activityindex} fields, its value, uncertainty and information pertaining to the input data.
For the stellar mass there is an upper and lower confidence level associated with the median value (three fields). Also, the mass was derived using two different sets of input data, to give a total of six \fieldName{mass} fields.
There are also some parameters that are produced by more than one \apsis\ module, such
as class probabilities (\fieldName{classprob}), \teff\ (\fieldName{teff}), and \mh\ (\fieldName{mh}).  
We refer to \fieldName{activityindex}, \fieldName{mass}, \fieldName{classprob}, \ldots as the fieldname root or \fieldName{parameter}.
To make it easier for the user to understand the AP content of \gdr{3} and to understand how each of these \totalfields
fields has been derived, the fieldname also includes the name of the \apsis\ module which
was responsible for deriving that parameter.  
We have adopted a general approach to naming the individual parameter fields and these mostly take the form of

\hspace{1.5cm}          \fieldName{parameter_module_variant_detail}\footnote{Some exceptions are: {equivalent width fields from \gspspec, \teff and \logg from \msc, \fieldName{classlabel_espels_flag}}}

Here, \fieldName{parameter} is one of the \numberuniqueparameters main parameters (not counting auxiliary data products) listed in Table~\ref{tab:dr3globalcontent}.
Next, \fieldName{module} is the name of the \apsis module that derived the AP, see Sect.~\ref{ssec:apsismodules}.
Then, \fieldName{variant} describes a variant of the method, models, or input data used to derive the AP. This part may be blank if only one method was used, or if the parameter value comes from the ``best'' of several methods.  
Finally, \fieldName{detail} may be blank if the field contains the value of the AP, otherwise it takes on values such as \fieldName{upper}, \fieldName{lower} or \fieldName{uncertainty}, where \fieldName{upper} and \fieldName{lower} imply upper and lower confidence interval (generally 68\%, but see data model descriptions), or an emission-line star type in case of a class probability field, such as \fieldName{ttauristar}.
As an example, \fieldName{teff_gspphot_marcs_upper}
is the upper confidence level of the \teff\ value estimated by the module \gspphot\ using the MARCS library of synthetic stellar spectra; 
\fieldName{classprob_dsc_combmod_quasar}
is the class probability value of being a quasar from the module \dsc using the \fieldName{combmod} method; or \fieldName{sife_gspspec_nlines} is the number of lines used to estimate the [Si/Fe] abundance from \gspspec.

Table~\ref{tab:dr3globalcontent} describes all of the unique parameters associated with the six categories  (classification, interstellar and distances, stellar-spectroscopic/evolutionary, extragalactic, outlier, auxiliary). 
The description of the unique parameter is given in the first column, and the field-name root (\fieldName{parameter}) used in the archive field name is given in the second column.
The third and fourth columns give the number of variants associated with a unique parameter and the total number of related fields, respectively. Using the above example, for \fieldName{mass} these numbers are 2 and 6, respectively. As another example, for \fieldName{classprob} there are four variants (three from \dsc\ and one from \espels) for a total of 24 related fields\footnote{including the two fields from \dsc\ reproduced in three archive tables}, so these numbers are 4 and 24, respectively, in the table.
The final column gives the maximum number of sources for which this parameter is available. 

In the next section we describe each of the \fieldName{parameter_module_variant_detail} fields grouped by category.  

\begin{table}[]
    \caption{Tables in the Gaia DR3 archive with  parameters from CU8.
 The last six are multi-dimensional data tables.
    The number of sources is approximate to the nearest million.}
    \label{tab:generaltableinfo}
    \centering
    \resizebox{\columnwidth}{!}{
    \begin{tabular}{lrr}
         table name & number of & number  \\
         &CU8 fields& of sources \\
         &&[millions] \\
         \hline\hline
         \aptable & 224 & 1\,590 \\
         \apsupptable & 173 & 470 \\
         \linktogaltable{galaxy_candidates} & 8 & 2 \\
         \linktogaltable{qso_candidates} &  11 & 6 \\
         \linktogstable{gaia_source} & 29 & 1\,800 \\
         \\
         &&dimensions\\
 \cline{3-3}\\
         \linktotable{oa_neuron_information} & 77 & 30x30 \\
         \linktotable{oa_neuron_xp_spectra} & 6 & 87x900 \\
         \linktotable{total_galactic_extinction_map} & 9  & 4xHP \\
         \linktotable{total_galactic_extinction_map_opt} & 6 & 1xHP9 \\
         \linktotable{mcmc_samples_gsp_phot} & 14 & 100x480M \\
         \linktotable{mcmc_samples_msc} & 10 & 100x480M \\
         \hline\hline
         \end{tabular}
         }
         \tablefoot{
    HP = 3\,145\,728 (=~HP9), 786\,432, 196\,608, and 49\,152 for \hpix\ level 9, 8, 7, and 6, respectively.  
     \fieldName{gaia_source} contains a subset of parameters from the \fieldName{astrophysical_parameters} table. 
     The \linktotable{mcmc_samples_gsp_phot} contains 2000 samples for sources with $G<9$ and for a random 1\% of all other sources.
}
\end{table}

\begin{table*}[]
    \caption{Overview of the contents of each table in the Gaia archive containing Apsis products, organised by product type.  A subset of the products from \dsc\ and \gspphot\ also appear in \linktogstable{gaia_source}.}
    \label{tab:overviewproducts-tables}
    \centering
    \begin{tabular}{lll}
    \hline\hline
    product type & Gaia DR3 archive tables & overview content\\
    \hline
        classification
        & \aptable & object class probabilities (quasar, galaxy, star, white dwarf, physical binary star), \\
        && emission-line class probabilities and label, spectral types \\
        &\linktogaltable{galaxy_candidates} & galaxy and QSO class probabilities and label, outlier class label \\
        &\linktogaltable{qso_candidates} & galaxy and QSO class probabilities and label, outlier class label \\
        interstellar 
        & \aptable & monochromatic extinction and extinction in $G_{\rm BP}$, $G_{\rm RP}$, $G$, \\
        && colour excess, distances, diffuse interstellar band characteristics \\
        & \apsupptable & monochromatic extinction and extinction in $G_{\rm BP}$, $G_{\rm RP}$, $G$, \\
        && colour-excess, distances \\
        & \linktotable{total_galactic_extinction_map}\tablefootmark{$a$} & total Galactic extinction 2D map at HEALpix levels 6, 7, 8, 9 \\
        & \linktotable{total_galactic_extinction_map_opt}\tablefootmark{$a$} & total Galactic extinction 2D map at HEALpix level 9 based on \\
        && the optimal \hpix\ level \\
        stellar
        & \aptable & atmospheric parameters for single and binary stars, \\
        && chemical abundances, equivalent widths, \\
        && rotation and activity parameters, evolutionary parameters \\
        & \apsupptable & atmospheric and evolutionary parameters \\
        & \linktotable{mcmc_samples_gsp_phot}\tablefootmark{$a$} & MCMC samples of stellar parameters from \gspphot \\
        & \linktotable{mcmc_samples_msc}\tablefootmark{$a$} & MCMC samples of stellar parameters from \msc \\ 
        extragalactic  
        & \linktogaltable{qso_candidates} & redshifts of qso candidates\\  
        & \linktogaltable{galaxy_candidates} & redshifts of unresolved galaxy candidates \\
                &  \aptable & class probabilities of a source being a galaxy or quasar \\
        outlier analysis 
        & \aptable & neuron membership and distance between source and neuron \\
        && prototype \bporrp spectra \\
        & \linktotable{oa_neuron_information}\tablefootmark{$a$} & self-organising map (SOM, 30x30) of outlier sources \\
        & \linktotable{oa_neuron_xp_spectra}\tablefootmark{$a$} & prototype \bporrp spectra corresponding to each SOM neuron \\
        auxiliary
        & \aptable & flags, convergence indicators, bolometric correction, library name\\
        & \apsupptable & flags, convergence indicators, bolometric correction, library name \\
        &  \linktogaltable{qso_candidates} & flags, quality indicators \\
        \hline\hline
    \end{tabular}
    \tablefoot{
     The links in the table point directly to the online data model description.
    \tablefoottext{$a$}{Multi-dimensional table.}
    }
\end{table*}

\begin{table*}
 \centering
 \caption{Summary of CU8 parameters in \gdr{3} source-based tables. For definition of 'variants' and 'related fields' see text (Sect.~\ref{sec:parameters_fields}).
   $^a$We use \fieldName{xfe} in this table to indicate abundances with respect to Fe -- these are in fact abundances of 11 different elements:
    N, Mg, Si, S, Ca, Ti, Cr, Ni, Zr, Ce, Nd, where x in the field name is replaced by the corresponding element symbol in lower-case.
      }
 \label{tab:dr3globalcontent}
 \begin{tabular}{llrrr}
                       &                &           & total   & maximum \\
                       &                & number of & related & sources \\
 parameter description & fieldname root & variants  & fields  & [million] \\
 \hline\hline
 classification \\
 \hline
 class probabilities & \fieldName{classprob} & 4 & 24 & 1\,591 \\
 class labels & \fieldName{classlabel} & 4 & 7 & 7\\
 spectral types & \fieldName{spectraltype} & 1 & 1 & 218 \\
 \hline
 interstellar \\
 \hline
 extinction in $G$ band & \fieldName{ag} & 7 & 20 & 470\\
 extinction at 541~nm & \fieldName{azero} & 7 & 20 & 470\\
 extinction in $G_{\rm BP}$ band & \fieldName{abp} & 5 & 15 & 470\\
 extinction in $G_{\rm RP}$ band & \fieldName{arp} & 5 & 15 & 470\\
 reddening & \fieldName{ebpminrp} & 6 & 17 & 470\\
 diffuse interstellar band & \fieldName{dib} & 1 & 9 & 0.5\\
 distance & \fieldName{distance} & 6 & 18 & 470\\
 \hline
 stellar-spectroscopic \\
 \hline
 effective temperature & \fieldName{teff} & 9 & 25 & 470\\
 surface gravity & \fieldName{logg} & 8 & 23 & 470\\
 global metallicity & \fieldName{mh} & 8 & 24 & 470\\
 $\alpha$-element over iron abundance & \fieldName{alphafe} & 2 & 6 & 5 \\
 iron and elemental abundances & \fieldName{fem}, \fieldName{feiim}, \fieldName{xfe}\tablefootmark{$a$} & 1 & 65 & 3\\
 activity index & \fieldName{activityindex} & 1 & 3 & 3 \\
 projected rotation velocity & \fieldName{vsini} & 1 & 2 & 2\\
 equivalent widths & \fieldName{ew} & 2 & 8 & 235\\
 binary component \teff &\fieldName{teff_msc1, teff_msc2} & 1 & 6 & 380\\
 binary component \logg & \fieldName{logg_msc1, logg_msc2} & 1 & 6 & 380 \\
 \hline
 stellar-evolutionary \\
 \hline
 luminosity & \fieldName{lum} & 2 & 6 & 280 \\
 absolute $G$ magnitude & \fieldName{mg} & 5 & 15 & 470 \\
 radius & \fieldName{radius} & 7 & 21 & 470 \\
 mass & \fieldName{mass} & 2 & 6 & 140\\ 
 age & \fieldName{age} & 2 & 6 & 130\\
 evolutionary stage & \fieldName{evolstage} & 2 & 2 & 170 \\
 gravitational redshift & \fieldName{gravredshift} & 2 & 6 & 280\\
 \hline
 extragalactic and outlier analysis\\
 \hline
 redshift & \fieldName{redshift} & 2 & 6 & 6\\
 characteristic of outlier neuron & \fieldName{neuron} & 1 & 3 & 56 \\
 \hline
 auxiliary \\
 \hline
 logarithm of posterior probability & \fieldName{logposterior} & 6 & 6 & \\
 logarithm of $\chi^2$ from fit & \fieldName{logchisq} & 2 & 2 & \\
 bolometric correction & \fieldName{bc} & 2 & 2 & \\
 processing and quality flags & \fieldName{flags} & 10 & 10 & \\
 MCMC acceptance rate & \fieldName{mcmcaccept} & 6 & 6 & \\
 drift of \msc\ MCMC chain & \fieldName{mcmcdrift} & 1 & 1 & \\
 name of best library & \fieldName{libname} & 1 & 2 & \\
 redshift CCF ratio & \fieldName{ccfratio} & 1 & 1 & \\
 redshift quality indicator & \fieldName{zscore} & 1 & 1 & \\
 \hline\hline 
 \end{tabular}
\end{table*}

\section{Catalogue results}\label{sec:content}

We describe all of the APs and other data products produced by CU8 and available in the \gdr{3} archive in this section ordered according to their category: classification (Sect.~\ref{sec:results-classification}), ISM and distances (Sect.~\ref{sec:results-ism}), stellar spectroscopic (Sect.~\ref{sec:results-spectroscopic}) and evolutionary (Sect.~\ref{sec:results-evolutionary}) parameters, extragalactic redshifts
(Sect.~\ref{sec:results-redshifts}), outliers
(Sect.~\ref{sec:results-outlieranalysis}), and auxiliary parameters
(Sect.~\ref{sec:results-auxiliary}).

\subsection{Classification}\label{sec:results-classification}
Class probabilities and class labels are provided by three \apsis\ modules for three categories of objects: \dsc\ provides the probabilities for all sources to belong to the classes quasar, galaxy, star, white dwarf, and physical binary star; \oa\ classifies sources with lower probabilities from \dsc; and \espels\ provides a spectral type classification and emission-line star types for stellar sources.

\subsubsection{\dsc}
\dsc\ provides normalized posterior probabilities for five classes from Specmod and Combmod, and for three classes (not white dwarfs or physical binaries) from Allosmod. These are all listed in the \aptable\ table. The Combmod probabilities for quasars and galaxies also appear in the \linktogaltable{qso_candidates} and \linktogaltable{galaxy_candidates} tables, and the Combmod quasar, galaxy, and star probabilities for all objects are duplicated in the \linktogstable{gaia_source} table.
Additionally, two class labels derived from these probabilities (defined in \linksec{}{Section 11.3.2 of the online documentation}),
\linktogalparam{qso_candidates}{classlabel_dsc} and \linktogalparam{qso_candidates}{classlabel_dsc_joint},
are listed in the \linktogaltable{qso_candidates} and \linktogaltable{galaxy_candidates} tables. 

\dsc\ Combmod and Specmod provide results for 1.59 billion sources. Allosmod has fewer, 1.37 billion sources, because some sources have only 2-parameter astrometric solutions (i.e.\ they have positions but lack parallaxes and proper motions). 
Users can classify sources using the probabilities, either by taking the class with the largest probability or that with the probability above some threshold (in the latter case multiple classifications or no classification is possible).

Taking a probability threshold of 0.5 on Combmod, we get around 5.2 million quasars and 3.6 million galaxies, although these samples have significant contamination. More complete numbers are given in \linksec{sec_cu8par_apsis/ssec_cu8par_apsis_dsc.html\#Ch11.T16}{Table 11.16 in the online documentation}. 
Most objects in \gaia\ are of course stars, so the star class is of little use in practice. Performance on white dwarfs and physical binaries is poor (the purities are low), and we recommend against using their probabilities  for building samples. 

\begin{figure}
\centering
\includegraphics[width=0.45\textwidth]{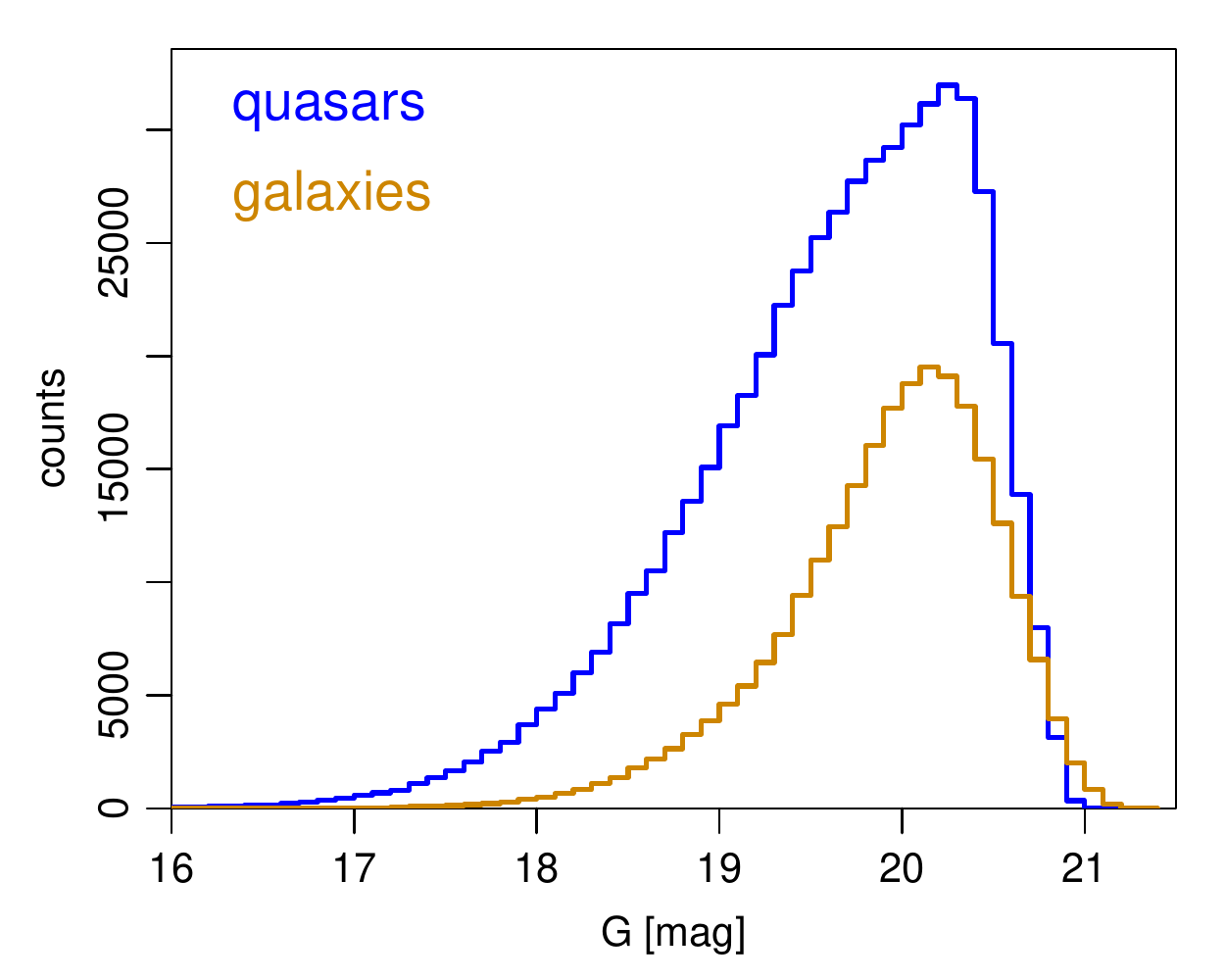}   
\caption{G-band magnitude distribution of the subset of candidates in the {\tt qso\_candidates} (blue) and {\tt galaxy\_candidates} (orange) tables identified using the {\tt classlabel\_dsc\_joint} field. These subsets comprise around 547 thousand quasars and 251 thousand galaxies.
\label{fig:dsc_joint_gmag_distribution}
}
\end{figure}

The purity and completeness of samples vary with probability threshold, as well as magnitude and Galactic latitude (and other parameters). Assessments of 
the purity and completeness are given in 
\linksubsec{sec_cu8par_apsis/ssec_cu8par_apsis_dsc}{Section 11.3.2 of the online documentation} 
(summarized in \linksec{sec_cu8par_apsis/ssec_cu8par_apsis_dsc.html\#Ch11.T17}{this table}), 
as well as in \cite{DR3-DPACP-101}, and in more detail in \cite{LL:CBJ-094}.
We see there that Specmod, Allosmod, and Combmod have rather different performances, so users may want to select on one or the other depending on their goals. More advice on the use of the \dsc\ results and the (non-trivial) interpretation of its performance can be found in \cite{DR3-DPACP-158} and
\linksubsec{sec_cu8par_apsis/ssec_cu8par_apsis_dsc}{Section 11.3.2 of the online documentation}.
The label \linktogalparam{qso_candidates}{classlabel_dsc_joint} in the \linktogaltable{qso_candidates} and \linktogaltable{galaxy_candidates} tables identify a set of extragalactic sources with purities of around 63\%, increasing to around 83\% for the subsets more than 11.5\deg\ from the Galactic plane. Their magnitude distributions are show in 
Fig.~\ref{fig:dsc_joint_gmag_distribution}.

\subsubsection{\oa}\label{ssec:classification-oa}
\begin{figure*}[h]
    \centering
    \includegraphics[width=0.75\textwidth]{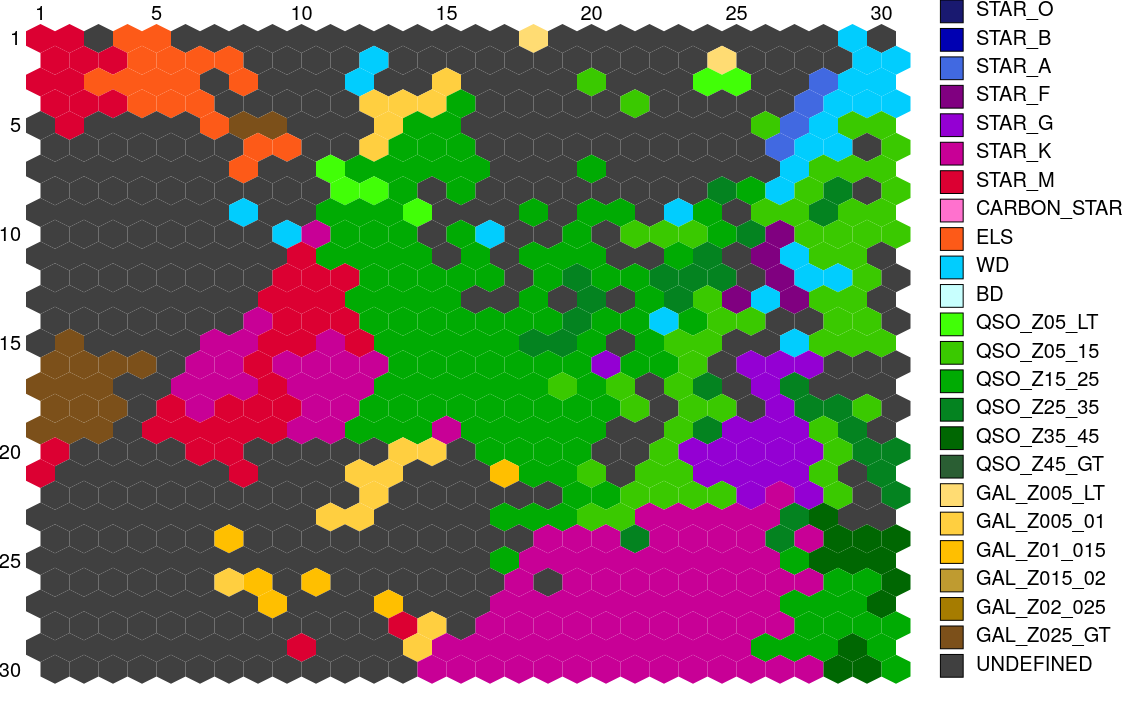}

    \caption[\oa SOM map visualised using GUASOM tool for specific class labels]{SOM map lattice visualised using the GUASOM tool \citep{Alvarez_2021} representing the specific class labels assigned to each neuron by the \oa module.  The \oa\ module  analysed the 56 million sources with the poorest classification probabilities from \dsc.  Those neurons for which such a label can not be attributed remain as ``undefined''.
    }
    \label{fig:oa_specific_label_map}
\end{figure*}

For \gdr{3}, \oa processed around 56 million objects whose \gmag\ magnitudes peaked around $20.8$ mag, in general faint stars and extragalactic objects. It provides an unsupervised classification that complements the one produced by \dsc, by analysing the sources with the lowest classification probability from \dsc.  
OA produces a  self-organising map (SOM, see Sect.~\ref{sec:results-outlieranalysis}) with 900 ($30\times 30$) neurons, see e.g. Fig~\ref{fig:oa_specific_label_map}.   An object belonging to any of the 900 neurons can be found in the \aptable\ table by the \linktoapparam{astrophysical_parameters}{neuron_oa_id}.  The associated parameters indicate how close the source is to the neuron prototype and its ranking in distance to that prototype,  
\linktoapparam{astrophysical_parameters}{neuron_oa_dist} and \linktoapparam{astrophysical_parameters}{neuron_oa_dist_percentile_rank}.  More information on OA and its multi-dimensional data is given in Sect.~\ref{sec:results-outlieranalysis}.
Some examples of exploiting these data are given in the Appendix~\ref{ssec:exploitoa_oa}

\subsubsection{\espels}\label{sec:results-classification-espels}

\begin{figure}
    \centering
   \includegraphics[width=0.5\textwidth]{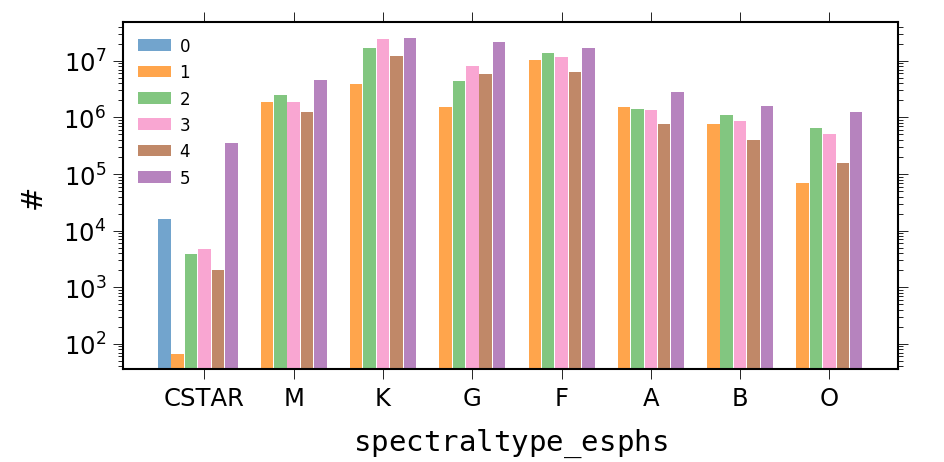}
    \caption{Histogram of the distribution of \linktoapparam{astrophysical_parameters}{spectraltype_esphs} which processed sources with $G \le 17.65$. 
     A coloured distinction is made between the different values taken by the quality assessment flag (second digit of \fieldName{flags_esphs}). Usually the flag takes values ranging from 1 to 5, with the lower value indicating higher quality. However, for the {\tt CSTAR} tag it can also be ``0''.
    }
    \label{fig:histogram_spectraltype_els}
\end{figure}

\espels\ provides for 218 million targets with $G \le 17.65$ one of the following spectral type tags
\linktoapparam{astrophysical_parameters}{spectraltype_esphs}\footnote{\esphs and \espels are related modules, and this field was originally produced by \esphs\ at the time of the archive data model definition, but it was finally produced by the upstream \espels\ module}: CSTAR, M, K, G, F, A, B, and O  (see Table\,\ref{tab:espels_elsrfc1}). An indicator of the spectral tag quality is stored in the second digit (reading from left to right) of \linktoapparam{astrophysical_parameters}{flags_esphs}. In most cases, its value ranges from 1 to 5 (the lower, the better) and is based on the relative value of the first and second highest probabilities. Value 0 was added during the validation to identify those candidate carbon stars ({\tt CSTAR} tag) having BP/RP spectra with significantly stronger C2 and CN molecular bands than in {\it normal} stars \citep{DR3-DPACP-123}.  The distribution of the spectral types according to the quality flag is shown in Fig.~\ref{fig:histogram_spectraltype_els}.  As can be seen, only the 'CSTAR' type has a value of 0. 

The module also identified 57\,511 emission-line stars, for which it suggests a stellar class (\linktoapparam{astrophysical_parameters}{classlabel_espels}) based on the combined probabilities (e.g. \linktoapparam{astrophysical_parameters}{classprob_espels_wcstar}) provided by two random forest classifiers. The ELS classes that were considered, as well as the corresponding {\tt classlabel} are: Be stars ({\tt beStar}), Herbig Ae/Be stars ({\tt HerbigStar}), T Tauri stars ({\tt TTauri}), active M dwarf stars ({\tt RedDwarfEmStar}), Wolf-Rayet WC ({\tt wC}) and WN ({\tt wN}), and planetary nebula ({\tt PlanetaryNebula})\footnote{The {\tt classprob\_espels\_*} corresponding names are {\tt bestar, herbigstar, ttauri, dmestar, wcstar, wnstar, pne}.}.

\subsection{Interstellar medium characterisation and distances}\label{sec:results-ism}

The second category of astrophysical parameters concerns the characterisation of the interstellar medium (ISM) and distances.   
Source-based ISM characterisation is provided by \gspphot, \esphs, and \msc, as one of the spectroscopic parameters estimated from BP/RP spectra (\azero, \ag, \abp, \arp, \ebpminrp) and by \gspspec\ based on the analysis of the $\lambda 862$~nm diffuse interstellar band  (DIB). 
The \tge\ module exploits individual source-based extinction from \gspphot\ to provide a 2D total Galactic extinction map.
Both \gspphot\ and \msc\ additionally estimate distances.  
Further details on most of these parameters is found in \cite{dr3-dpacp-160}, while \tge\ is discussed in \cite{DR3-DPACP-158}.

\subsubsection{\gspphot}\label{sec:results-ism-gspphot}

\gspphot\ estimates the monochromatic extinction \azero,  called 
\linktoapparam{astrophysical_parameters}{azero_gspphot},  
for all processed sources by fitting the observed BP/RP spectrum, parallax, and apparent $G$ magnitude. 
 \gspphot\ also estimates the broad-band extinctions \ag, \abp, and \arp. The latter are not free fit parameters but instead obtained from integrating an attenuated model SEDs (\linksubsec{sec_cu8par_data/ssec_cu8par_data_xp}{see Section 11.2.3 of the online documentation}). 
Using these extinction estimates, one can also compute reddenings, e.g.\ $\ebpminrp=\abp-\arp$. These extinction and reddening estimates along with upper and lower confidence levels are available in the \aptable\ table (\azero, \ag, \abp, \arp, \ebpminrp) from the best library i.e. the library that produced the highest posterior probability for that source, see \linktoapparam{astrophysical_parameters}{libname_gspphot}.
The \apsupptable\ table contains the five ISM parameters \azero, \ag, \abp, \arp, \ebpminrp\ for the individual library results (MARCS, PHOENIX, A, OB).
\gspphot\ additionally derives a  \fieldName{distance} estimate to be consistent with the inferred parameters.  
The parameters 
\linktoapparam{astrophysical_parameters}{azero_gspphot},
\linktoapparam{astrophysical_parameters}{ag_gspphot},
\linktoapparam{astrophysical_parameters}{ebpminrp_gspphot},
and 
\linktoapparam{astrophysical_parameters}{distance_gspphot}, 
and their upper and lower confidence levels
are copied from the \aptable\ table to the
\linktogstable{gaia_source} table, for convenience to the user. 
A sample of the MCMC from \gspphot\ inference is also made available as a datalink product.

\subsubsection{\msc}\label{ssec:ism_msc}
Like \gspphot, \msc also estimates the $A_0$ parameter but by assuming that the BP/RP spectrum is a composite of the two components of an unresolved binary i.e.\ two stars at the same distance with a common interstellar extinction.  
These parameters for sources with $G \le 18.25$ are found 
in the \aptable\ table:  
\linktoapparam{astrophysical_parameters}{azero_msc} and 
\linktoapparam{astrophysical_parameters}{distance_msc} along with their upper and lower 
confidence levels.
By assuming that the flux comes from a combined system, the distances  are necessarily larger than the \gspphot\ ones, \linksubsec{sec_cu8par_validation/ssec_cu8par_validation_qa_summary}{see Section~11.4.1 of the online documentation}.

\subsubsection{\esphs} \label{sec:results-spectroscopic-esphs}

\begin{figure}[!htbp]
\centering
\includegraphics[width=1\linewidth]{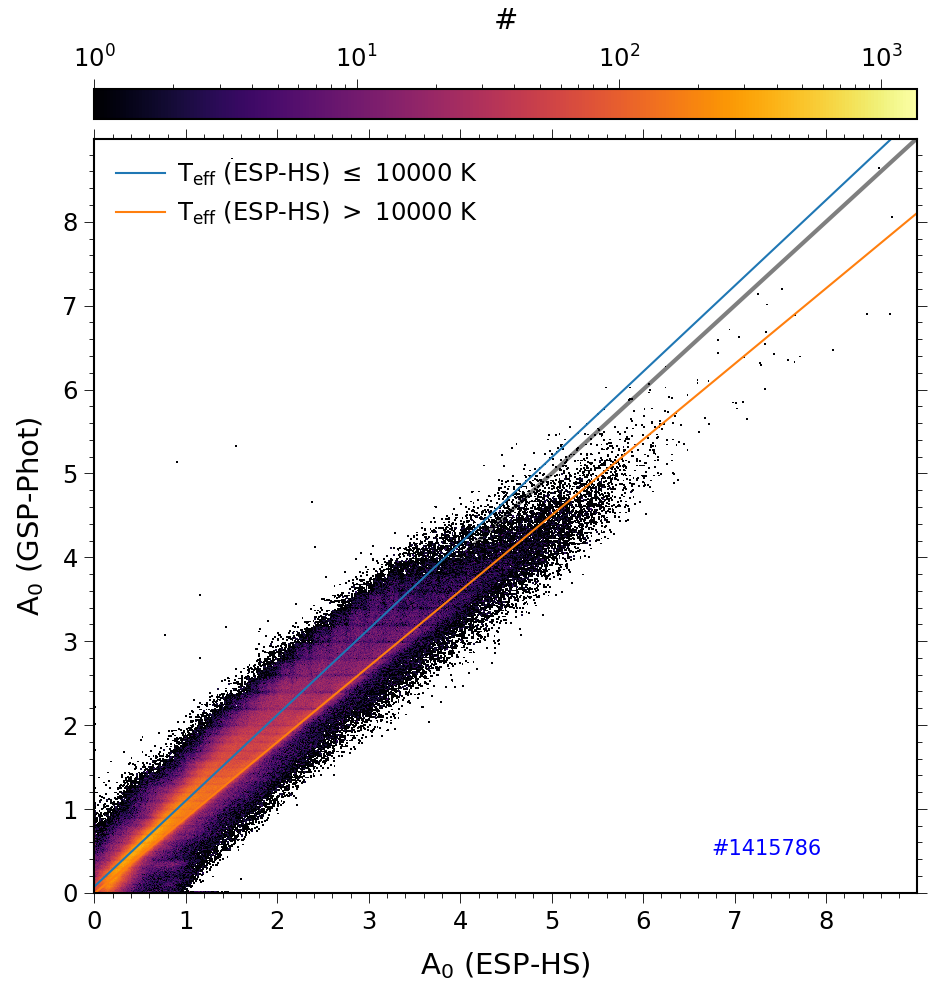}
\caption{ Distribution of the \azero\ derived by \gspphot\ vs. \azero\ derived by \esphs. The gray diagonal is the identity relation, while the blue and orange lines were fitted through the values obtained for targets cooler and hotter than 10\,000~K, respectively.}
\label{fig:gsp_esp_extinction}
\end{figure}

For stars hotter than 7\,500~K, and using a preliminary classification from \espels, \esphs\ measures  the \azero interstellar extinction by fitting the observed BP/RP and, where available, also the RVS data, called 
\linktoapparam{astrophysical_parameters}{azero_esphs}. 
While \azero\ is the free parameter representing interstellar absorption during the fit, the corresponding extinction in the $G$ band, \ag, and interstellar reddening, \ebpminrp, along with uncertainties are derived simultaneously and provided as well.
These results are found in the  
\aptable\ table for hot stars with $G<17.65$.
We show in Fig.\,\ref{fig:gsp_esp_extinction} a comparison between the \azero\ estimates for the 1\,433\,932 hot stars (\teff\ $>$ 7\,500~K) in common between \gspphot\ and \esphs. The synthetic spectra adopted by the modules are slightly different, as \esphs\ makes some corrections to account for systematics between the observations and simulations, and the wavelength range above 800 nm was not taken into account.   The impact of this is mostly seen in the B- and O-type star \teff\ range where \esphs\ estimates tend to be slightly larger than those obtained by \gspphot.

\subsubsection{\gspspec} 
\begin{figure}
    \centering
    \includegraphics[width=0.48\textwidth]{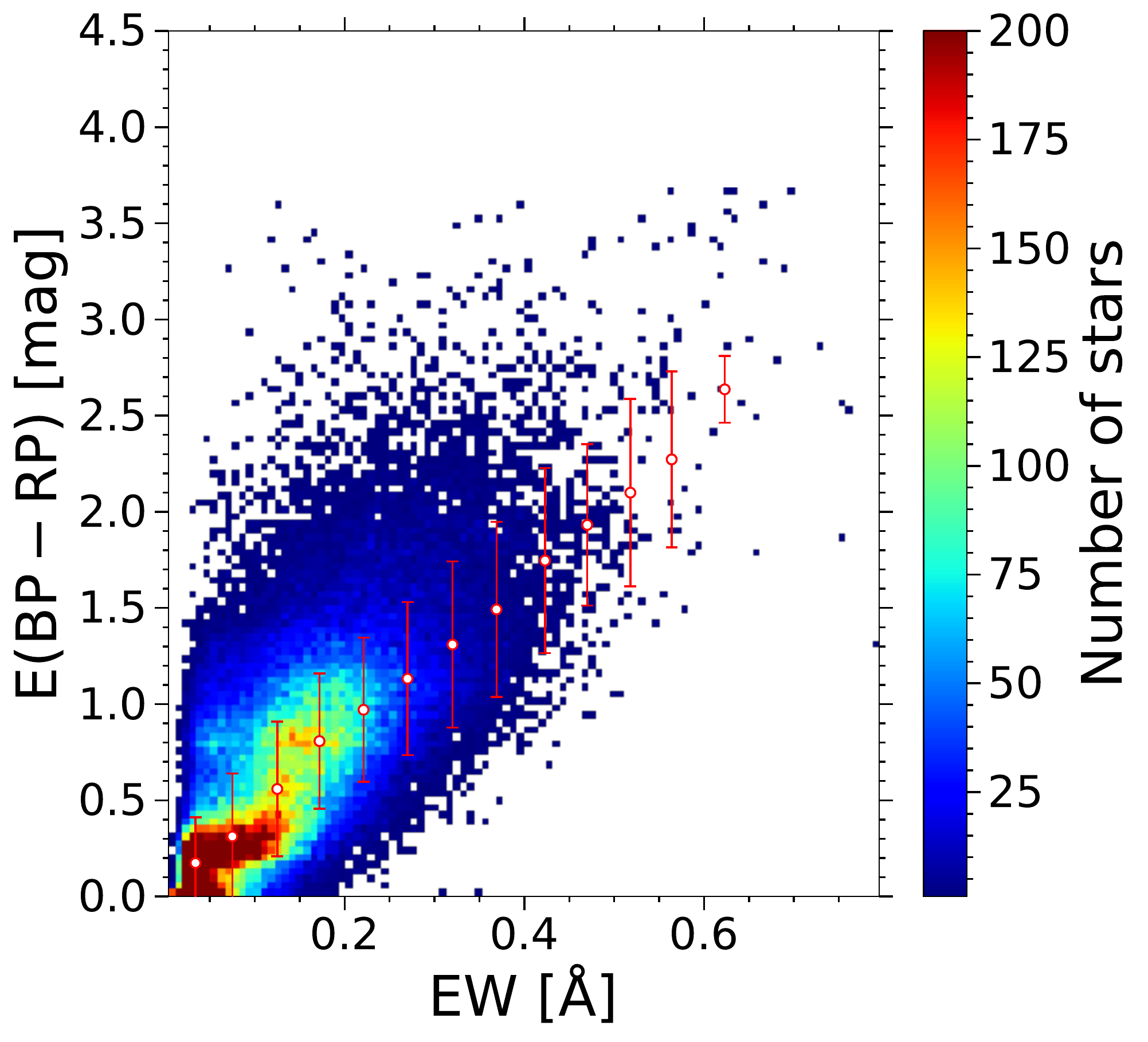}
    \caption{Comparison between \gspspec\ DIB equivalent width and \gspphot\ \ebpminrp.  The red dots are the median values of \ebpminrp\ taken in EW bins from 0.0 to 0.6\,{\AA} with a step of 0.05\,{\AA}. The error bars show the standard deviations of \ebpminrp\ for each EW bin. }
    \label{fig:dibvsebpminrp}
\end{figure}

For the sources where an analysis of the DIB in the RVS spectra 
is possible, see Fig.~\ref{fig:cu6_input_spectra} lower right panel, 
we provide a measurement of the DIB $\lambda 862$~nm equivalent width 
\linktoapparam{astrophysical_parameters}{dibew_gspspec},
and the modelled 
depth 
\linktoapparam{astrophysical_parameters}{dibp0_gspspec}
and width
\linktoapparam{astrophysical_parameters}{dibp2_gspspec} parameters, 
together with uncertainties.   
A quality flag 
\linktoapparam{astrophysical_parameters}{dibqf_gspspec}
is also available ranging from 0 (highest quality) to 5 (lowest quality).
Results for DIB measurements are available for 476\,117
stars, and are found in the  \aptable\ table for \teff\ ranging from $\sim$3000 - 50\,000 K. A comparison between \fieldName{dibew_gspspec} and \fieldName{ebpminrp_gspphot} is shown for a high-quality sub-sample in Fig.~\ref{fig:dibvsebpminrp} for stars with \linktoapparam{astrophysical_parameters}{dibqf_gspspec} $< 2$ , where  the median DIB EW  increases with \ebpminrp. A detailed discussion between the correlation of the DIB carrier and the dust extinction can be found in \citep{DR3-DPACP-144}.

\subsubsection{\tge} 

\begin{figure*}
    \centering
    \includegraphics[width=0.9\textwidth]{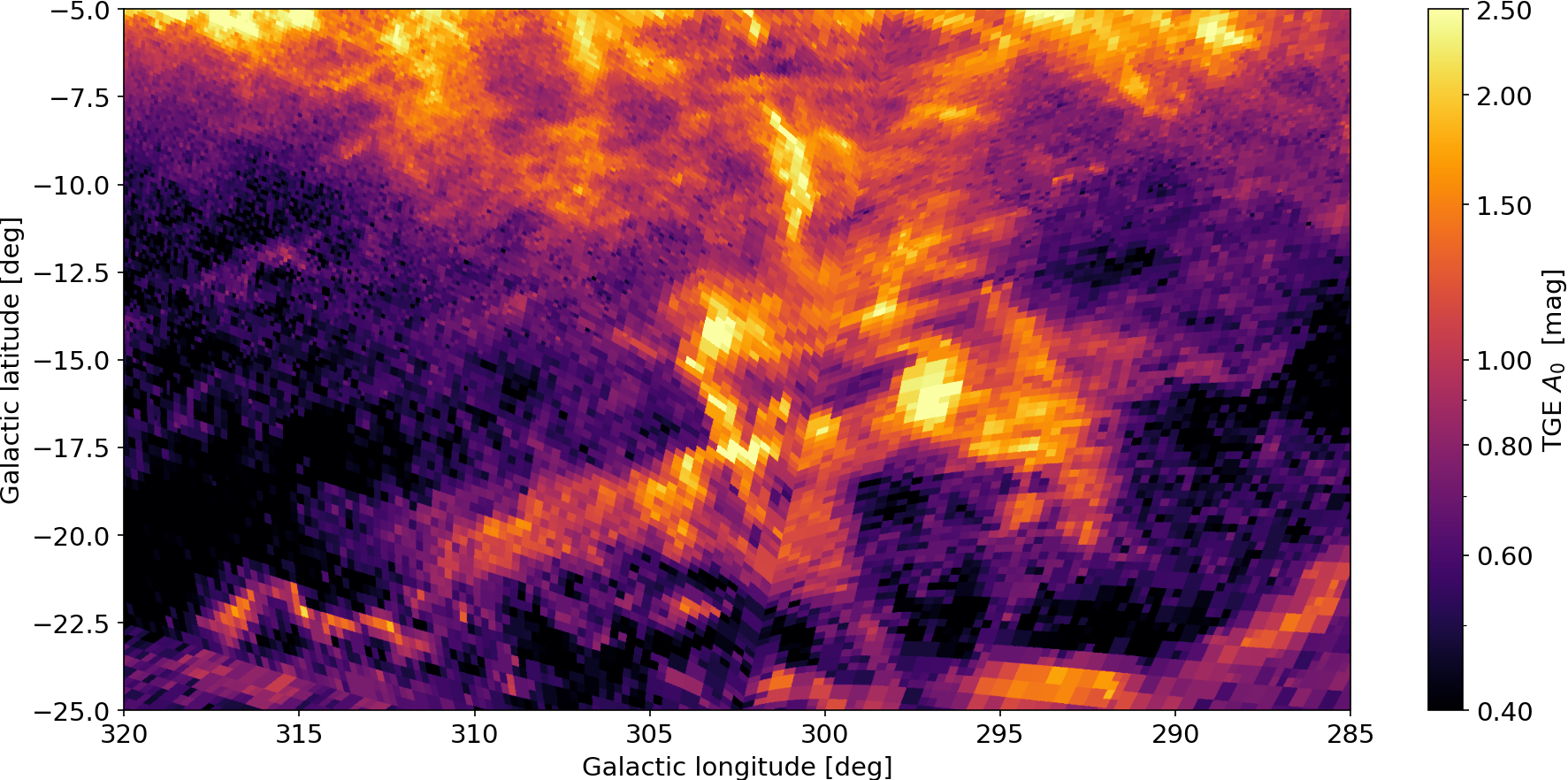}
    \caption{Total galactic extinction of the Chamaeleon region from the  \linktotable{total_galactic_extinction_map_opt}, showing the extinction at the optimal \hpix\ level (between 6 and 9). }
    \label{fig:tgemap}
\end{figure*}

All-sky \hpix\ maps of the total Galactic extinction are made available in two separate tables in the \gdr{3} archive at various resolutions (\hpix\ levels), namely the tables \linktotable{total_galactic_extinction_map} and  \linktotable{total_galactic_extinction_map_opt}.  
The estimation of the total Galactic extinction in each HEALPix is taken as the median \azero of the extinction tracers, as measured by \gspphot,  
where the tracers are giants outside the ISM layer of the disc of the Milky Way.  The first table, \linktotable{total_galactic_extinction_map}, contains HEALPix maps at levels 6 through 9 (corresponding to pixel sizes of 0.839 to 0.013 deg$^2$), with extinction estimates for all HEALPixes that have at least three extinction tracers. The second map is a reduced version of this first map, using a subset of the pixels to construct a map at variable resolution, using the highest HEALPix level available (6 through 9) that has at least ten tracers for that \hpix\ level.  
An example of the \tge\ map for the Chameleon region is shown in Fig.~\ref{fig:tgemap}.

\subsection{Stellar spectroscopic parameters}\label{sec:results-spectroscopic}

The BP/RP and RVS spectra contain information about atmospheric parameters of stars: \teff, \logg, \mh, along with chemical abundances, an activity index, equivalent widths and \vsini.
The parameters are derived by the two general stellar parametrizers: \gspphot\ and \gspspec\ based on the BP/RP and RVS spectra respectively, assuming a single source.  
Other estimates of these parameters are produced by modules working in specific stellar regimes, and depending on the scientific case the user may prefer to use these results: \esphs, \espcs, and \espucd are tailored to analyse hot stars, cool active stars, and ultra-cool dwarfs, respectively.  
Finally, \msc\ provides two \teff, two \logg, and one \mh parameter assuming that the BP/RP spectra are a combination of two components of an unresolved binary.
The quality, validation and use of the stellar spectroscopic and evolutionary parameters are described in the accompanying Paper II \citep{dr3-dpacp-160}.

\subsubsection{\gspphot}
\gspphot\ provides estimates of the \teff,  \logg, \mh, and upper and lower confidence intervals for 470 million sources. These parameters are estimated at the same time as extinction, see Sect.~\ref{sec:results-ism-gspphot} for details.  These parameters are also available in the \linktotable{mcmc_samples_gsp_phot}.
The values for the best library are provided in the \aptable\ table, and these are duplicated to  \linktogstable{gaia_source} for convenience to the user.  The auxiliary parameter  \linktoapparam{astrophysical_parameters}{logposterior_gspphot} indicates how well the data  fits the model. 
Results from individual libraries (MARCS, PHOENIX, A, OB) are available in the \apsupptable\ table.  
A HR diagram using \gspphot\ \teff\ and \flame\ \lum\ is shown in Sect.~\ref{ssec:flame} (Fig. \ref{fig:hrdiag-evol}), colour-coded by 
evolutionary stage.

\subsubsection{\gspspec}
\begin{figure*}
    \centering
    \includegraphics[width=0.98\textwidth]{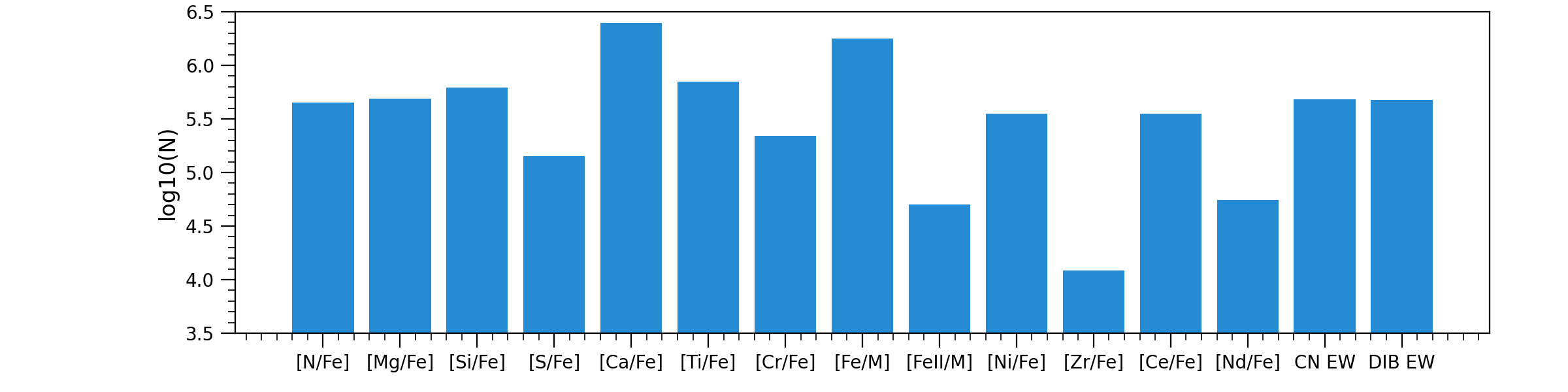}
    \caption{Histogram showing the number of sources for each chemical species with abundances or equivalent widths in \gdr{3} produced by the \gspspec\ \mgalgo\ method, in logarithmic scale.
    [$\alpha$/Fe] is derived at the same time as the atmospheric parameters (\teff, \logg, \mh) and is available for $\sim$5 million sources.  A quality flag 
    \fieldName{flags_gspspec} is provided for the best use of the elemental abundances.}
    \label{fig:abundances}
\end{figure*}

The \gspspec\ \mgalgo\ method provides \red{23} independent APs in the \aptable\  table for up to 6 million sources derived from the RVS spectra, see Fig.~\ref{fig:cu6_input_spectra} top panels. These include: \teff, log g, \mh, [$\alpha$/Fe], goodness-of-fit over the entire spectral range, individual chemical abundances of 12 elements, CN equivalent width and its fitting parameters, DIB equivalent width and its fitting parameters. For each chemical element abundance, the number of used spectral lines, as well as the line-to-line scatter are presented.  A histogram with the available chemical abundances and equivalent widths of the CN line and DIB is shown in Fig.~\ref{fig:abundances}.

A second method, \gspspec-ANN, based on the ANN method \citep{2016A&A...594A..68D,2010PASP..122..608M},  provides four APs in the \apsupptable\ table: \linktoapparam{astrophysical_parameters_supp}{teff_gspspec_ann}, \linktoapparam{astrophysical_parameters_supp}{logg_gspspec_ann}, \linktoapparam{astrophysical_parameters_supp}{mh_gspspec_ann}, 
\linktoapparam{astrophysical_parameters_supp}{alphafe_gspspec_ann}, 
and their upper and lower confidence values, along with a goodness-of-fit over the entire spectral range
\linktoapparam{astrophysical_parameters_supp}{logchisq_gspspec_ann}.

Finally, following the results of the internal \gspspec validation a long \gspspec catalogue flag has been implemented during the post-processing and published in both the \aptable\  and the \apsupptable\ tables, and the users should therefore check this flag depending on the use case of the parameters, see \linktoapparam{astrophysical_parameters}{flags_gspspec}, and 
\linktoapparam{astrophysical_parameters_supp}{flags_gspspec_ann} \citep[more details on the use of these flags are provided in][]{DR3-DPACP-186}. 
A HR diagram using \teff\ from \gspspec\ \mgalgo\ and the \flame\ luminosity is shown in Sect.~\ref{ssec:flame} (Fig. \ref{fig:hrdiag-evol}), colour-coded by 
stellar age.

\subsubsection{\espels}

\begin{figure*}
    \centering
   \includegraphics[width=0.8\textwidth]{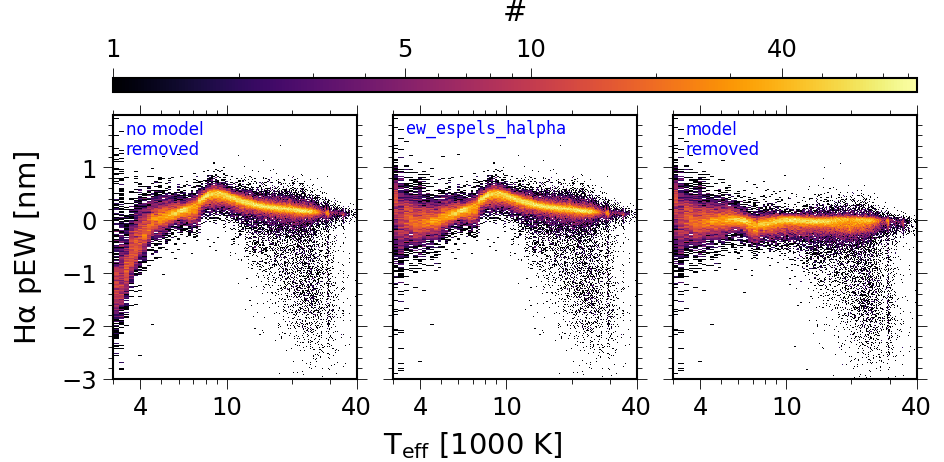}
    \caption{Distribution of the H$\alpha$ pseudo-equivalent width (pEW) obtained by \espels\ for 135\,258 targets chosen in order to homogeneously cover the temperature domain as a function of the effective temperature derived by \gspphot\ and \esphs (\teff\ $>$ 7500~K only for the latter). Left panel: We report the value obtained before the removal of the model estimate. Middle panel: We show the result saved in \linktoapparam{astrophysical_parameters}{ew_espels_halpha} (the model value is removed for sources with \teff\ $\le 5000$ K). Right panel: We show the result obtained when the model value (\linktoapparam{astrophysical_parameters}{ew_espels_halpha_model}) is also removed for stars hotter than 5000~K. 
    }
    \label{fig:halpha_els}
\end{figure*}

The \espels\ module identifies emission-line stars (ELS) in the H$\alpha$ wavelength domain. An estimate of the H$\alpha$ pseudo-equivalent width (pEW H$\alpha$), \linktoapparam{astrophysical_parameters}{ew_espels_halpha}, for 235 million stars is  provided in the catalogue. For stars having 
\linktoapparam{astrophysical_parameters}{teff_gspphot} ~$\le$~5000~K, a correction was applied to mitigate the impact of blends with spectral lines and molecular bands present in the spectra of cooler stars as follows 

\begin{equation}
\mathrm{ew\_espels\_halpha} =
\begin{cases}
\mathrm{pEW H}\alpha&\text{T$_\mathrm{eff} >$ 5\,000 K}\\
\mathrm{pEW H}\alpha - \mathrm{pEW H}\alpha^\mathrm{model} &\text{T$_\mathrm{eff} \le$ 5\,000 K}
\end{cases}
\label{eq:cu8par_apsis_espels_halpha_pew}
\end{equation}

\noindent where $\mathrm{pEW H}\alpha^\mathrm{model}$ is the pEW H$\alpha$ value as measured on the simulated/synthetic spectrum that best corresponds to the astrophysical parameters provided by \gspphot. The value of the correction  is provided by \linktoapparam{astrophysical_parameters}{ew_espels_halpha_model}. When the correction was applied, the value of the H$\alpha$ quality flag, \linktoapparam{astrophysical_parameters}{ew_espels_halpha_flag}, was set to one, if not it was set to zero. We show in Fig.\,\ref{fig:halpha_els} the temperature distribution of the pseudo-equivalent width (pEW). As expected, when the model estimate is subtracted for the cooler stars (middle panel), the H$\alpha$ pEW peaks in absorption (i.e. positive values) at temperatures between 8000 and 9000~K. When the model estimate is also applied for the hotter \teff\ (right panel), the negative estimates are expected to belong to emission-line stars.

\subsubsection{\esphs}
\begin{figure*}[!htbp]
\centering
\includegraphics[width=1\linewidth]{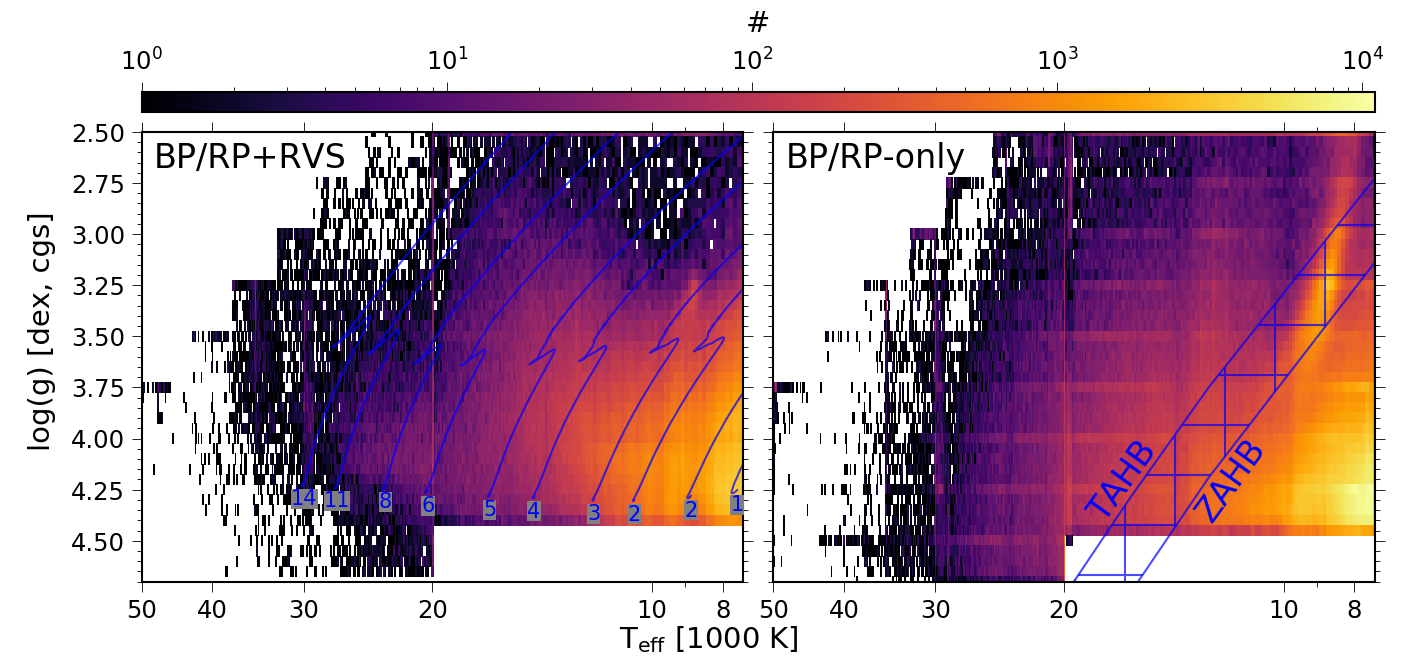}
\caption{Kiel diagram of \esphs\ results obtained in BP/RP+RVS (left panel) and BP/RP-only (right panel) processing modes. Left panel: The evolutionary tracks of \citet{2013A&A...553A..24G} for solar metallicity, and $\frac{\omega}{\omega_\mathrm{c}}$\,=\,0.8 (rotation at 80\% of its critical velocity) are shown in blue. 
The initial mass in solar masses is indicated at the start of each track. Right panel: The region occupied by hot horizontal branch (HB) stars is delimited by the expected zero age (ZAHB) and terminal age (TAHB) HB lines. ZAHB and TAHB boundaries are taken from \citet{1993ApJ...419..596D}.}
\label{fig:esphs_kiel}
\end{figure*}

\esphs\ determines the astrophysical parameters \teff (\linktoapparam{astrophysical_parameter}{teff_esphs}) and \logg (\linktoapparam{astrophysical_parameter}{logg_esphs}) of $\sim$2 million stars hotter than 7500~K according to the spectral type tag provided by \espels (\linktoapparam{astrophysical_parameters}{spectraltype_esphs}). 
These results are found in the \aptable\ table.  
The module assumes a solar chemical composition, and therefore no corresponding metallicity value is saved in the catalogue. The parameters are derived by fitting the BP/RP spectra and, when available, the RVS spectra, see Fig.~\ref{fig:cu6_input_spectra} lower panels. If RVS data are used, \esphs\ also estimates a line broadening term (i.e. aimed to take into account the broadening mechanisms not included when preparing the simulated/synthetic spectra) by assuming that it is only due to the axial rotation of the star (\vsini). Note that an attempt to measure it can only be made on the RVS data, when the instrumental broadening does not dominate. Therefore, a value of \vsini (\linktoapparam{astrophysical_parameter}{vsini_esphs}) is provided along with the spectroscopic parameters for the brighter targets (where RVS spectra was available for processing). The mode adopted to process the data is stored in the first digit reading from the left of the \esphs flag
\linktoapparam{astrophysical_parameters}{flags_esphs}.
Its value is 0 for BP/RP+RVS processing and 1 for BP/RP-only processing. 
We show in Fig.\,\ref{fig:esphs_kiel} the Kiel diagram obtained in both modes. In the fainter magnitude regime (i.e. BP/RP-only mode), the overdensity perpendicular to the main sequence is mainly due to hot horizontal branch stars as was confirmed by a systematic query in the Simbad database \citep{2000A&AS..143....9W}.

\subsubsection{\espucd}
\espucd provides \teff\ estimates, 
\linktoapparam{astrophysical_parameters}{teff_espucd},
and uncertainties for ultracool dwarfs (UCDs) 
for $\sim 94\,000$ sources in the \aptable\ table.
An input target list was provided in order to process UCDs, see Sect.~ \ref{sec:content-operations}, and these sources were 
selected according to the following criteria: 
    $\varpi > 1.7$ mas, 
    $G-G_{\rm RP} > 1.0$ mag, 
    $q_{33} > 60$, 
    $q_{50} > 71$, and
    $q_{67} > 83$, 
where $q_{33}$, $q_{50}$ and $q_{67}$ represent the pixel indices at which the 33.33, 50 and 66.67 percentiles of the total flux in the RP spectrum are attained. 
These criteria were defined using the Gaia Ultra Cool Dwarf Sample and include a safety margin to go as far as M6.

In order for the source to appear in the catalogue,  we required a \teff\ estimate in the 500~K to 2700~K range,  \linktoapparam{xp_summary}{rp_n_transits} $\ge 15$ and $\log_{10}(\sigma_\varpi) \geq -0.8 + 1.3  \log_{10}(\varpi)$, where $\varpi$ is the parallax.
We also imposed criteria 
on the RP flux and 
distance between the source RP spectrum and its nearest training set template in order to retain the source in the DR3 catalogue\footnote{These criteria are the following: 
 the normalized RP spectrum median curvature $\tau <2.0 \times 10^{-5}$ (see \linksubsec{sec_cu8par_apsis/ssec_cu8par_apsis_espucd}{Section 11.3.10 of the online documentation} for definition of $\tau$),
 $\log_{10}(d_{TS}) < -2.05$ where $d_{TS}$ is the distance to the template,
 the sum of negative normalised RP spectrum fluxes $\le -0.1$, and 
 the reddest flux corresponding to the 120-th pixel of the (normalized) RP spectrum is less  than 0.015.
}. 
Because the \teff\ is based on a regression module trained with empirical data, it should be noted by the user that results may be biased for sources with metallicity and gravity departing significantly from the training sample values 
 (solar metallicity and  
$5.0 \lesssim \log g \lesssim  5.5$).

\begin{figure}
    \centering
   \includegraphics[width=0.48\textwidth]{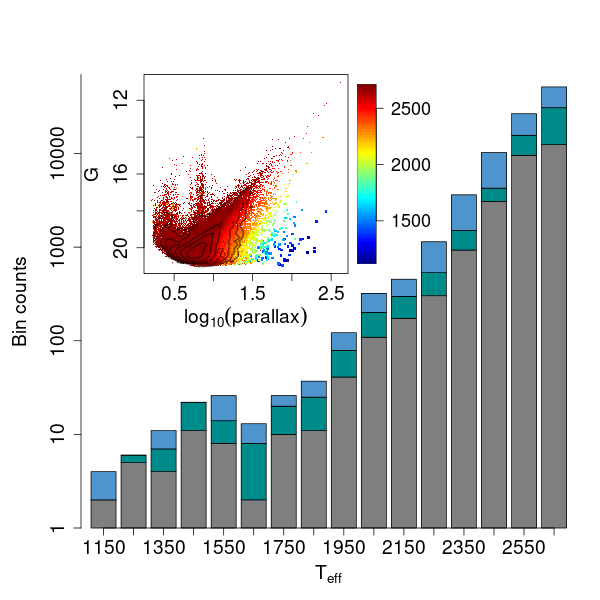}
    \caption{
    Distribution of the \teff\ of ultra-cool dwarfs from \espucd according to their quality flag (note the log scale), \linktoapparam{astrophysical_parameters}{flags_espucd} (0 = 40\,633, 1 = 26\,795, 2 = 26\,730 sources). 
    {\sl Inset:} Distribution of these same sources in \gmag\ and parallax, colour-coded by \teff.
    }
    \label{fig:ucd-teff}
\end{figure}

The final catalogue of \gaia~UCDs contains a total of { 94\,158} sources in three quality categories,  
\linktoapparam{astrophysical_parameters}{flags_espucd} = 
0 (best), 1, 2 (see \linksubsec{sec_cu8par_apsis/ssec_cu8par_apsis_espucd}{Section 11.3.10 of the online documentation} for a more detailed definition). 
In Fig.~\ref{fig:ucd-teff} we show the distribution of \teff\ for each of the quality levels across the full \teff\ range of ultra-cool dwarfs.  The inset shows the distribution of these sources in magnitude--parallax space, colour-coded by \linktoapparam{astrophysical_parameters}{teff_espucd}.

\begin{figure}
    \centering
    \includegraphics[width=0.48\textwidth]{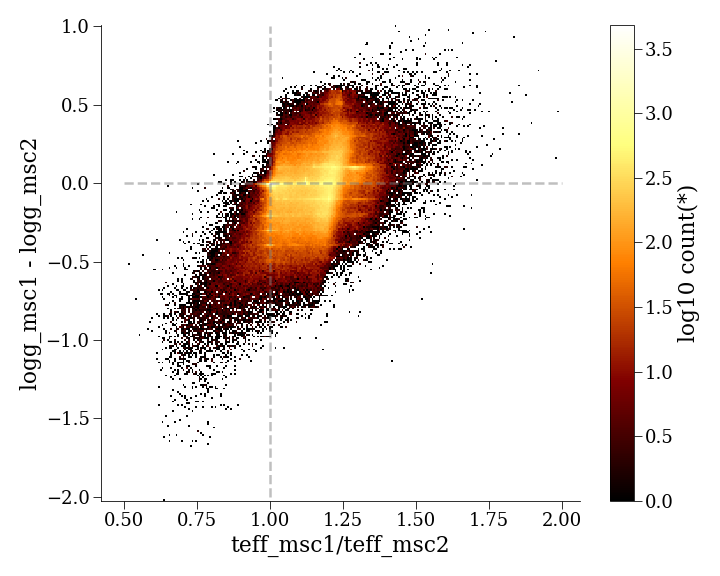}
    \caption{Distribution of difference in surface gravity ($\log g_1 - \log g_2$ versus effective temperature ratio $T_{\rm eff,1}/T_{\rm eff,2}$ of a half million random sources with results from \msc.  \msc\ assumes that each source is an unresolved binary with the same \mh, distance, and \azero. The peak uncertainty in $T_{\rm eff,1}/T_{\rm eff,2} \sim 0.2$ and  $\log g_1 - \log g_2 \sim 0.7$. } 
    \label{fig:msc_results}
\end{figure}

\subsubsection{\espcs}
In \gdr3, \espcs\ estimates a stellar activity index  \linktoapparam{astrophysical_parameters}{activityindex_espcs} and its uncertainties \linktoapparam{astrophysical_parameters}{activityindex_espcs_uncertainty}, in nm units, from the calcium infrared triplet (\cairt, at 849.8, 854.2, and 866.2~nm) in the RVS spectra, see .  
These parameters and 
a further parameter, \linktoapparam{astrophysical_parameters}{activityindex_espcs_input}
are found in the \aptable\ table for $\sim$2 million sources.
The latter parameter indicates whether the source APs  used in defining the purely photospheric spectrum to which the RVS spectrum is compared with are from \gspspec\ "M1" or \gspphot\ "M2". 
During the processing, the default value is to use the parameters from \gspspec\ because the activity index is derived from the same data as the atmospheric parameters, but when they are not available, the ones from \gspphot\ are used.

\begin{figure*}
\centering
\includegraphics[width=0.8\textwidth]{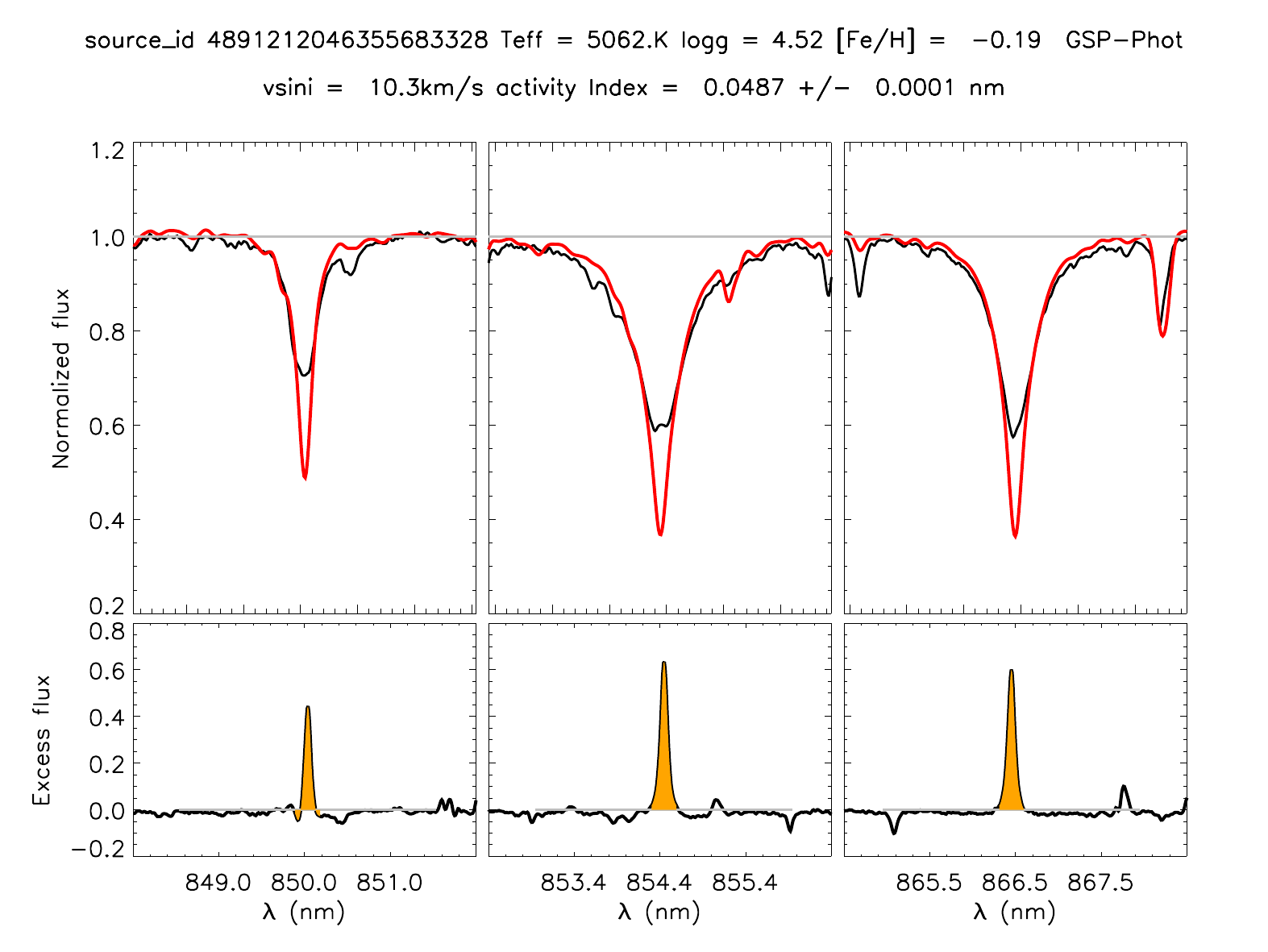}
\includegraphics[width=0.8\textwidth]{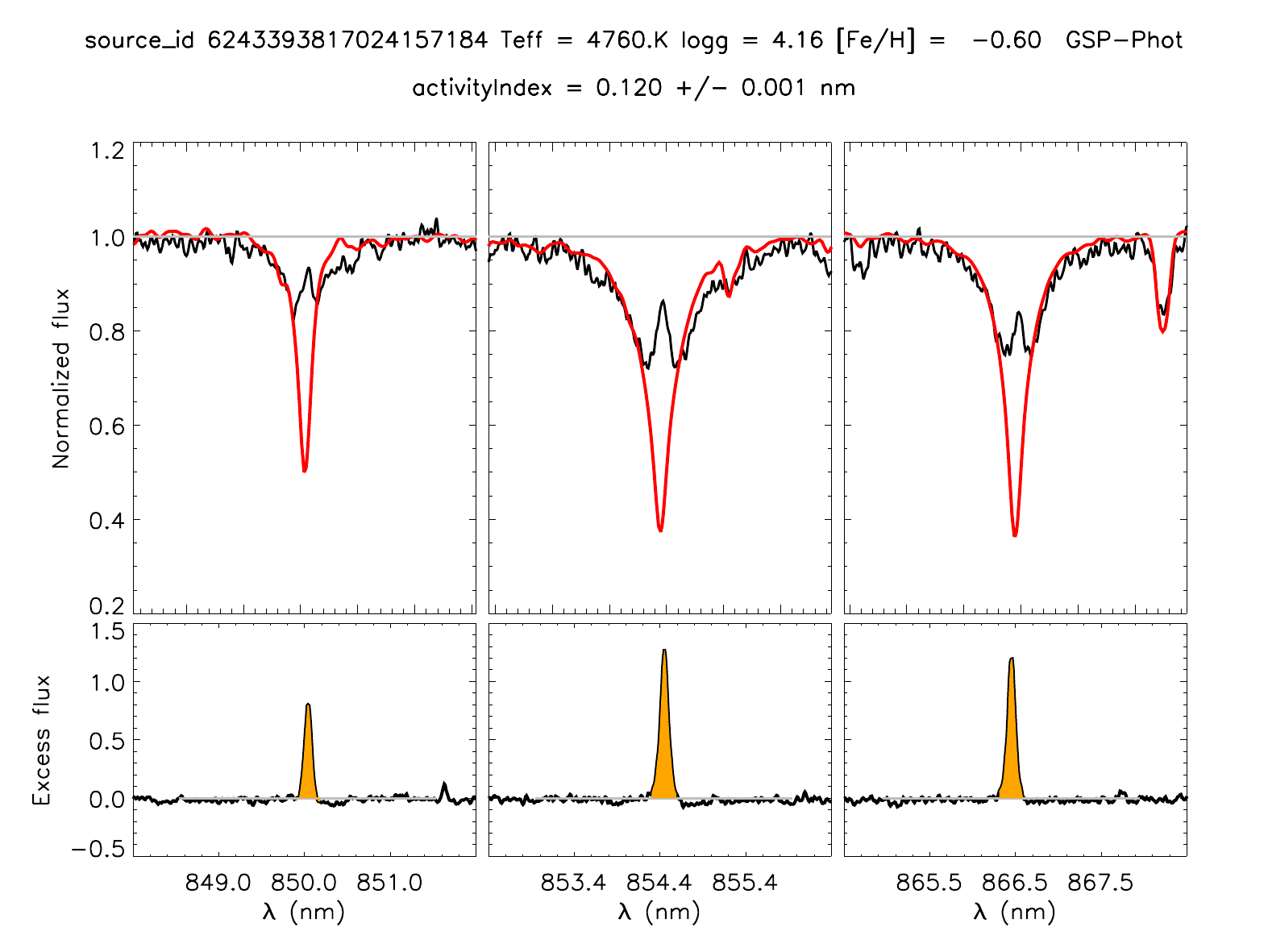}
\caption{Examples of the activity index derived by the \espcs\ module. Top panel: \cairt\ RVS spectrum of the chromospherically active star \gaia\,DR3\,4891212046355683328 (HIP\,20737) with a measured $R'_{HK} = -3.72$ from FEROS spectra using the Ca H\&K doublet. Bottom panel: RVS spectrum for the T\,Tauri star \gaia\ DR3 6243393817024157184 with a mass accretion rate $ \log \dot{M} = -10.51\, \mathrm{M_\odot yr^{-1}}$.  The same method is applied to measure these activity indices for both types of excess flux. 
Black lines are the observed spectra. Red lines are the purely photospheric spectrum template. The orange filled spectral regions are the area over which the integral of the excess flux is evaluated to produce the activity index.}
\label{fig:ESPCS_sample}
\end{figure*}

\espcs\ has processed stars with $G \lesssim 15$, \teff\ in the range (3000\,K, 7000\,K), \logg\ in the (3.0, 5.0) range, and \mh\ in the ($-$0.5, 1.0) range.
Only results for sources with the RVS spectrum $\mathrm{SNR} \ge 20$ are found in the archive.

In Fig.\,\ref{fig:ESPCS_sample} two examples of the \espcs\ analysis are shown. One is the case of the chromospherically active star \gaia\,DR3\,4891212046355683328 (HIP\,20737), with \fieldName{activityindex_espcs} $\approx 0.05$\,nm.
From the analysis of ESO-FEROS archive spectra, \cite{DR3-DPACP-175} derive a corresponding activity index  $R'_{HK} = -3.72$ \citep{1984ApJ...279..763N} from the \ion{Ca}{ii}\ H\&K doublet. 
The second example is the case of the T\,Tauri star \gaia\,DR3,6243393817024157184 with a mass accretion rate $ \log \dot{M} = -10.51\, \mathrm{M_\odot yr^{-1}}$ \citep{2020A&A...639A..58M}.

The \espcs\ activity index is given as an enhancement factor in the core of \cairt\ lines with respect to a synthetic template representing the spectrum of an inactive star with the same \teff, \logg, and \mh, see \cite{DR3-DPACP-175} for details.
Despite the fact that the method ensures that the photospheric contribution is removed from the activity index parameter, in principle, because of the contrast effect with the underlying continuum, the index derived gives a relative measure of the stellar activity at a given \teff. 
In practice, it can be used to compare stars with similar \teff\ or same spectral type, but it is unsuitable for comparing stars with very different \teff or spectral type.
\cite{DR3-DPACP-175} provides a method to derive an index $R'_{\rm IRT}$ from the \espcs activity index and \teff, which is analogous to the  $R'_{\rm HK}$ and largely independent from the contrast effect. 

In general, a value of the activity index around 0.03 -- 0.05 separates the regimes in which the chromospheric activity or mass accretion dominate. 
The separation in terms of $R'_{\rm IRT}$ is discussed in \cite{DR3-DPACP-175}.

\subsubsection{\msc}
\msc\ assumes that the \bporrp spectrum is a composition of two unresolved components of a binary system, and it estimates \teff, 
\linktoapparam{astrophysical_parameters}{teff_msc1} and
\linktoapparam{astrophysical_parameters}{teff_msc2}, 
and \logg,
\linktoapparam{astrophysical_parameters}{logg_msc1} and
\linktoapparam{astrophysical_parameters}{logg_msc2}, 
for the two components for 349 million sources, with upper and lower confidence intervals.  It assumes a solar metallicity prior and estimates one unique metallicity for each source
\linktoapparam{astrophysical_parameters}{mh_msc}.  
These parameters are inferred at the same time as \azero\ and distance, see Sect.~\ref{ssec:ism_msc}
In Fig.~\ref{fig:msc_results} we show the distribution of the temperature ratio and \logg\ differences of the individual components according to the \msc\ assumption of an unresolved binary.  The grey dashed lines indicate where two sources are of equal mass i.e. same \teff\ and \logg.   
Results from \msc\ are in \gdr{3} if  $G<18.25$.  However, a user will need to construct a binary sample using external literature sources, or other indicators of binarity, such as {\tt classprob\_dsc\_combmod\_binary} or the other tables in the archive indicating binarity, see \cite{DR3-DPACP-100}.

\subsection{Stellar evolutionary parameters}\label{sec:results-evolutionary}

By stellar evolutionary parameters we imply the following: mass \mass\footnote{Mass is technically not an evolution parameter but it is the most important physical property responsible for evolution.}, luminosity \lum, absolute magnitude \mg, 
radius \radius, radial velocity\footnote{The  \linktogsparam{gaia_source}{radial_velocity} parameter produced by CU6 is found in \linktogstable{gaia_source}.  In \gdr{3} it is not corrected for the gravitational redshift nor the convective shift.} correction for the stellar gravitational redshift, \gravshift, the age of a star $\tau$, and the evolutionary stage ({\tt evolstage}).
The parameters \gravshift\ and age are in units of kilometers per second and gigayears, respectively.
The parameter {\tt evolstage} is
an integer from 100 to 1300 indicating the phase the star is in, in the evolution sequence, see \cite{2018ApJ...856..125H}.
Most of these parameters are derived by \flame\ but \gspphot\ also derives \mg\ and \radius. These parameters show good agreement between the two modules in most parameter ranges, see Paper II \citep{dr3-dpacp-160} and \linksec{sec_cu8par_validation/ssec_cu8par_qa_evolution_aps}{Sect.~11.4.5 of the online documentation} for these comparisons.

\begin{figure}
    \centering
    \includegraphics[width=0.48\textwidth]{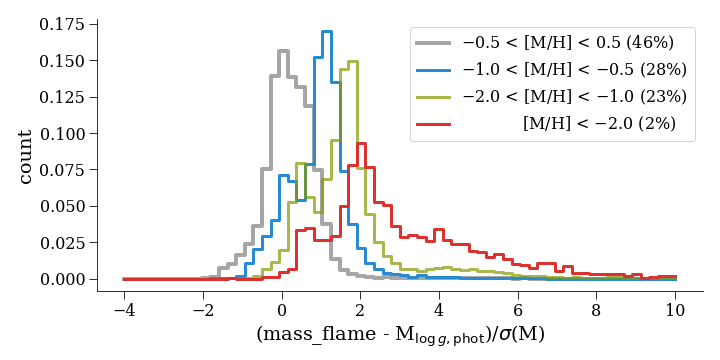}
    \caption{Comparison of the mass derived from \gspphot\ using \logg\ and \radius, M$_{\rm \log g,phot}$, and \linktoapparam{astrophysical_parameters}{mass_flame} for stars of different metallicities.  The histograms are normalised for visual purposes and the  relative number of stars in each sample is indicated in the label.}
    \label{fig:mass_flame_gspphot}
\end{figure}

\subsubsection{\flame}\label{ssec:flame_evol_mrl}
\flame\ produces all of the evolutionary parameters except for \mg, although this can be derived directly using \lum\ and \bcg.
Two separate results are provided in the archive: the first in the \aptable\ table is based on the "best" library
from \gspphot\ for  $\sim${280} million sources. 
A second set of results are based on the \mgalgo\ \gspspec\ parameters for approximately {5} million sources, and these are found in the supplementary table 
\apsupptable.  Not only are the data found in separate tables, but the field names are also distinguishable with the latter containing {\tt spec} e.g. 
\linktoapparam{astrophysical_parameters}{mass_flame} and 
\linktoapparam{astrophysical_parameters}{mass_flame_spec}. 

The values of \mass, \radius, \lum, \gravshift, and age are accompanied by an upper and lower confidence level encompassing a confidence interval of 68\% e.g. \linktoapparam{astrophysical_parameters}{mass_flame_upper}.
To derive \lum\ a bolometric correction for the $G$ band is needed,
see Sect.~\ref{ssec:bolometric_correction}, and this is provided as an auxiliary parameter 
\fieldName{bc_flame(_spec)}\footnote{A tool has been provided to calculate this, see \url{https://gitlab.oca.eu/ordenovic/gaiadr3_bcg}}.  
An estimate of the distance is also needed: for the results in the \linktotable{astrophysical_parameters} table either the \linktogsparam{gaia_source}{parallax} or 
\linktoapparam{astrophysical_parameters}{distance_gspphot} is used.   This processing information
is provided as the second character of \linktoapparam{astrophysical_parameter}{flags_flame} where '0' implies the use of the parallax, '1' is the use of \fieldName{distance_gspphot}, while '2' is also parallax but where convergence issues with \fieldName{distance_gspphot} have been reported, see \linksec{}{the online documentation for details}.

A solar metallicity-prior was assumed for deriving 
\mass, age and \fieldName{evolstage}, due to known but unquantified issues with \mh\ at the time of operations, see Sect.~\ref{sec:discussioncaveats}.  This assumption does have an impact on the results for non-solar metallicity stars, in particular for the age, 
where metal-poor/metal-rich stars will have a biased age towards younger/older ages, see \cite{LL:OLC-035} for further discussion.  One should therefore be cautious when using the age value outside of the --0.5~$<$~\mh~$<$~+0.5 regime.     

As we can derive a mass using \linktoapparam{astrophysical_parameters}{logg_gspphot} and \linktoapparam{astrophysical_parameters}{radius_gspphot} we can investigate the impact of the solar-metallicity assumption on the masses by comparing these to \linktoapparam{astrophysical_parameters}{mass_flame}.
In Fig.~\ref{fig:mass_flame_gspphot} we show the differences in the two mass determinations, normalised by their joint uncertainties, for solar-metallicity stars (grey), and then for non-solar metallicity stars (blue, green, red).  The histogram is normalised for visual purposes and the percentage of stars in each histogram out of the total is indicated on the label.  One can see that for low-metallicity stars ($<-2.0$), the mass from \gspphot\ differs by typically 2$\sigma$ or more from \flame.  One may prefer to use such an estimate of mass for the roughly 2\% of lower-metallicity stars.

Determining the masses and ages of giants is a delicate task compared to the less evolved stars, and our validation showed that the masses should be used with caution for evolved stars.
We have therefore added quality information in
\linktoapparam{astrophysical_parameters}{flags_flame} and
\linktoapparam{astrophysical_parameters_supp}{flags_flame_spec} which takes a  value of '1' as the first character to indicate
that the star is a giant with a published mass (and usually an age) and that these corresponding parameters should be used with caution.  
Additional validation showed that results for giants with \mass\ $>2$ \Msun\ are mis-classified and should not be used.

\begin{figure}
    \centering
    \includegraphics[width=0.49\textwidth]{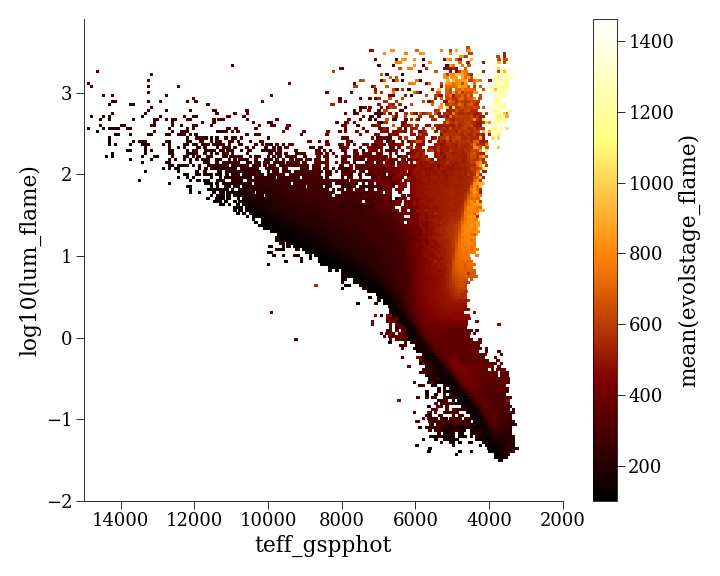}
    \includegraphics[width=0.49\textwidth]{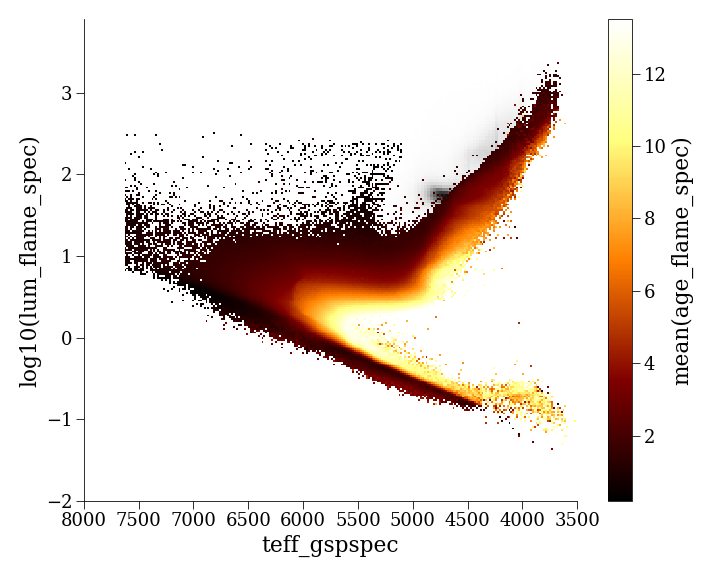}
    \caption{HR diagrams colour-coded by \flame\ parameters.  Note that the colour-scale is linear and not a density plot.  
    {\sl Top:} \fieldName{lum_flame} versus \fieldName{teff_gspphot} colour-coded by \fieldName{evolstage_flame} for stars with relative parallax errors better than 10\%. We have applied the recommended \flame\ filter for giants.   
    {\sl Lower:}
    \fieldName{lum_flame_spec} versus \fieldName{teff_gspspec} colour-coded by \fieldName{age_flame_spec} for stars with \fieldName{flags_gspspec} {\tt like '0000000000000\%'}.  In the background the red clump can be seen in grey.  These have no age values associated with them.  
    Even though we have made selections on quality on certain parameters, there are still some artefacts that can be seen, such as the high-luminosity, low \teff\ giants in the upper panel colour-coded by yellow, or the high-luminosity low mass main sequence stars in the lower panel.  These artefacts can be removed by filtering on luminosity uncertainty, or requiring that the \teff\ from both \gspspec\ and \gspphot\ agree to e.g. 300 K, or filtering on spectra SNR.}
    \label{fig:hrdiag-evol}
\end{figure}

HR diagrams are shown in Fig.~\ref{fig:hrdiag-evol}
using a random subset of 2 million sources from \gdr{3}: the top panel shows \linktoapparam{astrophysical_parameters}{lum_flame} versus \linktoapparam{astrophysical_parameters}{teff_gspphot} 
colour-coded by 
\linktoapparam{astrophysical_parameters}{evolstage_flame} while the bottom panel shows \linktoapparam{astrophysical_parameters_supp}{lum_flame_spec} versus \linktoapparam{astrophysical_parameters}{teff_gspspec} colour-coded by 
\linktoapparam{astrophysical_parameters}{age_flame_spec}. 

\subsubsection{\gspphot}

Given \gspphot's use of isochrones in a forward model context, see
 Sect.~\ref{sec:results-ism-gspphot}, \gspphot\ also provides estimates of absolute magnitude \mg\ and radius \radius,
\linktoapparam{astrophysical_parameters}{mg_gspphot} and 
\linktoapparam{astrophysical_parameters}{radius_gspphot}, and 
upper and lower confidence levels. These parameters are found in the \aptable\ table for 470 million sources, and the results for the individual libraries are found in the \apsupptable\ table. 
From these \gspphot\ results, the user could also compute \gspphot\ estimates of the (bolometric) luminosity,
\begin{equation}
    \frac{L}{L_{\odot}} = \left(\frac{R}{R_\odot}\right)^2\left(\frac{\teff}{5772{\rm K}}\right)^4
    \,\textrm{.}
\end{equation}
While \gspphot\ does not directly provide absolute magnitudes in the BP or RP bands, \gspphot\ does estimate the distance and extinctions \abp\ and \arp. Therefore, the user can compute absolute magnitudes $M_\textrm{XP}$ using the observed apparent magnitudes $G_\textrm{XP}$ via
\begin{equation}
    M_\textrm{XP}=G_\textrm{XP} - A_\textrm{XP} - 5\log_{10}d + 5
    \,\textrm{,}
\end{equation}
where XP stands for BP or RP, and $d$ is the distance in parsec. Given those, the user can then compute bolometric corrections in those bands via
\begin{equation}
    \textrm{BC}_\textrm{XP}=4.74-2.5\log_{10}(L/L_\odot) - M_\textrm{XP}
    \,\textrm{.}
\end{equation}
Uncertainties on those additional quantities can be obtained from the \gspphot\ MCMC samples (see Appendix~\ref{ssec:mcmcchains}) by processing all samples through those equations and then computing, e.g., their median values and quantiles.

\subsection{Extragalactic redshifts}\label{sec:results-redshifts}

The redshifts of extragalactic objects are produced by two modules, \qsoc\ and \ugc, which analyse BP/RP spectra of quasars and galaxies, respectively.   The selection of the processed sources uses the Combmod class probabilities of \dsc, 
\linktoapparam{astrophysical_parameters}{classprob_dsc_combmod_quasar} and
\linktoapparam{astrophysical_parameters}{classprob_dsc_combmod_galaxy}.  
The CU8 extragalactic parameters are found in the \linktogaltable{qso_candidates} and \linktogaltable{galaxy_candidates} tables.
More details on the quality and processing of these parameters are given in the accompanying Paper III \citep{DR3-DPACP-158}.

\subsubsection{\qsoc}

\begin{figure}
    \centering
    \includegraphics[width=0.49\textwidth]{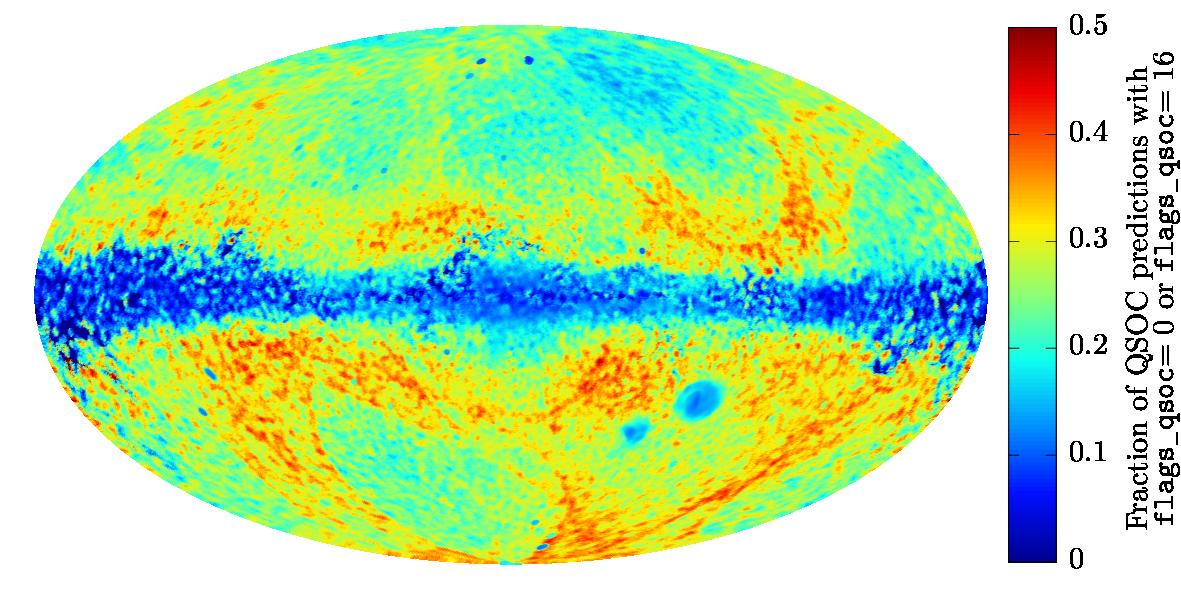}
    \caption{Galactic sky distribution of the fraction of \qsoc predictions that do not raise warning flags (i.e. \linktoapparam{qso}{flags_qsoc}$=0$), even if they are based on BP/RP spectra of lower quality (i.e. flag \texttt{Z\_BAD\_SPEC}$=16$ can be set). Fractions are computed within radii of $1\deg$ over the whole celestial sphere.}
    \label{fig:qsoc_gal_zwarn}
\end{figure}

\qsoc predicts quasar redshifts \linktoapparam{qso_candidate}{redshift_qsoc}  and associated confidence levels  \linktoapparam{qso_candidate}{redshift_qsoc_lower} and \linktoapparam{qso}{redshift_qsoc_upper} 
in the range $0.0826 < z < 6.1295$. 
Intentionally, the module chose to be complete and produced results on 6.4 million 
sources, which is three times the expected number of quasars that \gaia\ should theoretically observe. Although this choice may seem  questionable, it gives the final user a much higher chance of finding the redshift of the sources they are interested in, at the expense of having many contaminating stars amongst these predictions \citep[see however][their Section {8} for the selection of purer samples]{DR3-DPACP-101}.

In order to help discriminating valuable redshift predictions from those where potential processing issues may arise, we defined two quality measurements: \linktogalparam{qso_candidates}{ccfratio_qsoc} and \linktogalparam{qso_candidates}{zscore_qsoc}. The \linktogalparam{qso_candidates}{ccfratio_qsoc} field is associated with the $\chi^2$ resulting from the fit of the BP/RP spectra to the templates at the predicted redshift. Predictions whose redshift is associated with a minimal $\chi^2$ (compared to the $\chi^2$ resulting from alternative redshifts) have \linktogalparam{qso_candidates}{ccfratio_qsoc} $ = 1$ and less than one otherwise. The \linktogalparam{qso_candidates}{zscore_qsoc} field is associated with the successful modelling of common quasar emission lines. We have that \linktogalparam{qso_candidates}{zscore_qsoc} $ = 1$ if all covered quasar emission lines appear in the spectrum whereas the miss of a single emission line often leads to very low values of \linktogalparam{qso_candidates}{zscore_qsoc}. These quality measurements are summarised in the \linktogalparam{qso_candidates}{flags_qsoc} field where boolean flags are set that principally depend on the values of the \linktogalparam{qso_candidates}{ccfratio_qsoc} and \linktogalparam{qso_candidates}{zscore_qsoc} fields.

To illustrate the potential filtering that can be done using \linktogalparam{qso_candidates}{flags_qsoc}, in Fig.~\ref{fig:qsoc_gal_zwarn}  we show the sky distribution of the fraction of \qsoc predictions for which all flags other than the \texttt{Z\_BAD\_SPEC}\footnote{The \texttt{Z\_BAD\_SPEC} flag is raised for sources of lower quality, and rely on a combination of the source \gmag\ magnitude and number of \bporrp spectral transits.} flag are set to zero. These correspond to sources where no processing error occurs, even though some predictions are based on spectra of lower quality (i.e. predictions with either \linktogalparam{qso_candidates}{flags_qsoc}~$=0$ or \linktogalparam{qso_candidates}{flags_qsoc}~$=16$). We can clearly see that high stellar density regions (Galactic plane, Magellanic Clouds, globular clusters and nearby galaxies) usually have a lower fraction of predictions with \linktogalparam{qso_candidates}{flags_qsoc}~$=\lbrace0, 16\rbrace$. Imprints of the scanning law are also seen. These arise from the higher or lower number of spectral transits that leads to a higher or lower SNR of the spectra and hence more or less confident predictions by \qsoc.

\subsubsection{\ugc}

The \ugc\ module provides galaxy redshift parameters  \linktogalparam{galaxy_candidates}{redshift_ugc},  with $0.0 \le z \le 0.6$, and associated uncertainties \linktogalparam{galaxy_candidates}{redshift_ugc_lower} and  \linktogalparam{galaxy_candidates}{redshift_ugc_upper}, for $1,367,153$ sources in the \linktogaltable{galaxy_candidates} table. 
Note that these uncertainties are computed from the standard deviation of the SVM predictions of sources with known redshift and should accordingly not be considered as per-source confidence intervals but rather as a measure of the SVM performance.

\begin{figure}
    \centering
     \includegraphics[width=0.5\textwidth]{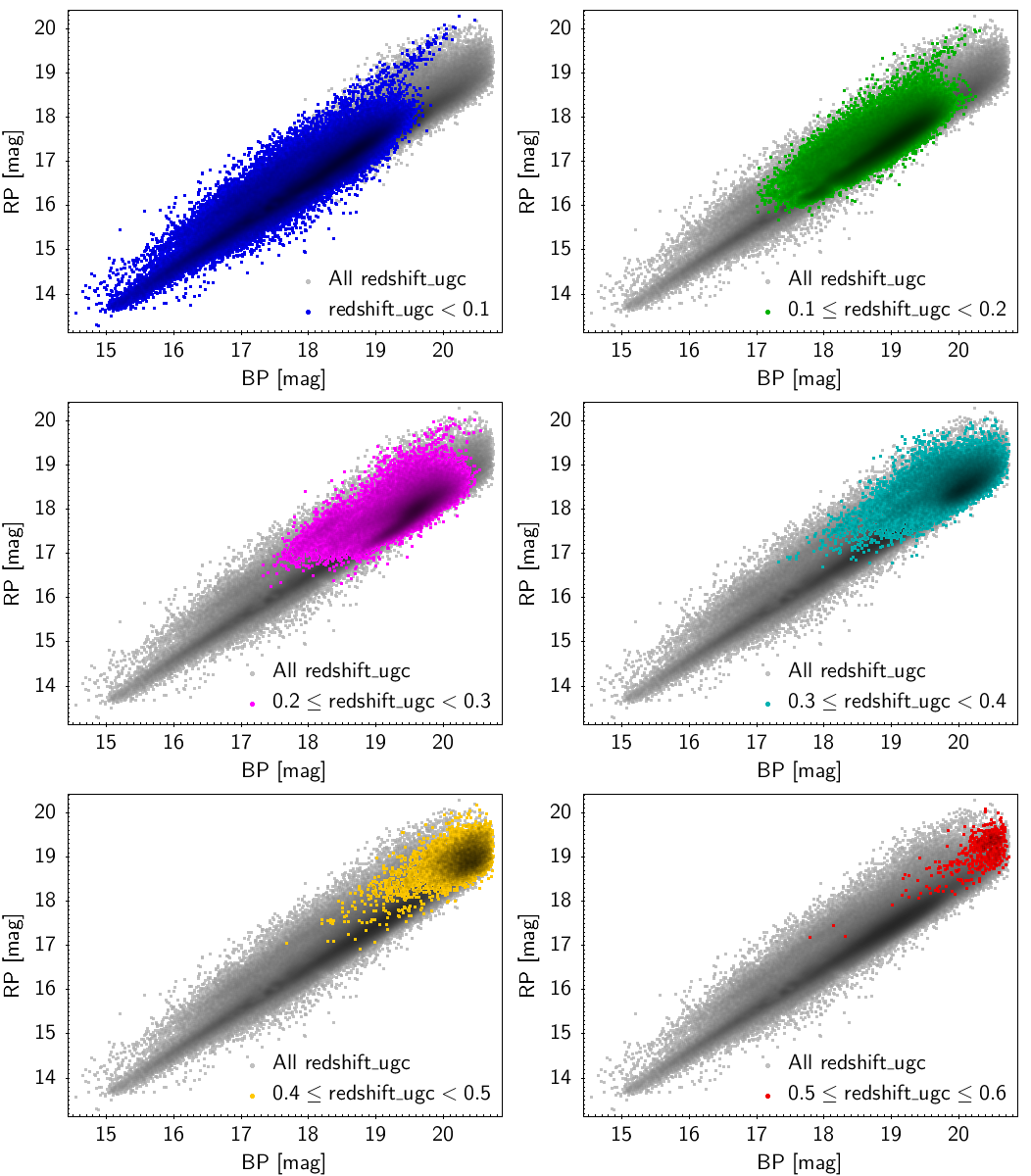}
    \caption{Magnitude-magnitude diagrams for the 1.4 million galaxies for which redshifts are provided by \ugc.  Each panel shows the density distribution of a different redshift range in a different colour, while the distribution in grey shows the whole sample.}
    \label{fig:ugc_redshifts}
\end{figure}

As the sources in \ugc may have a relatively low probability to be galaxies (our selection is \linktoapparam{astrophysical_parameters}{classprob_dsc_combmod_galaxy} $>0.25$), we expect a number of misclassified quasars to contaminate the \linktogalparam{galaxy_candidates}{redshift_ugc} results. Potentially, about $1\%$ of these are true
quasars with redshifts  $z>0.6$. It is also expected that some true high-redshift galaxies are erroneously processed by \ugc, however  their number is estimated to be negligibly small.  Their predicted redshifts would therefore be underestimated by \ugc. 
Estimated redshifts below 0.02 or larger than 0.40 are not well constrained, and there is a suspicious peak of sources in a very narrow bin at $0.0707<z<0.0709$ 
(for details see \linksec{ssec:cu8par_apsis_dsc}{Section~11.3.13 of the online documentation}).
In Fig.~\ref{fig:ugc_redshifts} we show \rpmag\ versus \bpmag\  diagrams illustrating the magnitude ranges for which we find galaxies with specific redshifts. Higher-redshift galaxies ($z\sim 0.6$) are only found at the very faint end, as expected.

\subsection{Outlier analysis}\label{sec:results-outlieranalysis}
The results produced by \oa can be helpful to perform an extensive analysis of those sources that were assigned a lower classification probability by \dsc. Such sources are usually faint stars or extragalactic objects, which can be studied through the parameters generated by \oa, but they also contain known objects such as white dwarfs and brown dwarfs.
The sources classified as outliers are given in the \aptable\ table, as explained in Sect.~\ref{ssec:classification-oa}.  
Two multi-dimensional tables are also available to further interpret these data.  

The \linktotable{oa_neuron_information} table is a self-organising map (SOM) arranged in a rectangular lattice composed of 900 neurons, where each neuron groups similar objects that are described by means of different statistical parameters.   
    In order to assess the quality of the clustering, different indices are available in this same table, so that high quality or low quality neurons can be identified and filtered as required by the user to perform their own analysis or to isolate specific types or groups of objects.   To ease such an analysis, an indication of the astronomical type of the sources is also provided for the best quality neurons. 
 The SOM is shown in Fig.~\ref{fig:oa_specific_label_map} and the coloured neurons are the best quality ones with class labels.
 
Each neuron is associated with a synthetic BP/RP spectrum, the so-called prototype, which is representative of the spectra of the sources that are assigned to a certain neuron.      In addition, for those neurons where a class label is provided, the BP/RP spectrum of the template is found in the \linktotable{oa_neuron_xp_spectra} table.  Further information is provided in the accompanying Paper III \citep{DR3-DPACP-158}.

\subsection{Auxiliary data products}\label{sec:results-auxiliary}
The auxiliary data products comprise quality metrics, convergence indicators, flags, the name of the 
best library for the \gspphot\ results, and the bolometric correction.   Most of these auxiliary data products have been described in one of the above subsections.   These are also listed towards the bottom of Table~\ref{tab:dr3globalcontent}.


\section{Validation of results}\label{sec:validation}

The aim of this paper is to explain the production of and overall content of the astrophysical parameters from CU8 in \gdr{3}.   The accompanying Papers II and III focus on the validation and quality of the results for stellar-based APs \citep{dr3-dpacp-160} and the non-stellar content and source classification \citep{DR3-DPACP-158}.   Additional validation results are also given in the dedicated \linksec{}{online documentation} chapter.  
Validation on \gspphot, \gspspec, and \espcs-specific
products are also found in their dedicated papers \citep{DR3-DPACP-156,DR3-DPACP-175,DR3-DPACP-186}.
Here we briefly describe our validation procedures.

The validation of the CU8 data products included several steps.  
At a first level, many \apsis\ test runs were performed prior to receiving the upstream data during the DR3 development stage (2018 -- 2020).
This repeated validation was done on a module-by-module basis for a limited number of mostly random sources (10 million).  The teams compiled many validation tests to ensure that the software performed as intended.  Such tests comprise checking the astrophysical content of the data, e.g. HR diagrams such as Fig.~\ref{fig:hrdiag-evol}, which helped to point out weaknesses in the codes in certain parameter spaces.  Comparisons with external data allowed the teams to check whether their results are consistent with what is already known in the literature.
It is important to point out, however, that \apsis\ does not do any calibration of its APs to mimic external catalogues.  The external catalogues were merely used as a consistency check.
Once the final input data had been received (six months before operations), further test runs were performed to refine the codes and to adapt parameter settings to the final data.   

A  higher level of validation was performed using a validation database hosted at the ESAC operations center in Madrid.  This allowed the CU8 team to perform many cross-checks between the individual parameters, e.g. Fig.~\ref{fig:gsp_esp_extinction}, and provided important feedback for necessary modifications to the code before the full operations sequence.  It also allowed a statistical check on the full dataset, which then allowed the post-processing codes to be prepared for filtering results and setting archive flags.  

A third level of validation was then performed by the Coordination Unit 9 (CU9) archive team once the operational data had been delivered.   In essence they were the first users of the data, and had an external view of the full results in the archive just as a user would, see \cite{DR3-DPACP-127}.  
Once the CU8 results were final, the validation by CU9 only allowed us to perform minor updates on parameters through post-processing, e.g. removing results for some sources or removing fields from the \gaia\ archive. There are, nonetheless, some issues that are now known that could not be corrected, and these along with some caveats are summarised in the next section.


\section{Caveats and known issues}\label{sec:discussioncaveats}

There are several caveats that the user should be aware of before using the data.  Additionally, a number of issues have been found following extensive validation.
We list both caveats and the main issues known to us at the time of writing, starting with general comments on variability and crowding, followed by a discussion on a module-by-module basis.  
The user should consider these issues when using the APs in \gdr{3}.

\paragraph*{Variability:} Apsis processes the mean BP/RP and RVS spectra, astrometry, and mean magnitudes provided by upstream processing systems.  Therefore, we advise users to consider the variability of their source before deciding whether the APs from Apsis are adapted to their specific science case.   As a concrete example, RR Lyrae stars have large amplitude variability, and a mean spectrum for these stars will not necessarily represent the mean state of such a star.  Additionally, as these spectra vary significantly, the concept of a \teff\ from one mean spectrum does not make astrophysical sense, see e.g. \cite{DR3-DPACP-168}.

\begin{figure*}
\centering
\includegraphics[width=0.98\textwidth]{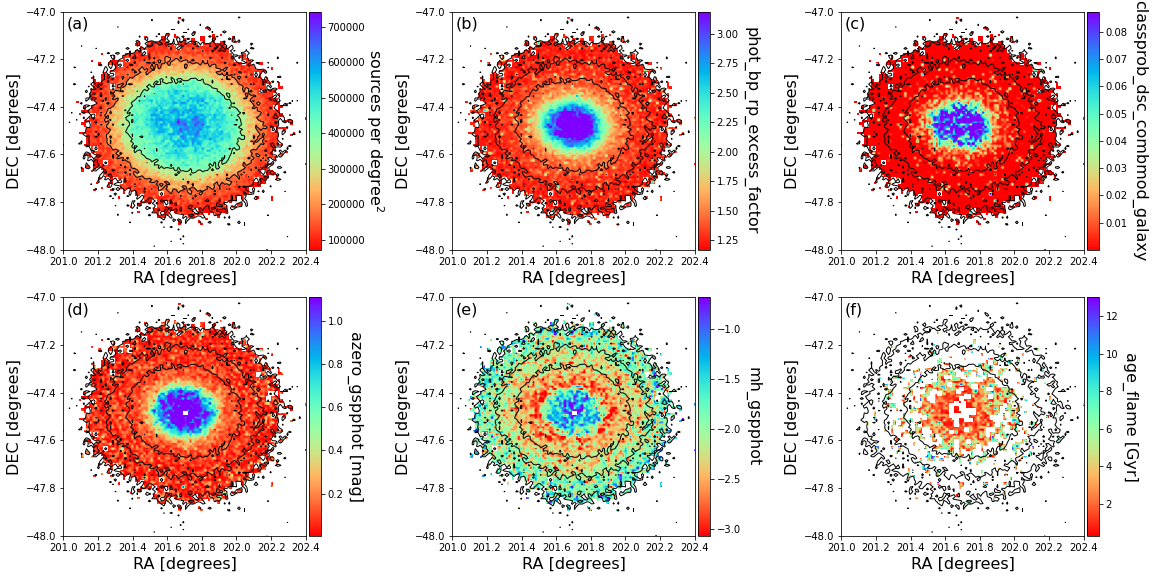}
\caption{Crowding effects in the globular cluster Omega Centauri. Black contours are identical in all panels and indicate source density dropping by factors of 2. Panel a: source density. Panel b: excess factor in photometric flux. Panel c: galaxy class probability from \dsc-Combmod. Panel d: extinction estimate \azero\ from \gspphot. Panel e: metallicity estimate from \gspphot. Panel f: age estimate from \flame.}
\label{fig:crowding-in-Omega-Cen}
\end{figure*}

\paragraph*{Crowding:}
Crowding is a major limitation in dense regions such as stellar clusters. As an example, Fig.~\ref{fig:crowding-in-Omega-Cen} shows that CU8 results differ significantly between the dense core and the outer regions of the globular cluster Omega Centauri \citep{omegacen}. For low-resolution BP/RP spectra, the allocated CCD window is 3.5~arcsec $\times$ 2.1~arcsec \citep{2021A&A...652A..86C}, i.e.\ theoretically $\sim$1.76 million windows would fit into one square degree. In practice though, windows on a source over a range of observation epochs will have quasi-random orientations on the sky and Fig.~\ref{fig:crowding-in-Omega-Cen}a suggests that CCD windows already start to overlap at $\sim$600\,000 sources per square degree, thereby producing blended BP/RP spectra (and photometry, see Fig.~\ref{fig:crowding-in-Omega-Cen}b) that lead to systematically incorrect CU8 results. For RVS spectra, the window size is much larger than for BP/RP spectra \citep[74.2~arcsec $\times$ 1.8~arcsec prior to June 2015, 75.3~arcsec $\times$ 1.8~arcsec after June 2015, see][]{DR2-DPACP-46} but RVS spectra are deblended \citep{DR3-DPACP-154}.

\paragraph*{\dsc:} Performance on white dwarfs and physical binaries is poor, and in general the probabilities may not be well calibrated.
The purity of quasars and galaxies on the full sample is low, but this improves (to $\sim$80\%) when excluding low Galactic latitudes ($|b| < 10^{\circ}$) and using {\tt classlabel\_dsc\_joint}. Note that these purities account for the expected dominance of contaminating stars in a random sample selected from \gaia.

\paragraph*{\gspphot:} Distance estimates tend to be underestimated beyond 2--3 kpc due to a harsh extinction prior. Yet, distances remain reliable for high-quality parallax measurements ($\frac{\varpi}{\sigma_\varpi}>20$) even out to 10kpc. 
{Metallicities show an offset of about --0.2 dex compared to external literature sources with \mh$>-1$ dex, and additional systematics exist below --1.0 dex}. 
We recommend correcting these metallicities using the empirical correction that has been made available to the community, see Appendix.~\ref{ssec:tools-for-community}.  
Also, the uncertainties are known to be underestimated.  This is most probably due to ignoring the off-diagonal elements of the covariance matrix\footnote{Ignoring the covariance matrix could also lead to overestimated uncertainties, but tests with \gspphot\ have shown that the effects are primarily to underestimate them.}.  Another possible explanation could be in the mismatches between model SEDs and observed BP/RP spectra (see Fig. 1 in \citealt{DR3-DPACP-156}). This is the subject of ongoing investigations.
Comparisons with external data show median absolute differences on the order of 120~K for FGK-type stars, 340 K for A stars and 1600 K for B stars, see \linksubsec{sec_cu8par_apsis/ssec_cu8par_apsis_gspphot}{Table~11.19 in the online documentation}.   

\paragraph*{\gspspec:} Quality flags with up to 41 characters have been provided for the best use of the data; users are strongly encouraged to use these for selecting best sample stars. In the case of \mgalgo\ (\aptable\ table), the parameters \logg\ and \mh\ show some biases with respect to the literature, and these depend on \logg.  \cite{DR3-DPACP-186} have proposed a \logg\ and \mh-calibration procedure, to be applied only if the user-specific science case requires it. 
ANN results (\apsupptable) also show biases with respect to the literature, as discussed in \citeauthor{{DR3-DPACP-186}}.

\paragraph*{\msc} uses a solar-metallicity prior for deriving \teff\ and \logg\ of binary components, and \logg\ values are in general overestimated with respect to external catalogues.  Additionally, as \msc\ treats all stars as binaries, one would  need an external catalogue to identify a reliable set of binary stars.
We also note that the value reported as the parameter in the \aptable\ table is the median value of the last 100 values available in the \linktotable{mcmc_samples_msc} table, while
the 16$^{\rm th}$ and 84$^{\rm th}$ percentiles come from the full MCMC chain from the processing.

\paragraph*{\flame} uses a solar-metallicity prior for deriving masses, ages, and evolutionary stage, therefore for stars with known metallicities $< -0.5$ the age should be used with caution.  
The uncertainties in the \flame\ masses and ages are also underestimated, mainly due to the underestimated uncertainties in \teff.
For the use of the masses of giants i.e. with the first digit of \fieldName{flags_flame(_spec)} = 1, the published results should only be used within the range 1--2 \Msun (approximately 14 million of 27 million sources).

\paragraph*{\esphs} uses a solar-metallicity prior to derive spectroscopic parameters of hot stars. The validity of this assumption should be considered in the context of the user-specific science case.

\paragraph*{\espucd} uses empirical training data for the prediction of \teff, so sources deviating significantly from the median metallicities or gravities of the training set, solar metallicities and $\logg\approx5-5.5$,  may have biased estimates of \teff. 
Also, the list of UCD candidates in the quality class 2 contains some contaminants due to incorrect astrometry. This is visible in the distribution of UCD candidates on the celestial sphere as overdensities in the Galactic disk plane (see \linksubsec{sec_cu8par_apsis/ssec_cu8par_apsis_espucd}{Sect.~11.3.10 of the online documentation}). Finally, comparisons with effective temperatures of limited samples in the literature show a systematic difference in the sense that \espucd estimates are $\sim 65$ K lower than the literature values for the hot end of the sample \cite[$T_{\rm eff} > 2300$ K; see ][]{DR3-DPACP-123}.

\paragraph*{\qsoc} aims for completeness, not purity, and accordingly processed a large fraction of stars. The prediction of the redshifts of $0.9 < z < 1.3$ quasars are complicated by the sole detectable presence of the \ion{Mg}{ii} emission lines in the BP/RP spectra over this redshift range. \qsoc is designed to process Type-I/core-dominated quasars with broad emission lines in the optical and accordingly yields only poor predictions on galaxies, type-II AGN, and BL Lacertae/Blazar objects.  Use of the flags is also encouraged.  

\paragraph*{\ugc:} Some contamination by high redshift galaxies and high redshift quasars is expected in the sample.    There is also a suspicious peak of sources in the  $0.0707<z<0.0709$ range.

\paragraph*{\tge:} While the \tge\ extinction maps show excellent agreement with comparable extinction maps, there is a small bias at very small extinctions (\azero < 0.1 mag), and possibly at large extinctions (\azero > 4 mag).  It is not advisable to use the \tge\ maps at very low ($|b| < 5\deg$) Galactic latitudes,  see \citet{DR3-DPACP-158} for further details.


\section{Conclusions}\label{sec:conclusions}

\gdr{3} contains one of the most extensive catalogues of astrophysical parameters to be exploited by the community, and it is based on \gaia-only data.  It contains valuable information on stellar and non-stellar sources, and these parameters appear in ten main archive tables.  A minor subset of APs also appear in {\tt gaia\_source} to simplify querying for a new user of ADQL.
There are up to 1.6 billion classifications of objects (star, galaxy, ...), along with 470 million stellar-based APs and 6 million extragalactic redshifts of quasar (6M) and galaxy (2M) candidates, along with a self-organising map of outliers, total Galactic extinction maps and MCMC samples.  All of these were produced by the \apsis\ analysis system (Sect.~\ref{ssec:apsismodules}).

Stellar-based analyses were performed using {\it general} methods by the \apsis\ modules \gspphot\ and \gspspec\ which derived the spectroscopic parameters (\teff, \logg, ...) using  the BP/RP and RVS spectra, respectively, assuming each source to be a single star. The \msc\ module also analysed the BP/RP spectra in a {\it general} way but assumed the BP/RP spectrum of each source to be a composition of  two unresolved components of a physical binary. The \flame\ module derived evolutionary parameters (\radius, \mass, ...) using the spectroscopic parameters from \gspphot\ and \gspspec together with distance measures and photometry.   
Furthermore, specialised modules produced results focussing on specific types of stars:  \espcs, \esphs, \espucd, and \espels\ analysed cool active stars, hot stars, ultra cool dwarfs, and emission-line stars, respectively to provide class probabilities and labels of emission-line stars, activity index, \teff, \vsini, H-$\alpha$ equivalent widths, and spectral types.    All of the parameters are found in the main \aptable\ table, with supplementary results from \gspphot, \gspspec, and \flame\ in the \apsupptable\ table.  Samples of the MCMC chains are also available for \gspphot\ and \msc, see Appendix~\ref{ssec:mcmcchains}.

Redshifts of galaxies and quasars are available in the \linktogaltable{qso_candidates} and \linktogaltable{galaxy_candidates} tables.  
Two-dimensional total Galactic extinction maps are provided, which are based on the extinction tracers provided by \gspphot, at four different HEALPix levels and an optimal one.  The results of an unsupervised analysis of outliers are found in the  \linktotable{oa_neuron_information} and 
\linktotable{oa_neuron_xp_spectra} 
tables. 

As with any extensive catalogue analysing all sources in a homogenous way, there are both high quality as well as low quality results.  Caveats and known issues with the catalogue are summarised in Sect.~\ref{sec:discussioncaveats}.  Further information 
on the quality, validation and use of the data is provided in the accompanying Papers II and III which target the stellar parameters and the non-stellar content, respectively \citep{dr3-dpacp-160,DR3-DPACP-158}.   
More technical details are available \linksec{}{in the online documentation}.  
Dedicated papers focusing on specific \apsis\ modules are also available, for \gspphot\ \citep{DR3-DPACP-156}, \gspspec\ \citep{DR3-DPACP-186}, \espcs\ \citep{DR3-DPACP-175} and 
\qsoc\ \citep{2018MNRAS.473.1785D}. 
In addition to these data, several tools are made available to the community to aid their exploitation, see Appendix~\ref{sec:cu8tools}.

Examples of the use and performance of the astrophysical parameters are shown in several papers accompanying \gdr{3}.  
\cite{DR3-DPACP-101} explores the extra-galactic content of \gaia, by combining results from Apsis with other \gaia\ results.
\cite{DR3-DPACP-75} exploits the Apsis results along with kinematics and known variable sources to map the spiral arms of the Milky Way. 
\cite{DR3-DPACP-123} focusses on several regions of the HR diagram to provide golden samples of APs, including a focus on Carbon stars and solar-analogues.
\cite{DR3-DPACP-104}  illustrates the chemo-dynamical analysis of disc and halo populations using the chemical analysis from \apsis, and 
\cite{DR3-DPACP-144}  traces the spatial structure of the Galactic interstellar medium using the $\lambda$862~nm DIB.

\gdr{4} (2025) will be based on the analysis of 66 months of \gaia\ data which is almost twice as long as the data in the current data release.  This fact along with the application of improved data processing methods will ensure a significant improvement in the quality of the astrometric, photometric and spectroscopic data.
This, in turn, will impact the quality of the astrophysical parameters with improved control of sources of systematic errors, to deliver an even more extensive catalogue of \gaia-based astrophysical parameters to the community.


\begin{acknowledgements}
We thank the referee for their careful reading of the paper and for providing constructive feedback.
This work presents results from the European Space Agency (ESA) space mission \gaia. \gaia\ data are processed by the \gaia\ Data Processing and Analysis Consortium (DPAC). Funding for the DPAC is provided by national institutions, in particular the institutions participating in the \gaia\ MultiLateral Agreement (MLA). The \gaia\ mission website is \url{https://www.cosmos.esa.int/gaia}. The \gaia\ archive website is \url{https://archives.esac.esa.int/gaia}. Acknowledgments from the financial institutions are in Appendix~\ref{dpac_acknowledgements}. 

We thank our DPAC colleague Hector Canovas for providing Python routines to download spectra and MCMC samples.
    The data analysis made use of
Vaex \citep{Breddels2018},
        TOPCAT \citep{2005ASPC..347...29T},
        and R \citep{RManual}.\\
\end{acknowledgements}

\bibliographystyle{aa}
\bibliography{dpac,additional_references}

\appendix

\section{Empirical training in Apsis modules}\label{ssec:empiricaltraining}

Here we provide details of algorithm training using empirical data for the modules that take this approach, namely \dsc, \ugc, \espucd, \espels, and \msc.

\subsection{\dsc}\label{ssec:empiricaltraining-dsc}

\dsc\ classifies sources into one of five classes:
quasar, galaxy, star, white dwarf, physical binary star.
\dsc\ is trained empirically, meaning it is trained on a labelled subset of the \gaia\ data it is later applied to. We select an external catalogue for each class of objects (e.g.\ quasars), assumed to have high purity, and cross-match this to \gaia\ sources. The BP/RP spectra, photometry, and astrometry of those matched sources (of order $10^5$) are then used as the training data for that class. Once trained, the machine learning models -- ExtraTrees and Gaussian Mixture Models -- map the \gaia\ data to the class probability for each source.

For the quasar and galaxy classes, sources were selected from  spectroscopically-confirmed objects in SDSS catalogues
\citep{2018A&A...613A..51P,2019ApJS..240...23A}. White dwarfs were taken from the \href{http://www.montrealwhitedwarfdatabase.org/tables-and-charts.html}{Montreal White Dwarf Database}\footnote{\url{http://www.montrealwhitedwarfdatabase.org/tables-and-charts.html}}. Binaries were constructed artificially by combining real BP/RP spectra of single stars; hence this class (only) is defined by objects that do not appear explicitly in the \gaia\ data. The final class in \dsc, `star',  is simply a sample of objects drawn at random from \gdr{3} that are not in the other training sets. This class is, therefore, strictly an anonymous class of objects. But as the vast majority of sources in \gaia\ are stars (well over 99\%), and most of these appear single in \gaia\ data, this class is essentially equivalent to apparently single stars.

As usual in empirical training, the selection of the training objects defines the \dsc\ classes. Note, therefore, that the galaxy class does not include all conceivable galaxies, but only that subset of galaxies that both \gaia\ and SDSS observe. Note further that not all types of galaxies have complete data in \gaia\ (many lack parallaxes and proper motions, for example, meaning they cannot be classified by \dsc-Allosmod). \dsc\ is a posterior probabilistic classifier, meaning that its published probabilities depend both on a likelihood model, and on a prior for each class. This prior is set explicitly to reflect their expected global occurrence in \gdr{3}, in particular the rareness of extragalactic objects. Such a prior must be used in any classification project to avoid misleading performance, as explained in more detail in \citet{2019MNRAS.490.5615B}.
More details of the training sets and the \dsc\ models can be found in the \linksec{}{online documentation for \gdr{3}}.

\subsection{\ugc}\label{ssec:training-ugc}
\ugc uses support vector machines (SVM) through the LIBSVM \citep{CC01a} package for the SVM model development, configuration, tuning, training, and testing. The SVM-models for redshift prediction are trained empirically, using a set of BP/RP spectra of unresolved galaxies observed by \gaia\ and sampled by \smsgen. The selection of sources suitable for this was taken from the SDSS DR16 archive \citep{2020ApJS..249....3A,2017AJ....154...28B}. There are $2\,714\,637$ sources with SDSS class ``GALAXY'' that have reliable photometry as well as spectroscopic redshifts. Of these sources, $1\,189\,812$ have been observed by \gaia\ while BP/RP spectra are available for $711\,600$ of them. These spectra, along with the corresponding SDSS redshifts, were used for training and testing the SVM models. Because of the very small number of galaxies with redshifts lower than 0.01 and higher than 0.6, the redshift range for which the SVMs were trained was limited to $0.01\leq{}z\leq{}0.60$, leaving $709\,449$ galaxies, of which $6000$ were selected for the SVM models training set, while the rest were used for performance testing. 

The suitability of the galaxies selected for training has been established via a number of rules and conditions involving various source parameters. The allowed range of magnitudes was limited to $13.0\leq{}\gmag\leq{}21.0$, and the galaxy class probability as estimated by \dsc-Combmod had to be larger than $0.25$. Moreover, the galaxy image size, as defined in SDSS by the Petrosian radius at 50\% of the total flux in the r-band, was restricted to $0.5\arcsec \leq{}petroRad50\_r\leq5.0\arcsec$ so as to avoid both suspiciously compact sources and very extended ones. An interstellar extinction (as defined by the SDSS r-band) upper limit of 0.5  was
also applied, to exclude significantly reddened sources. Finally, it was required that the mean BP/RP spectrum was constructed from at least six epoch spectra, and that the mean fluxes lie within limits $0.3\leq$ \fieldName{bp_flux_mean} $\leq 100$ e$^-$s$^{-1}$ and $0.5\leq$ \fieldName{rp_flux_mean} $\leq 200$ e$^-$s$^{-1}$, thus removing spectra with low signal-to-noise ratios or with suspiciously high signals. Details on the SVM model preparation can be found in \citet{DR3-DPACP-158}.

\subsection{\espucd}\label{ssec:empiricaltraining-espucd}

The effective temperatures inferred by the \espucd\ module come from a Gaussian Process regression model \citep{GPs4ML} trained on the set of \gaia\ RP spectra described below.
Empirical training was adopted for \espucd\ because systematic differences were found between the simulated RP spectra (see Section \ref{ssec:MIOG-simulations} above) obtained for the synthetic library of BT Settl spectra \citep{2013MmSAI..84.1053A} and the observed RP spectra of ultracool dwarfs in the \gaia\ UltraCool Dwarf Sample \citep[hereafter GUCDS]{2017MNRAS.469..401S,2019MNRAS.485.4423S}.      
The empirical training set is composed of a total of 995 observed \gaia\ RP spectra. The first 36 are GUCDS sources defined as spectral standards and their effective temperatures are derived from the calibration by \cite{2009ApJ...702..154S}. We then added 282 sources with temperature determinations from high resolution spectra \citep{2018A&A...615A...6P} or derived using interferometric radii \citep[see e.g.][]{2019MNRAS.484.2674R,2014AJ....147...94D}, most (but not all) of them are hotter than the UCD regime but included to allow for extrapolation. These 36+282 examples with effective temperature estimates were not enough and did not cover homogeneously the range of effective temperatures of the UCD regime. In order to complete the training set, we added 679 sources selected from the \gaia\ Catalogue of Nearby Stars \citep[GCNS][]{EDR3-DPACP-81}, lying outside the high source density regions of the sky (Galactic bulge and disk, and the Magellanic Clouds), with $\gmag<19$\,mag and $G-\rpmag < 1.8$\,mag. Their RP spectra were visually inspected to confirm their UCD nature. Since these 679 sources did not have effective temperature estimates, we had to infer them -- as described below -- in order to use them as part of the training set. 

\begin{figure}
    \centering
    \includegraphics[width=0.47\textwidth]{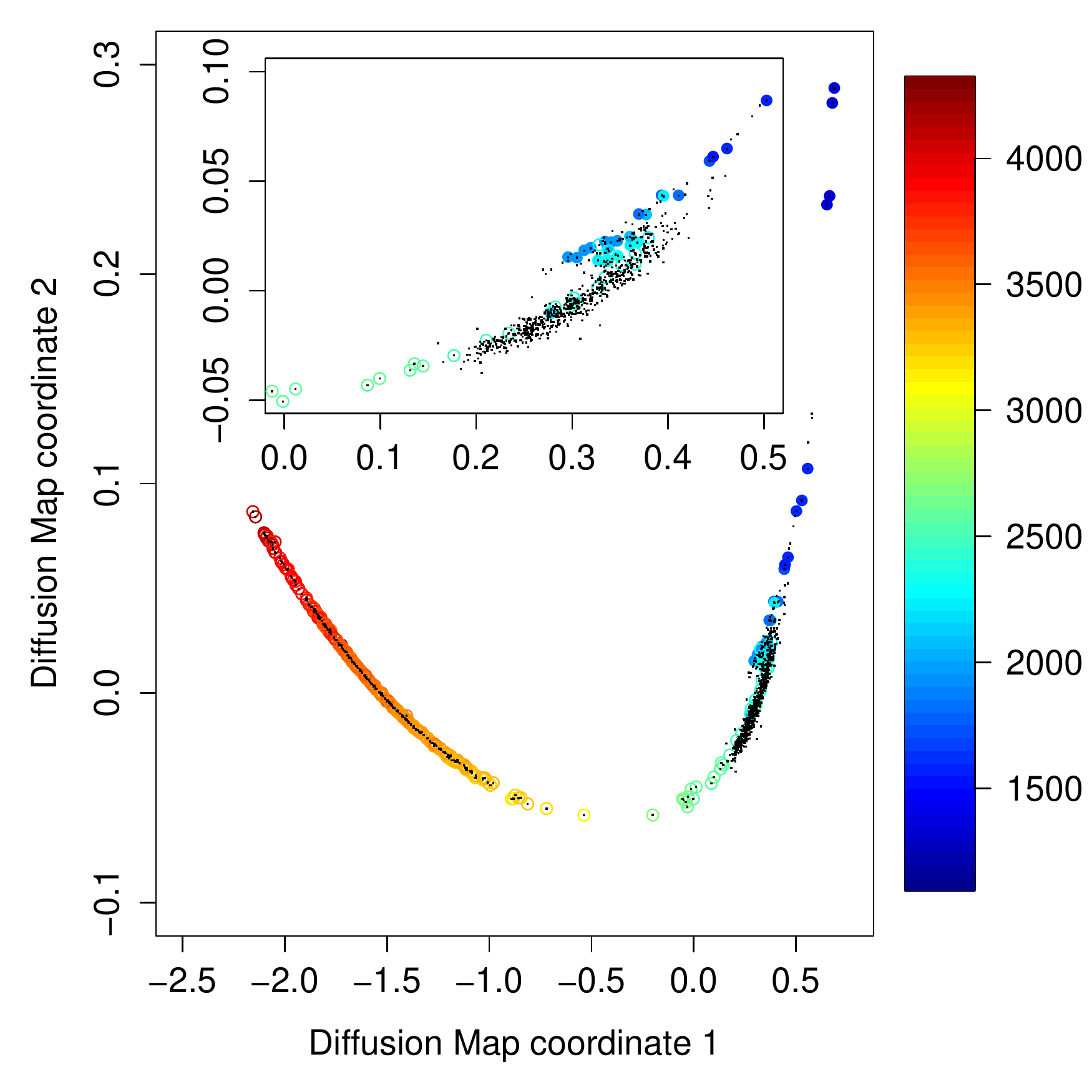}
    \caption{Projection onto the first two diffusion map coordinates of the set of 995 spectra defined in the text. The colour code reflects the temperatures assigned according to their spectral types (filled circles) or literature values (empty circles). Black dots represent the entire set of sources used to construct the Diffusion Map. The inset plot is a zoom of the problematic zone around 2000 K where temperature variations causes only very subtle changes in the RP spectrum.}  
    \label{fig:espucd-empirical-DM}
\end{figure}

The data set then consists of 995 RP spectra with 120 flux values each. The features of these RP spectra were assumed to vary smoothly as a function of temperature so that we could use the labelled spectra (those of the 36+282 sources with effective temperatures in the literature) to calibrate the relation and assign temperatures to the unlabelled sources. If this assumption holds, it is justified to add the 679 sources to the training set. Quantifying the relation between the 120 fluxes of each spectrum and the effective temperatures is however a high-dimensional problem.  In order to simplify the task we constructed a Diffusion Map \citep{COIFMAN20065} to reduce the dimensionality of the data set and found that, as hypothesized, the 995 RP spectra trace a curve in the first two Diffusion Map coordinates with a very small scatter around it. The two-dimensional curve shown in Figure \ref{fig:espucd-empirical-DM} represents in practice an ordering of the 995 RP spectra where the position along the curve (non-linearly) parametrizes the temperature as shown by the coloured circles. 

\begin{figure}
    \centering
    \includegraphics[width=0.45\textwidth]{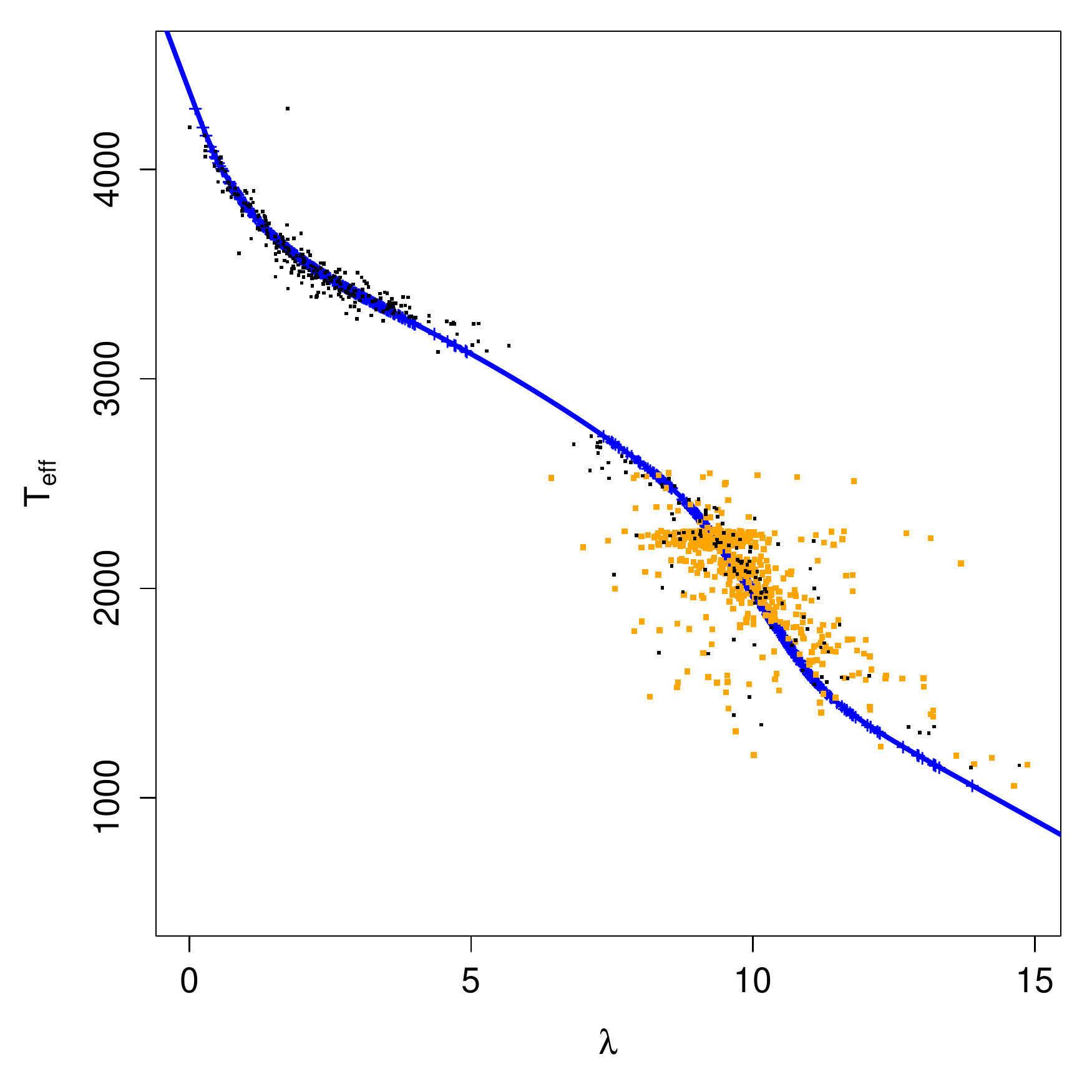}
    \caption{Calibration in temperature of the position of sources along the principal curve fit in the 3D space comprised of the two first Diffusion Map coordinates and the absolute magnitude, \mg. The position along the curve is parametrized with $\lambda$ and artificial scatter is added to help visualize the number of sources used. Black points represent sources with effective temperatures derived from high-resolution spectra and GUCDS standards while orange points represent the rest of GUCDS sources.} 
    \label{fig:espucd-empirical-lambda-teff}
\end{figure}

We used the position of the sources with effective temperature estimates from the literature to calibrate the relation between the Diffusion Map coordinates and temperature. This was done by fitting a Principal Curve \citep{10.2307/2289936} to the first two Diffusion Map coordinates and the $\gmag+5 \log_{10}(\varpi)+5$ (=\mg) values of the 995 sources. The third coordinate was introduced to avoid the non-monotonicity of the curve in the Diffusion Map coordinates. The Principal Curve represents the minimum scatter maximum-likelihood fit and implicitly defines a parameter $\lambda$ along it.  We then calibrated the relation between the curve parameter $\lambda$ and the effective temperature using the labelled examples and a spline regression model. The resulting calibration is shown in Figure \ref{fig:espucd-empirical-lambda-teff}. Finally, we used this regression model to infer effective temperatures for the set of 679 sources which were then included in the training set. 

\subsection{\espels}\label{sect:training.espels}

Three Random Forest classifiers were used during the processing of \espels: ELSRFC1, ELSRFC2, and ELSRFC3.\par  

\noindent
\paragraph{ELSRFC1} provides a spectral type tag to each star in order to avoid processing carbon stars and to allow \esphs\ to preselect O-, B- and A-type stars. It was trained on the MARCS, OB, A, and BTSettl synthetic spectra, see Sect.~\ref{ssec:synthetic-spectra}) with spectral type tags: ``O'', ``B'', ``A'', ``F'', ``G'', ``K'', ``M'' (see Table\,\ref{tab:espels_elsrfc1}) and on the observed BP/RP spectra of the Galactic carbon N stars compiled by \citet{2020A&A...633A.135A} with spectral type tag ``CSTAR''. The wavelength domain considered during the training varied from 340 to 600~nm in BP, and from 640 to 850~nm in RP. Both passbands were normalized individually to their respective integrated flux, while their colour indices (\bpmag$-$\gmag) and (\gmag$-$\rpmag) were added to the flux arrays.

\begin{table}[!htp]
\caption{Adopted effective temperature range covered by the spectral type tag assigned to each synthetic spectrum during the training of ELSRFC1.}\label{tab:espels_elsrfc1}
\begin{tabular}{c|rcr}
 tag & \multicolumn{3}{c}{\teff\ range [K]} \\
 \hline
 O & &$>$& 30\,000  \\
 B & 9\,790 &-& 30\,000 \\
 A & 7\,300 &-& 9\,790 \\
 F & 5\,940 &-& 7\,300 \\
 G & 5\,150 &-& 5\,940 \\
 K & 3\,840 &-& 5\,150 \\
 M & &$\le$& 3\,840
\end{tabular}
\end{table}

\noindent
\paragraph{ELSRFC2} identifies those BP/RP spectra belonging to the Wolf-Rayet stars, WC and WN, and planetary nebula, PN. To these three classes we added the tag ``unknown'' to be given to all targets not showing features expected in Wolf Rayet stars or planetary nebulae. The classification is based on the flux measured in various wavelength domains extracted from the BP/RP spectra and normalized at their edges, as well as on the astrophysical parameters (i.e. \teff, \logg, and \azero). These wavelength ranges were selected following the line features generally expected to be seen in emission in various classes of ELS (see Fig.\,\ref{fig:els_features}). 
Before the extraction and normalisation of the features, the spectra are divided by the instrument response provided by the {\tt MIOG} simulator. For the training, we adopted the available MARCS, OB, A, and BTSettl synthetic libraries as  representative of the spectra of non-emission line stars. We further added the observed BP/RP spectra of Be, Herbig Ae/Be, T Tauri, and dMe stars (i.e.\ as a representation of targets that are not Wolf-Rayet nor planetary nebulae), and observed BP/RP spectra of WC, WN, and PNe.
The observed data of targets with known stellar classification were carefully inspected to only keep those spectra with striking and unambiguous emission features. The number of targets finally considered is given in the online documentation.

\noindent
\paragraph{ELSRFC3} only processed targets with previously detected H$\alpha$ emission. It was trained on the same features as ELSRFC2, but this time only extracted and normalized for the reference Be, Herbig Ae/Be, T Tauri, and active M dwarf stars. To the spectroscopic information, we further also added the astrophysical parameters derived by \gspphot\ during the processing (i.e. no filters and correction applied).

\subsection{\msc}\label{ssec:empiricaltraining-msc}

\msc\ uses an ExtraTrees algorithm \citep{Geurts2006} to initialise its MCMC chain. The algorithm was trained on stellar parameters from a wide binary sample \citep{2018MNRAS.480.4884E} for which we artificially summed the BP/RP spectra\footnote{The stellar parameters were inferred from \gaia\ parallaxes and photometry using \url{https://github.com/jan-rybizki/isochrone_fitting_example} with PARSEC isochrones \citep{2017ApJ...835...77M} under the assumption of equal age, extinction, metallicity and distance. The extinction was fixed using a 3D extinction map from \citet{2020PASP..132g4501R}.}. The forward model used in the MCMC inference is based on an empirical BP/RP spectra grid (next paragraph).
The distance and extinction prior use data from \citet{2020PASP..132g4501R} and the flux ratio and HR-diagram prior are based on the wide binary sample's parameter distribution.

\msc\ uses a model grid of empirical BP/RP spectra instead of simulated spectra. We thereby circumvent the problem of instrument modelling and the unavoidable mismatch between simulated and real spectra. The grid is a function of \teff, \logg, [M/H], and \azero in the space of absolute BP/RP spectra (i.e.\ flux at 10\,pc distance). We used the ExtraTrees machine learning algorithm \citep{Geurts2006} on data from a sample of 80\,000 APOGEE \citep{2015AJ....150..148H} stars for which distance and extinction estimates are available from StarHorse \citep{2020A&A...638A..76Q} and 
\teff, \logg, and [M/H]
estimates from the ASPCAP pipeline \citep{2020AJ....160..120J} crossmatched with \gaia\ BP/RP spectra. 

\section{Accessing MCMC chains for \gspphot\ and \msc}\label{ssec:mcmcchains}

MCMC chains are provided for \gspphot\ and \msc\ through the Gaia Archive DataLink. There are two methods to access them.

First, for any ADQL query results, the user can click on the right-most symbol showing two interlocked links. This will open a pop-up window where the user needs to select ``Gaia DR3'' as data release and ``RAW'' as data structure, and the MCMC data available for download will be listed (together with, e.g., BP/RP spectra also available for download). However, this is limited to a maximum of 5000 MCMC chains.

Second, MCMC chains (and also BP/RP or RVS spectra) can be downloaded without the 5000 limit using Python. A tutorial with example Python scripts for downloads can be found here: \url{https://www.cosmos.esa.int/web/gaia-users/archive/datalink-products\#datalink_jntb_get_above_lim}

\section{Accessing outlier results}\label{ssec:exploitoa_oa} 

The \oa products are essentially contained in three different tables of the archive:
\begin{itemize}
\item \linktotable{oa_neuron_information} table contains non-multidimensional parameters related to the neurons, such as statistical descriptions for different Gaia products (\gmag, \bpmag, \rpmag, \parallax, \bpminrp, etc.), and some quality measurements about the clustering itself. Moreover, a class label is also provided for the best quality neurons.

\item \linktotable{oa_neuron_xp_spectra} table contains multidimensional data related to BP/RP spectrophotometry of the neurons, so that the preprocessed spectra for both prototypes (\linktoapparam{oa_neuron_xp_spectra}{xp_spectrum_prototype_flux}) and templates (if available, \linktoapparam{oa_neuron_xp_spectra}{xp_spectrum_template_flux}) for each neuron can be retrieved.

\item \linktotable{astrophysical_parameters} table contains information about the correspondence between the sources and the neurons, among other results produced by many different modules from \apsis. Regarding \oa parameters, these are the identification of the neuron within which a source lies (if it was processed by \oa, \linktoapparam{astrophysical_parameters}{neuron_oa_id}), distance between the source BP/RP spectra and the neuron prototype (\linktoapparam{astrophysical_parameters}{neuron_oa_dist}, and  \linktoapparam{astrophysical_parameters}{neuron_oa_dist_percentile_rank}). Additionally, it also provides a processing flag (\linktoapparam{astrophysical_parameters}{flags_oa}).
\end{itemize}

The following examples can guide the interested users to access \oa results in the Gaia Archive:

\begin{enumerate}
    \item Retrieve the identifications of all the sources that belong to a certain neuron. In this example the first neuron (located at $(0,0)$, and identified by $\linktoapparam{oa_neuron_information}{neuron_id} = 202105281205440000$) is used.
    
    {\bf Query:}\\
\texttt{ SELECT a.source\_id\\
FROM gaiadr3.oa\_neuron\_information n\\
	JOIN gaiadr3.astrophysical\_parameters a ON n.neuron\_id = a.neuron\_oa\_id\\
WHERE n.neuron\_id = 202105281205440000;\\
}

    \item Retrieve the identifications of all the sources that belong to a SOM neuron that were assigned a specific class label. For this example all galaxy labels were considered. Please, note that this type of query could potentially lead to a huge amount of data being accessed.
    
    {\bf Query:}\\
\texttt{ SELECT a.source\_id\\
FROM gaiadr3.oa\_neuron\_information n\\
	JOIN gaiadr3.astrophysical\_parameters a ON n.neuron\_id = a.neuron\_oa\_id\\
WHERE n.class\_label ILIKE '\%GAL\%';\\
}

    \item  Obtain all the neurons that achieve a certain quality category threshold. In this example, just high quality neurons are retrieved.
    
    {\bf Query:}\\
\texttt{ SELECT *\\
FROM gaiadr3.oa\_neuron\_information\\
WHERE quality\_category < 4;\\
}
\end{enumerate}

\section{Selection function}\label{sec:selectionfunction}
The processing limits and filtering performed during post-processing can lead in some cases to a complicated selection function, see also Fig.~\ref{fig:cmd-stellar}.  We summarise the filters that were imposed on the results for each module in the \apsis\ pipeline in \linksec{sec_cu8par_validation/}{Section 11.4 of the online documentation}.

\section{CU8 tools for the community}\label{sec:cu8tools}
\label{ssec:tools-for-community}

CU8 has made available a number of tools and datasets for the community to accompany \gdr{3} APs.

The datasets are made available on the Gaia cosmos webpage at \url{https://www.cosmos.esa.int/web/gaia/dr3-auxiliary-data}.  
The first of these is a set of \href{https://www.cosmos.esa.int/web/gaia/dr3-astrophysical-parameter-inference}{parameter simulation files}\footnote{\url{https://www.cosmos.esa.int/web/gaia/dr3-astrophysical-parameter-inference}} used in \apsis.  These comprise the full set of parameters that defines the BP/RP spectral simulations of stellar sources, along with extinction in \gmag, \bpmag, and \rpmag\ passbands, and the corresponding $V$ and $E(B-V)$, see Sections~\ref{ssec:synthetic-spectra} and \ref{ssec:extinction} for details.  
We also provide the \href{https://www.cosmos.esa.int/web/gaia/dr3-aps-wavelength-sampling}{wavelength sampling scheme}\footnote{\url{https://www.cosmos.esa.int/web/gaia/dr3-aps-wavelength-sampling}} that was used in \apsis\ processing.  

Information on the tools provided by CU8 can be found on the Gaia DR3 tools pages \url{https://www.cosmos.esa.int/web/gaia/dr3-software-tools}.
These are the following:\\
 -- A \href{https://www.cosmos.esa.int/web/gaia/dr3-oa-self-organising-map-tool}{visualisation tool}\footnote{\url{https://www.cosmos.esa.int/web/gaia/dr3-oa-self-organising-map-tool}} {\it GUASOM flavour DR3}\footnote{\url{https://guasom.citic.udc.es}} has been developed to analyse the outputs produced by \oa\ \citep{Alvarez_2021}, and it allows the user to interactively explore the SOM maps through several visualisations. \\ 
 -- The results from \tge\ have been implemented into the python dustmaps package\footnote{\url{https://dustmaps.readthedocs.io}}, and allows one to retrive the \href{https://www.cosmos.esa.int/web/gaia/dr3-extinction-as-function-of-l-b}{total Galactic extinction based on coordinates}\footnote{\url{https://www.cosmos.esa.int/web/gaia/dr3-extinction-as-function-of-l-b}}. \\
 -- A \href{https://www.cosmos.esa.int/web/gaia/dr3-bolometric-correction-tool}{$G$-band bolometric correction function}\footnote{\url{https://www.cosmos.esa.int/web/gaia/dr3-bolometric-correction-tool}} has been provided in python\footnote{\url{https://gitlab.oca.eu/ordenovic/gaiadr3_bcg}} based on the method that was used in \flame.\\
 -- A \href{https://www.cosmos.esa.int/web/gaia/dr3-gspphot-metallicity-calibration}{\gspphot\ metallicity calibration}\footnote{\url{https://www.cosmos.esa.int/web/gaia/dr3-gspphot-metallicity-calibration}} tool has been made available in python\footnote{\url{https://github.com/mpi-astronomy/gdr3apcal}} based on LAMOST DR6\footnote{\url{http://dr6.lamost.org/v2/catalogue}}\\
 -- A set of tools to \href{https://www.cosmos.esa.int/web/gaia/dr3-extinction-coefficients-in-various-passbands}{compute photometric extinction coefficients}\footnote{\url{https://www.cosmos.esa.int/web/gaia/dr3-extinction-coefficients-in-various-passbands}} has been made available through the \href{https://mfouesneau.github.io/dustapprox/}{{\tt dustapprox}}\footnote{\url{https://mfouesneau.github.io/dustapprox/}} python package.

\section{DPAC Acknowledgements}\label{dpac_acknowledgements}
The \gaia\ mission and data processing have financially been supported by, in alphabetical order by country:
\\ -- the Algerian Centre de Recherche en Astronomie, Astrophysique et G\'{e}ophysique of Bouzareah Observatory;
\\ -- the Austrian Fonds zur F\"{o}rderung der wissenschaftlichen Forschung (FWF) Hertha Firnberg Programme through grants T359, P20046, and P23737;
\\ -- the BELgian federal Science Policy Office (BELSPO) through various PROgramme de D\'{e}veloppement d'Exp\'{e}riences scientifiques (PRODEX)
      grants, the Research Foundation Flanders (Fonds Wetenschappelijk Onderzoek) through grant VS.091.16N,
      the Fonds de la Recherche Scientifique (FNRS), and the Research Council of Katholieke Universiteit (KU) Leuven through
      grant C16/18/005 (Pushing AsteRoseismology to the next level with TESS, GaiA, and the Sloan DIgital Sky SurvEy -- PARADISE);  
\\ -- the Brazil-France exchange programmes Funda\c{c}\~{a}o de Amparo \`{a} Pesquisa do Estado de S\~{a}o Paulo (FAPESP) and Coordena\c{c}\~{a}o de Aperfeicoamento de Pessoal de N\'{\i}vel Superior (CAPES) - Comit\'{e} Fran\c{c}ais d'Evaluation de la Coop\'{e}ration Universitaire et Scientifique avec le Br\'{e}sil (COFECUB);
\\ -- the Chilean Agencia Nacional de Investigaci\'{o}n y Desarrollo (ANID) through Fondo Nacional de Desarrollo Cient\'{\i}fico y Tecnol\'{o}gico (FONDECYT) Regular Project 1210992 (L.~Chemin);
\\ -- the National Natural Science Foundation of China (NSFC) through grants 11573054, 11703065, and 12173069, the China Scholarship Council through grant 201806040200, and the Natural Science Foundation of Shanghai through grant 21ZR1474100;  
\\ -- the Tenure Track Pilot Programme of the Croatian Science Foundation and the \'{E}cole Polytechnique F\'{e}d\'{e}rale de Lausanne and the project TTP-2018-07-1171 `Mining the Variable Sky', with the funds of the Croatian-Swiss Research Programme;
\\ -- the Czech-Republic Ministry of Education, Youth, and Sports through grant LG 15010 and INTER-EXCELLENCE grant LTAUSA18093, and the Czech Space Office through ESA PECS contract 98058;
\\ -- the Danish Ministry of Science;
\\ -- the Estonian Ministry of Education and Research through grant IUT40-1;
\\ -- the European Commission’s Sixth Framework Programme through the European Leadership in Space Astrometry (\href{https://www.cosmos.esa.int/web/gaia/elsa-rtn-programme}{ELSA}) Marie Curie Research Training Network (MRTN-CT-2006-033481), through Marie Curie project PIOF-GA-2009-255267 (Space AsteroSeismology \& RR Lyrae stars, SAS-RRL), and through a Marie Curie Transfer-of-Knowledge (ToK) fellowship (MTKD-CT-2004-014188); the European Commission's Seventh Framework Programme through grant FP7-606740 (FP7-SPACE-2013-1) for the \gaia\ European Network for Improved data User Services (\href{https://gaia.ub.edu/twiki/do/view/GENIUS/}{GENIUS}) and through grant 264895 for the \gaia\ Research for European Astronomy Training (\href{https://www.cosmos.esa.int/web/gaia/great-programme}{GREAT-ITN}) network;
\\ -- the European Cooperation in Science and Technology (COST) through COST Action CA18104 `Revealing the Milky Way with \gaia (MW-Gaia)';
\\ -- the European Research Council (ERC) through grants 320360, 647208, and 834148 and through the European Union’s Horizon 2020 research and innovation and excellent science programmes through Marie Sk{\l}odowska-Curie grant 745617 (Our Galaxy at full HD -- Gal-HD) and 895174 (The build-up and fate of self-gravitating systems in the Universe) as well as grants 687378 (Small Bodies: Near and Far), 682115 (Using the Magellanic Clouds to Understand the Interaction of Galaxies), 695099 (A sub-percent distance scale from binaries and Cepheids -- CepBin), 716155 (Structured ACCREtion Disks -- SACCRED), 951549 (Sub-percent calibration of the extragalactic distance scale in the era of big surveys -- UniverScale), and 101004214 (Innovative Scientific Data Exploration and Exploitation Applications for Space Sciences -- EXPLORE);
\\ -- the European Science Foundation (ESF), in the framework of the \gaia\ Research for European Astronomy Training Research Network Programme (\href{https://www.cosmos.esa.int/web/gaia/great-programme}{GREAT-ESF});
\\ -- the European Space Agency (ESA) in the framework of the \gaia\ project, through the Plan for European Cooperating States (PECS) programme through contracts C98090 and 4000106398/12/NL/KML for Hungary, through contract 4000115263/15/NL/IB for Germany, and through PROgramme de D\'{e}veloppement d'Exp\'{e}riences scientifiques (PRODEX) grant 4000127986 for Slovenia;  
\\ -- the Academy of Finland through grants 299543, 307157, 325805, 328654, 336546, and 345115 and the Magnus Ehrnrooth Foundation;
\\ -- the French Centre National d’\'{E}tudes Spatiales (CNES), the Agence Nationale de la Recherche (ANR) through grant ANR-10-IDEX-0001-02 for the `Investissements d'avenir' programme, through grant ANR-15-CE31-0007 for project `Modelling the Milky Way in the \gaia era’ (MOD4Gaia), through grant ANR-14-CE33-0014-01 for project `The Milky Way disc formation in the \gaia era’ (ARCHEOGAL), through grant ANR-15-CE31-0012-01 for project `Unlocking the potential of Cepheids as primary distance calibrators’ (UnlockCepheids), through grant ANR-19-CE31-0017 for project `Secular evolution of galxies' (SEGAL), and through grant ANR-18-CE31-0006 for project `Galactic Dark Matter' (GaDaMa), the Centre National de la Recherche Scientifique (CNRS) and its SNO \gaia of the Institut des Sciences de l’Univers (INSU), its Programmes Nationaux: Cosmologie et Galaxies (PNCG), Gravitation R\'{e}f\'{e}rences Astronomie M\'{e}trologie (PNGRAM), Plan\'{e}tologie (PNP), Physique et Chimie du Milieu Interstellaire (PCMI), and Physique Stellaire (PNPS), the `Action F\'{e}d\'{e}ratrice \gaia' of the Observatoire de Paris, the R\'{e}gion de Franche-Comt\'{e}, the Institut National Polytechnique (INP) and the Institut National de Physique nucl\'{e}aire et de Physique des Particules (IN2P3) co-funded by CNES;
\\ -- the German Aerospace Agency (Deutsches Zentrum f\"{u}r Luft- und Raumfahrt e.V., DLR) through grants 50QG0501, 50QG0601, 50QG0602, 50QG0701, 50QG0901, 50QG1001, 50QG1101, 50\-QG1401, 50QG1402, 50QG1403, 50QG1404, 50QG1904, 50QG2101, 50QG2102, and 50QG2202, and the Centre for Information Services and High Performance Computing (ZIH) at the Technische Universit\"{a}t Dresden for generous allocations of computer time;
\\ -- the Hungarian Academy of Sciences through the Lend\"{u}let Programme grants LP2014-17 and LP2018-7 and the Hungarian National Research, Development, and Innovation Office (NKFIH) through grant KKP-137523 (`SeismoLab');
\\ -- the Science Foundation Ireland (SFI) through a Royal Society - SFI University Research Fellowship (M.~Fraser);
\\ -- the Israel Ministry of Science and Technology through grant 3-18143 and the Tel Aviv University Center for Artificial Intelligence and Data Science (TAD) through a grant;
\\ -- the Agenzia Spaziale Italiana (ASI) through contracts I/037/08/0, I/058/10/0, 2014-025-R.0, 2014-025-R.1.2015, and 2018-24-HH.0 to the Italian Istituto Nazionale di Astrofisica (INAF), contract 2014-049-R.0/1/2 to INAF for the Space Science Data Centre (SSDC, formerly known as the ASI Science Data Center, ASDC), contracts I/008/10/0, 2013/030/I.0, 2013-030-I.0.1-2015, and 2016-17-I.0 to the Aerospace Logistics Technology Engineering Company (ALTEC S.p.A.), INAF, and the Italian Ministry of Education, University, and Research (Ministero dell'Istruzione, dell'Universit\`{a} e della Ricerca) through the Premiale project `MIning The Cosmos Big Data and Innovative Italian Technology for Frontier Astrophysics and Cosmology' (MITiC);
\\ -- the Netherlands Organisation for Scientific Research (NWO) through grant NWO-M-614.061.414, through a VICI grant (A.~Helmi), and through a Spinoza prize (A.~Helmi), and the Netherlands Research School for Astronomy (NOVA);
\\ -- the Polish National Science Centre through HARMONIA grant 2018/30/M/ST9/00311 and DAINA grant 2017/27/L/ST9/03221 and the Ministry of Science and Higher Education (MNiSW) through grant DIR/WK/2018/12;
\\ -- the Portuguese Funda\c{c}\~{a}o para a Ci\^{e}ncia e a Tecnologia (FCT) through national funds, grants SFRH/\-BD/128840/2017 and PTDC/FIS-AST/30389/2017, and work contract DL 57/2016/CP1364/CT0006, the Fundo Europeu de Desenvolvimento Regional (FEDER) through grant POCI-01-0145-FEDER-030389 and its Programa Operacional Competitividade e Internacionaliza\c{c}\~{a}o (COMPETE2020) through grants UIDB/04434/2020 and UIDP/04434/2020, and the Strategic Programme UIDB/\-00099/2020 for the Centro de Astrof\'{\i}sica e Gravita\c{c}\~{a}o (CENTRA);  
\\ -- the Slovenian Research Agency through grant P1-0188;
\\ -- the Spanish Ministry of Economy (MINECO/FEDER, UE), the Spanish Ministry of Science and Innovation (MICIN), the Spanish Ministry of Education, Culture, and Sports, and the Spanish Government through grants BES-2016-078499, BES-2017-083126, BES-C-2017-0085, ESP2016-80079-C2-1-R, ESP2016-80079-C2-2-R, FPU16/03827, PDC2021-121059-C22, RTI2018-095076-B-C22, and TIN2015-65316-P (`Computaci\'{o}n de Altas Prestaciones VII'), the Juan de la Cierva Incorporaci\'{o}n Programme (FJCI-2015-2671 and IJC2019-04862-I for F.~Anders), the Severo Ochoa Centre of Excellence Programme (SEV2015-0493), and MICIN/AEI/10.13039/501100011033 (and the European Union through European Regional Development Fund `A way of making Europe') through grant RTI2018-095076-B-C21, the Institute of Cosmos Sciences University of Barcelona (ICCUB, Unidad de Excelencia `Mar\'{\i}a de Maeztu’) through grant CEX2019-000918-M, the University of Barcelona's official doctoral programme for the development of an R+D+i project through an Ajuts de Personal Investigador en Formaci\'{o} (APIF) grant, the Spanish Virtual Observatory through project AyA2017-84089, the Galician Regional Government, Xunta de Galicia, through grants ED431B-2021/36, ED481A-2019/155, and ED481A-2021/296, the Centro de Investigaci\'{o}n en Tecnolog\'{\i}as de la Informaci\'{o}n y las Comunicaciones (CITIC), funded by the Xunta de Galicia and the European Union (European Regional Development Fund -- Galicia 2014-2020 Programme), through grant ED431G-2019/01, the Red Espa\~{n}ola de Supercomputaci\'{o}n (RES) computer resources at MareNostrum, the Barcelona Supercomputing Centre - Centro Nacional de Supercomputaci\'{o}n (BSC-CNS) through activities AECT-2017-2-0002, AECT-2017-3-0006, AECT-2018-1-0017, AECT-2018-2-0013, AECT-2018-3-0011, AECT-2019-1-0010, AECT-2019-2-0014, AECT-2019-3-0003, AECT-2020-1-0004, and DATA-2020-1-0010, the Departament d'Innovaci\'{o}, Universitats i Empresa de la Generalitat de Catalunya through grant 2014-SGR-1051 for project `Models de Programaci\'{o} i Entorns d'Execuci\'{o} Parallels' (MPEXPAR), and Ramon y Cajal Fellowship RYC2018-025968-I funded by MICIN/AEI/10.13039/501100011033 and the European Science Foundation (`Investing in your future');
\\ -- the Swedish National Space Agency (SNSA/Rymdstyrelsen);
\\ -- the Swiss State Secretariat for Education, Research, and Innovation through the Swiss Activit\'{e}s Nationales Compl\'{e}mentaires and the Swiss National Science Foundation through an Eccellenza Professorial Fellowship (award PCEFP2\_194638 for R.~Anderson);
\\ -- the United Kingdom Particle Physics and Astronomy Research Council (PPARC), the United Kingdom Science and Technology Facilities Council (STFC), and the United Kingdom Space Agency (UKSA) through the following grants to the University of Bristol, the University of Cambridge, the University of Edinburgh, the University of Leicester, the Mullard Space Sciences Laboratory of University College London, and the United Kingdom Rutherford Appleton Laboratory (RAL): PP/D006511/1, PP/D006546/1, PP/D006570/1, ST/I000852/1, ST/J005045/1, ST/K00056X/1, ST/\-K000209/1, ST/K000756/1, ST/L006561/1, ST/N000595/1, ST/N000641/1, ST/N000978/1, ST/\-N001117/1, ST/S000089/1, ST/S000976/1, ST/S000984/1, ST/S001123/1, ST/S001948/1, ST/\-S001980/1, ST/S002103/1, ST/V000969/1, ST/W002469/1, ST/W002493/1, ST/W002671/1, ST/W002809/1, and EP/V520342/1.

The GBOT programme  uses observations collected at (i) the European Organisation for Astronomical Research in the Southern Hemisphere (ESO) with the VLT Survey Telescope (VST), under ESO programmes
092.B-0165,
093.B-0236,
094.B-0181,
095.B-0046,
096.B-0162,
097.B-0304,
098.B-0030,
099.B-0034,
0100.B-0131,
0101.B-0156,
0102.B-0174, and
0103.B-0165;
and (ii) the Liverpool Telescope, which is operated on the island of La Palma by Liverpool John Moores University in the Spanish Observatorio del Roque de los Muchachos of the Instituto de Astrof\'{\i}sica de Canarias with financial support from the United Kingdom Science and Technology Facilities Council, and (iii) telescopes of the Las Cumbres Observatory Global Telescope Network.

\end{document}